\pdfoutput=1
\documentclass{article} % default is 10 pt
\usepackage{subfigure}
\usepackage{graphicx} % needed for including graphics e.g. EPS, PS
\usepackage{pdfpages} % needed for including multi-page pdf documents
\usepackage{amssymb}
\usepackage{amsmath}
\usepackage{url}
\usepackage{appendix}
\usepackage{booktabs}
\usepackage{carlito}
\usepackage{comment}
\usepackage{float}
\usepackage{tikz}
\usepackage{verbatim}

\makeatletter
\def\@normalsize{\@setsize\normalsize{10pt}\xpt\@xpt
\abovedisplayskip 10pt plus2pt minus5pt\belowdisplayskip 
\abovedisplayskip \abovedisplayshortskip \z@ 
plus3pt\belowdisplayshortskip 6pt plus3pt 
minus3pt\let\@listi\@listI}

\def\subsize{\@setsize\subsize{12pt}\xipt\@xipt}
\def\section{\@startsection {section}{1}{\z@}{1.0ex plus
1ex minus .2ex}{.2ex plus .2ex}{\large\bf}}
\def\subsection{\@startsection 
   {subsection}{2}{\z@}{.2ex plus 1ex} {.2ex plus .2ex}{\subsize\bf}}
\makeatother
\begin{document}
\date{}

\title{\bf Physical and electrical analysis of LSST sensors}

\author{Craig Lage\\
  Department of Physics\\
  University of California - Davis\\
cslage@ucdavis.edu}

\maketitle

\subsection*{\centering Abstract}

% >>>>>>>>>>>>>>>>>>>>>>>>> Keywords and Abstract <<<<<<<<<<<<<<<<<<<<<
% Replace with your own keywords and abstract.  Text will be in italics
{\em Keywords: 
LSST, modeling, camera, CCD, simulation, diffusion, image processing.  
}
\\
Removing systematic effects from astronomical images taken with CCDs requires a detailed understanding of the physics of the imaging process.  To aid in this understanding, we have built detailed electrostatic simulations of the LSST CCDs.  In order to build an electrostatic model of the LSST CCDs, physical information about the CCDs is required.  These details include things such as the physical dimensions of the components of the CCD, dopant profiles, and in some cases, electrical measurements of the CCD.  This work documents the results of these physical and electrical measurements on LSST CCDs.

\section{Introduction}
The Large Synoptic Survey Telescope (LSST) is an innovative, large, fast survey telescope currently under construction at Cerro Pachon in Chile \cite{LSST_2019}.  The digital camera for the LSST, also currently under construction, will consist of approximately 3.2 gigapixels and will be the largest digital camera ever constructed.  The camera uses fully-depleted silicon Charge Coupled Devices (CCDs) which are back illuminated and 100 microns thick in order to optimize quantum efficiency in the near infrared.  The imaging area consists of 189 CCDs, with each CCD containing 16 imaging regions laid out in an 8x2 array.  Each imaging region has a pixel array with approximately 500x2000 10 micron square pixels, giving 16 Megapixels total.  Each imaging region also has its own independent amplifier (\cite{oconnor2019uniformity}, \cite{oconnor_2016}).  The LSST focal plane contains CCDs from two different vendors, the ITL STA3800C from the  University of Arizona Imaging Technology Laboratory \cite{ITL_website}, and the E2V CCD250 from Teledyne E2V \cite{E2V_website}.  

In order to produce high-quality scientific data from the LSST survey, it is important to have a detailed understanding of the CCD detectors.   It is also possible that during the decade long duration of the LSST survey unforeseen problems may occur, and having detailed models of the detectors can aid in diagnosing and solving these problems.  To further these goals, we have built electrostatic models of the CCDs from both CCD vendors.  Construction of these models requires physical information about the CCDs, including things such as the physical dimensions of the components of the CCD, dopant profiles, and electrical properties of the devices which make up the CCD.   This work documents the measurements which have been made to gather this data.

This work is divided into several sections.  The first section gives the results of the physical and electrical analysis of the ITL STA3800C CCD, and the second section gives the results of the physical analysis of the E2V CCD250.  Each section is broken into sub-sections giving the results of the different analysis techniques.  Because of more time and the availability of samples, more detailed electrical analysis was performed on the ITL CCD.  The final section shows the results of applying the measured physical parameters to modeling the brighter-fatter effect on CCDs from both vendors.  

\section{ITL STA3800C}
For the physical analysis of the ITL STA3800C CCD, the vendor kindly supplied an unpackaged CCD die.  For this CCD, there was no need to deprocess a packaged component to do the analysis.  The vendor was also interested in the SIMS analysis, and so provided feedback on the best places to measure, as detailed in Section \ref{ITL_SIMS_Section} and Figure \ref{ITL_SIMS}.

\subsection{Optical Micrographs}
\begin {figure}[H]
  \centering
  \includegraphics[trim=6.0in 3.0in 6.0in 2.0in,clip,width=1.15\textwidth,angle=90]{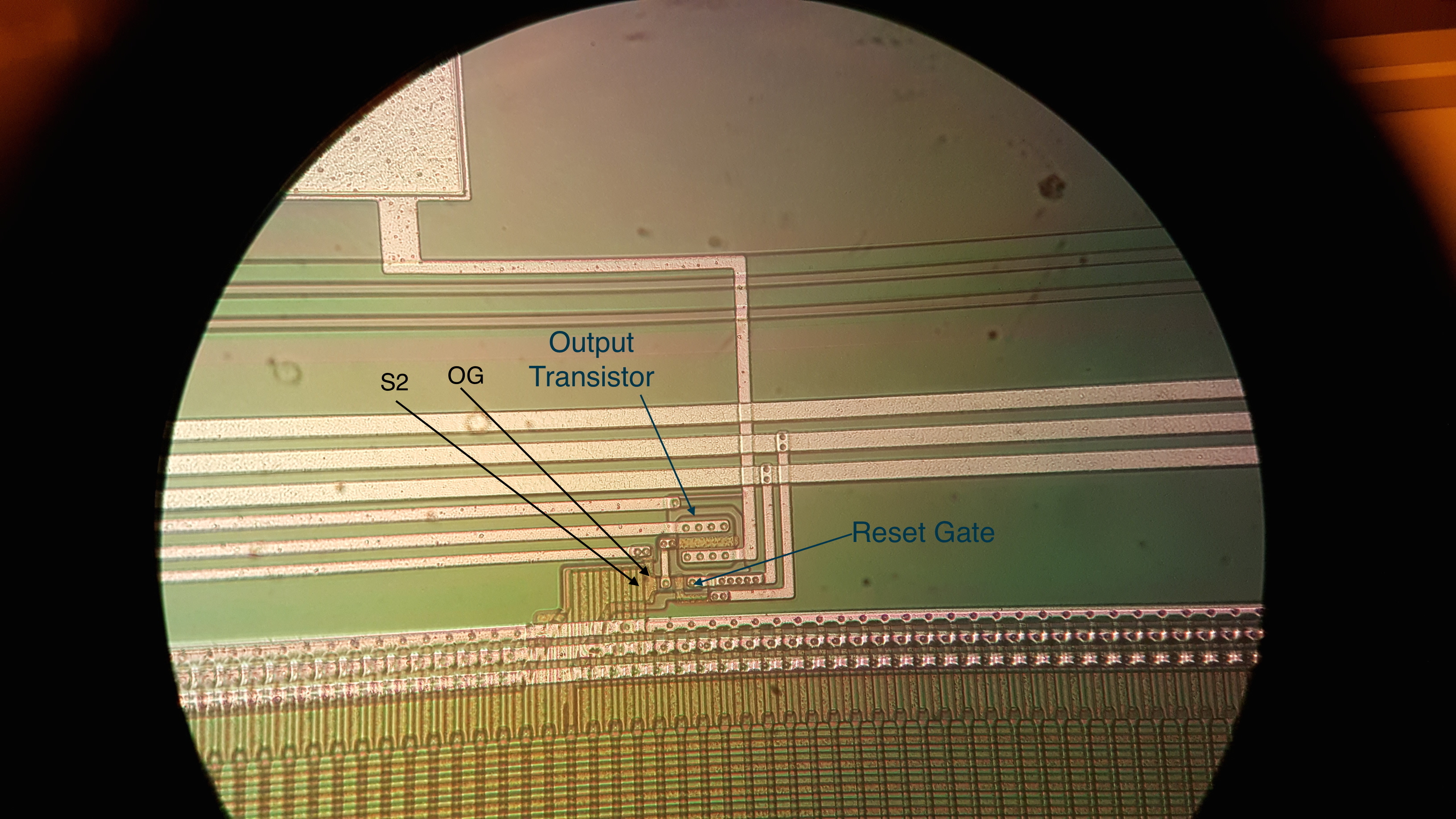}
  \caption{Optical micrograph of the output device chain on the ITL STA3800C device.  This shows the serial chain with the 45 degree bend of the pre-scan pixels, the single-stage output transistor, and the reset gate.}
  \label{ITL_Output}
  % Trim is Left Bottom Right Top
\end{figure}

Optical micrographs were taken of the ITL STA3800C CCD.  Figure \ref{ITL_Output} shows the serial chain as it bends at 45 degrees to escape the regular array and then feeds into the output amplifier. The STA3800C uses a single MOS transistor as the output amplifier, in contrast to the E2V CCD250, which uses a two-stage output amplifier.  To compensate for this difference and make the two CCDs present to the control circuitry in a similar way, the LSST design adds a JFET second stage amplifier to the STA3800C CCDs.  This is positioned on the flex cable which attaches to the CCD, as near to the CCD as practical.  A schematic for this circuitry is shown in Figure \ref{Schem_Net1} and the circuitry is discussed in more detail in Section \ref{SPICE_Model}.

\subsection{SEM Micrographs}
\begin {figure}[H]
	\centering
	\subfigure[b][SEM cross section parallel to the parallel gates]{\includegraphics[trim=0.0in 0.0in 0.0in 2.5in,clip,width=0.70\textwidth,angle=0]{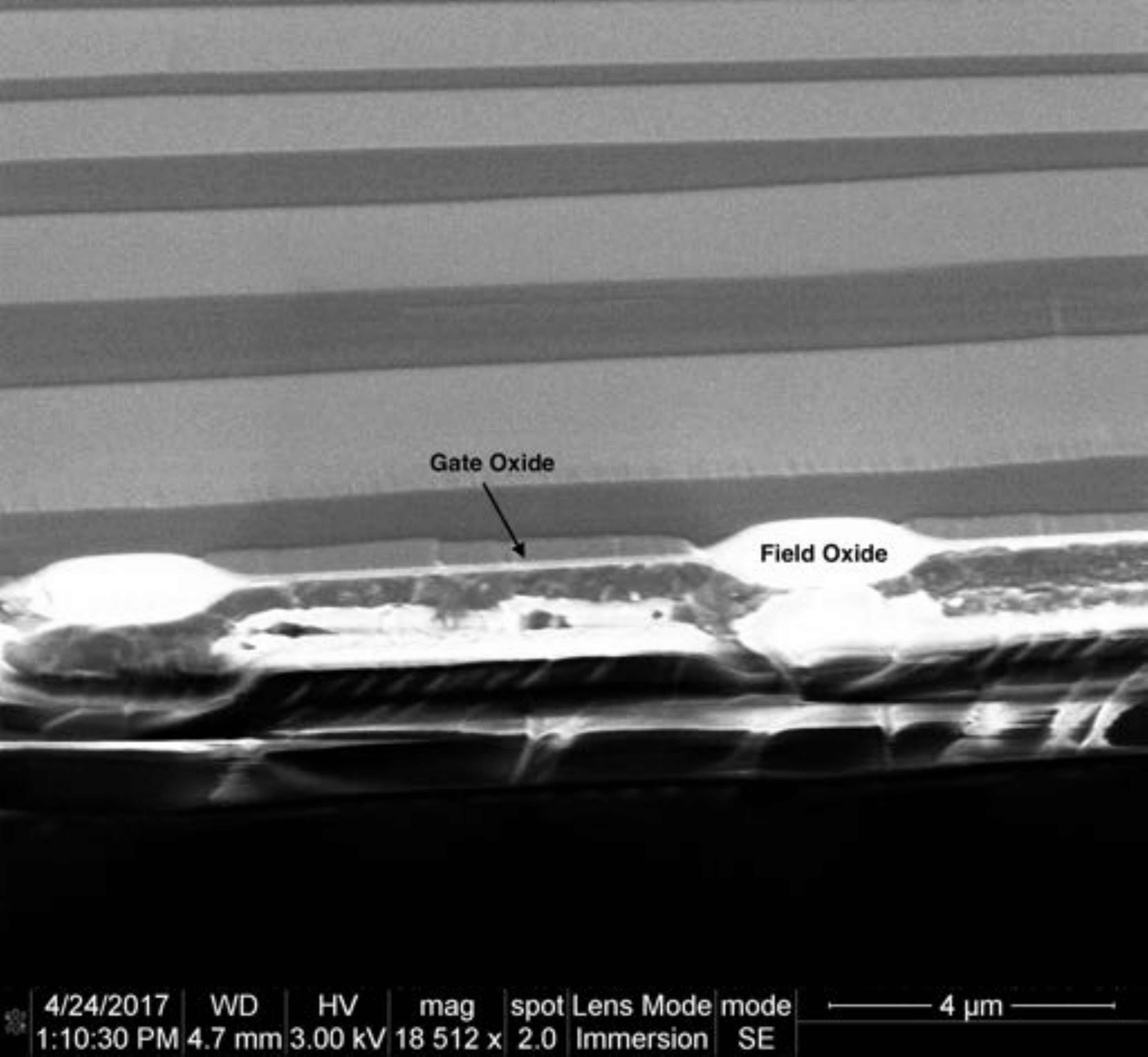}}
	\subfigure[b][SEM Cross-section perpendicular to the parallel gates]{\includegraphics[trim=0.0in 0.0in 0.0in 2.5in,clip,width=0.70\textwidth,angle=0]{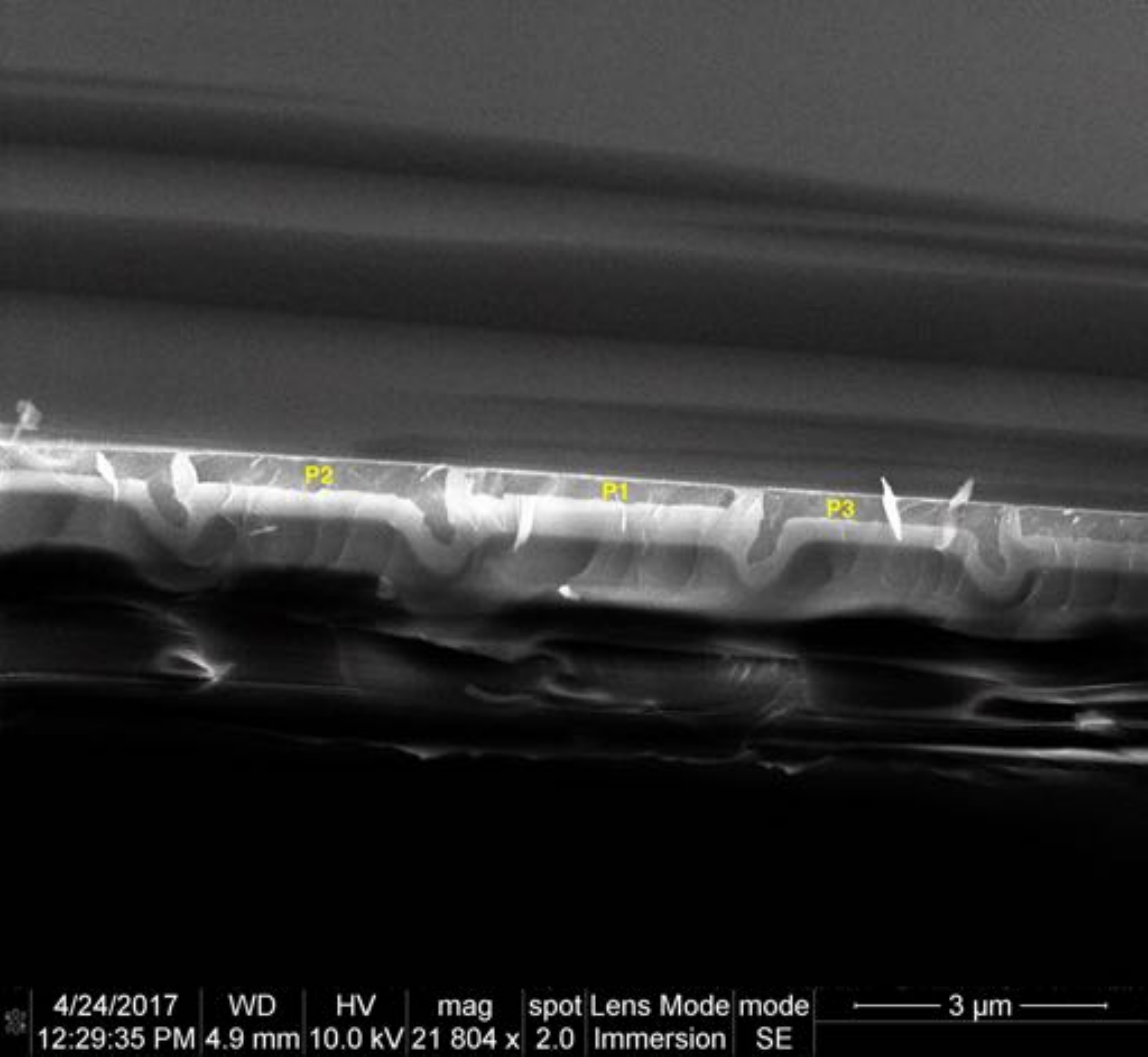}}

  \caption{SEM cross sections of the ITL STA3800C device.  The top image shows the thick field oxide in the channel stop region and the thin gate oxide in the channel region.  the bottom image shows the three overlapping polysilicon layers which make up the parallel gates.}
  \label{ITL_SEMS}
  % Trim is Left Bottom Right Top
\end{figure}

SEM cross-sectional micrographs were taken of the ITL STA3800C CCD in the array region in both directions.  These are shown in Figure \ref{ITL_SEMS}.  Figure \ref{ITL_SEMS} (a) shows a cross section parallel to the parallel gates.  This CCD uses a conventional LOCOS (LOCal Oxidation of Silicon) process to define the channel and channel stop regions.  The thick field oxide region is the channel stop region, and the thin gate oxide region is the channel region where the electrons are stored.  The dopant levels in these regions are also different, as discussed in Section \ref{ITL_SIMS_Section}.  Figure \ref{ITL_SEMS} (b) shows a cross section perpendicular to the parallel gates.  The three levels of polysilicon making up the parallel gate structure can be seen.

\subsection{SIMS Analysis}
\label{ITL_SIMS_Section}
\begin {figure}[H]
	\centering
	\subfigure[b][STA3800C layout showing regions profiled]{\includegraphics[trim=0.0in 0.0in 0.0in 0.0in,clip,width=0.45\textwidth]{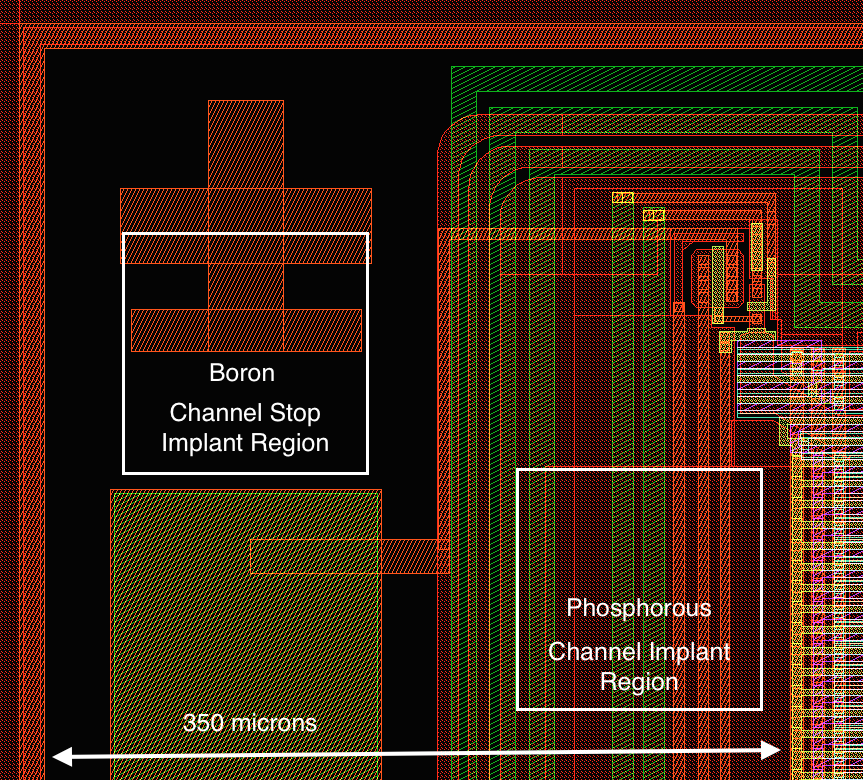}}
	\subfigure[b][Optical micrograph of the regions profiled.]{\includegraphics[trim=0.0in 0.0in 0.0in 0.0in,clip,width=0.54\textwidth]{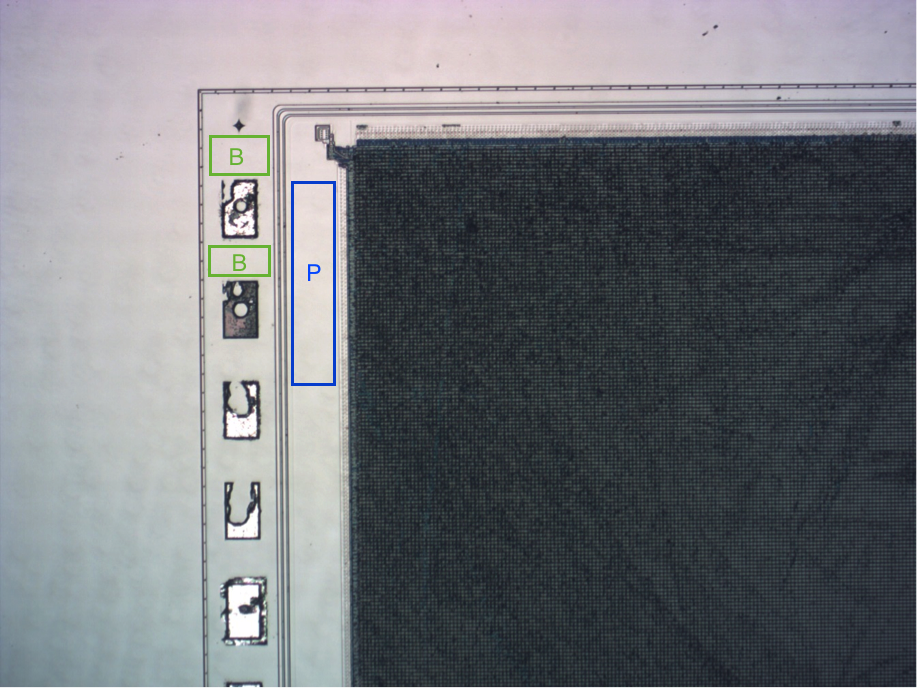}}
	\subfigure[b][Phosphorous channel profile.]{\includegraphics[trim=0.0in 0.0in 0.0in 0.0in,clip,width=0.49\textwidth]{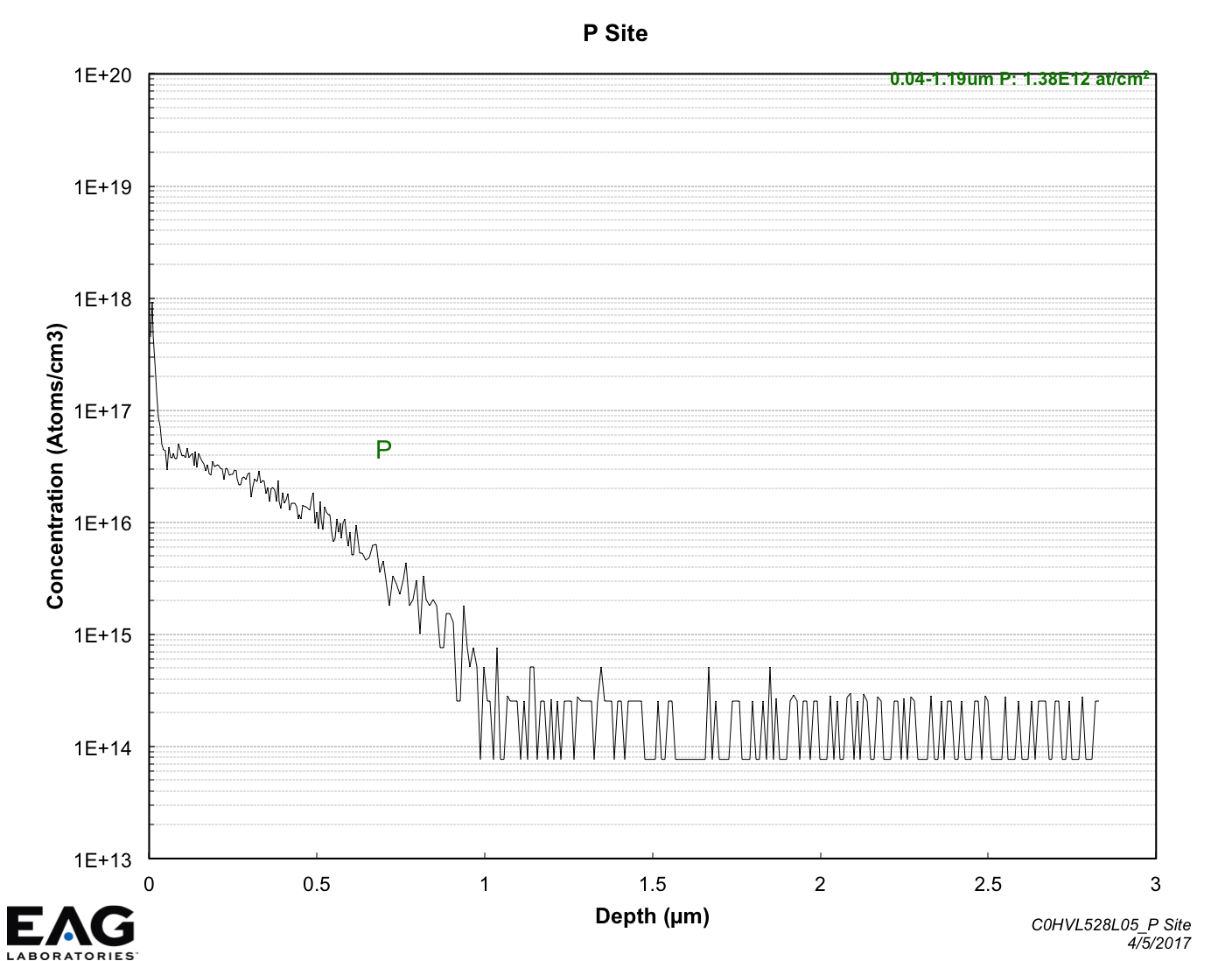}}
	\subfigure[b][Boron channel stop profile.]{\includegraphics[trim=0.0in 0.0in 0.0in 0.0in,clip,width=0.49\textwidth]{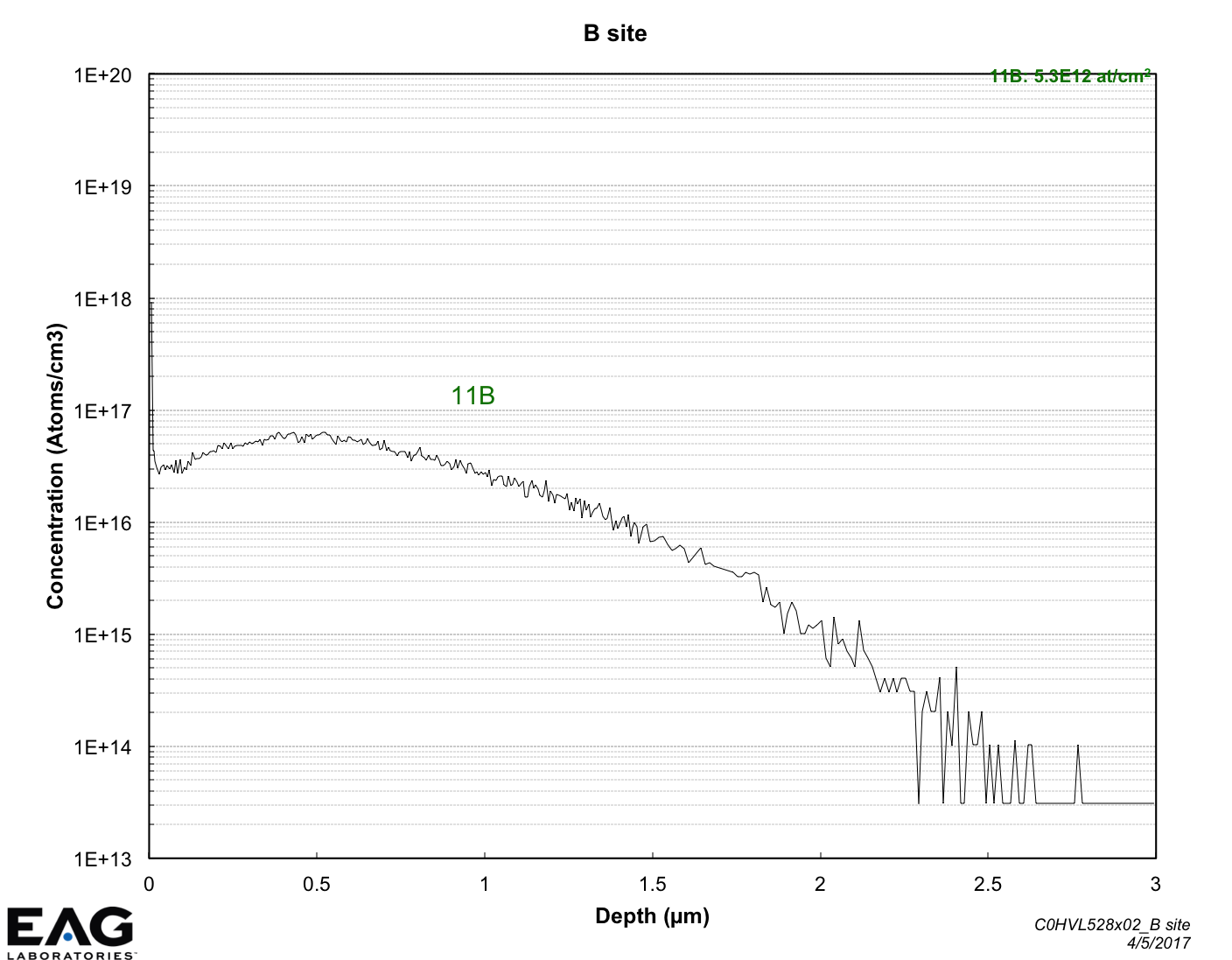}}

  \caption{SIMS dopant profiles of the ITL STA3800C device as measured by EAG laboratories \cite{EAG_website}.  While we are interested primarily in the effect of these implants in the imaging array, there are regions outside the imaging array which receive the same implants and are easier to measure.  The top two panels show these regions on the chip where the measurements were made, based on input from the device vendor.  The bottom two panels show the measured dopant profiles.}
  \label{ITL_SIMS}
  % Trim is Left Bottom Right Top
\end{figure}

Knowing the dopant profiles in the CCD is an important part of developing an electrostatic model.  To measure the dopant profiles, the technique of Secondary Ion Mass Spectrometry (SIMS) was used.  In this technique, a region of silicon is subjected to an accelerated ion beam which removes material from the silicon substrate at a steady rate, and the ions thus removed are sent through a mass spectrometer to determine their atomic type.  This technique is widely used in the semiconductor industry to measure dopant profiles.  We used EAG Laboratories in Sunnyvale, Ca. \cite{EAG_website} to make these measurements.  The results of the measurements are shown in Figure \ref{ITL_SIMS}, and the results of using these measurements to build the electrostatic model are discussed in Section \ref{Poisson_Model}.

\subsection{Electrical Measurements of the Output Transistor}
\begin {figure}[H]
  \centering
  \subfigure[b][Id-Vd characteristics of the output transistor]{\includegraphics[trim=0.0in 0.0in 0.0in 1.0in,clip,width=0.70\textwidth]{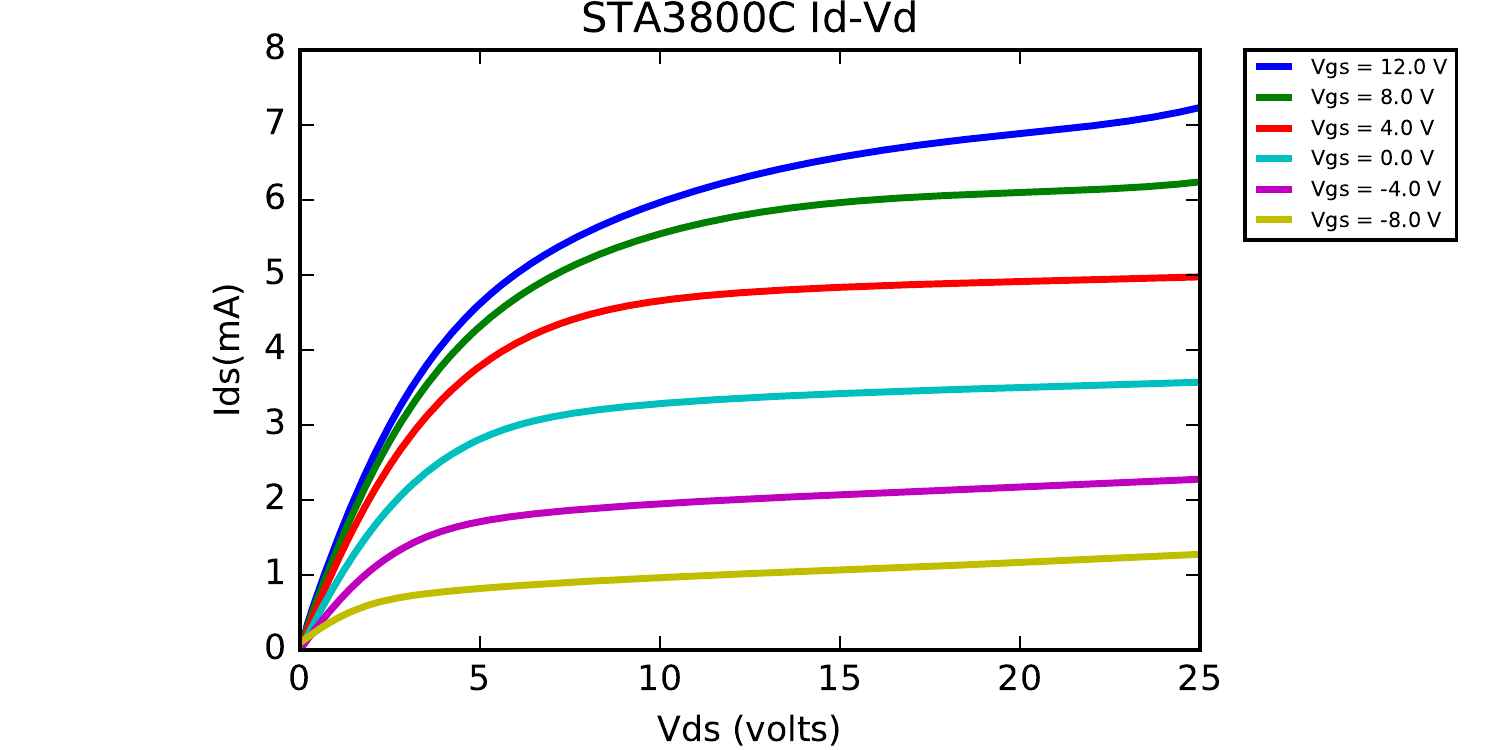}}
  \subfigure[b][Id-Vg characteristics of the output transistor]{\includegraphics[trim=0.0in 0.0in 0.0in 1.0in,clip,width=0.70\textwidth]{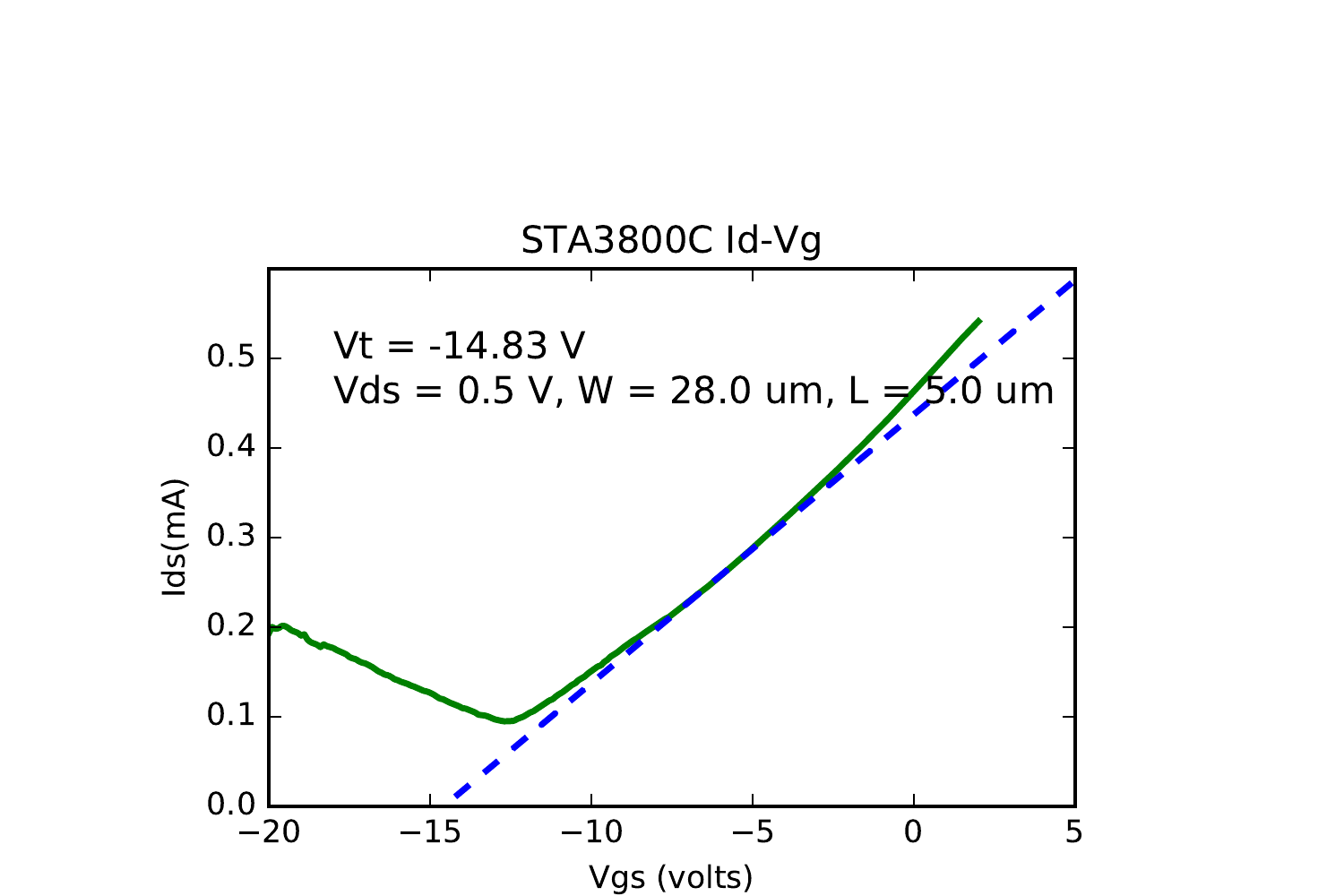}}
  \subfigure[b][Id-Vg characteristics of the output transistor with varying Vbs]{\includegraphics[trim=0.0in 0.0in 0.0in 1.0in,clip,width=0.70\textwidth]{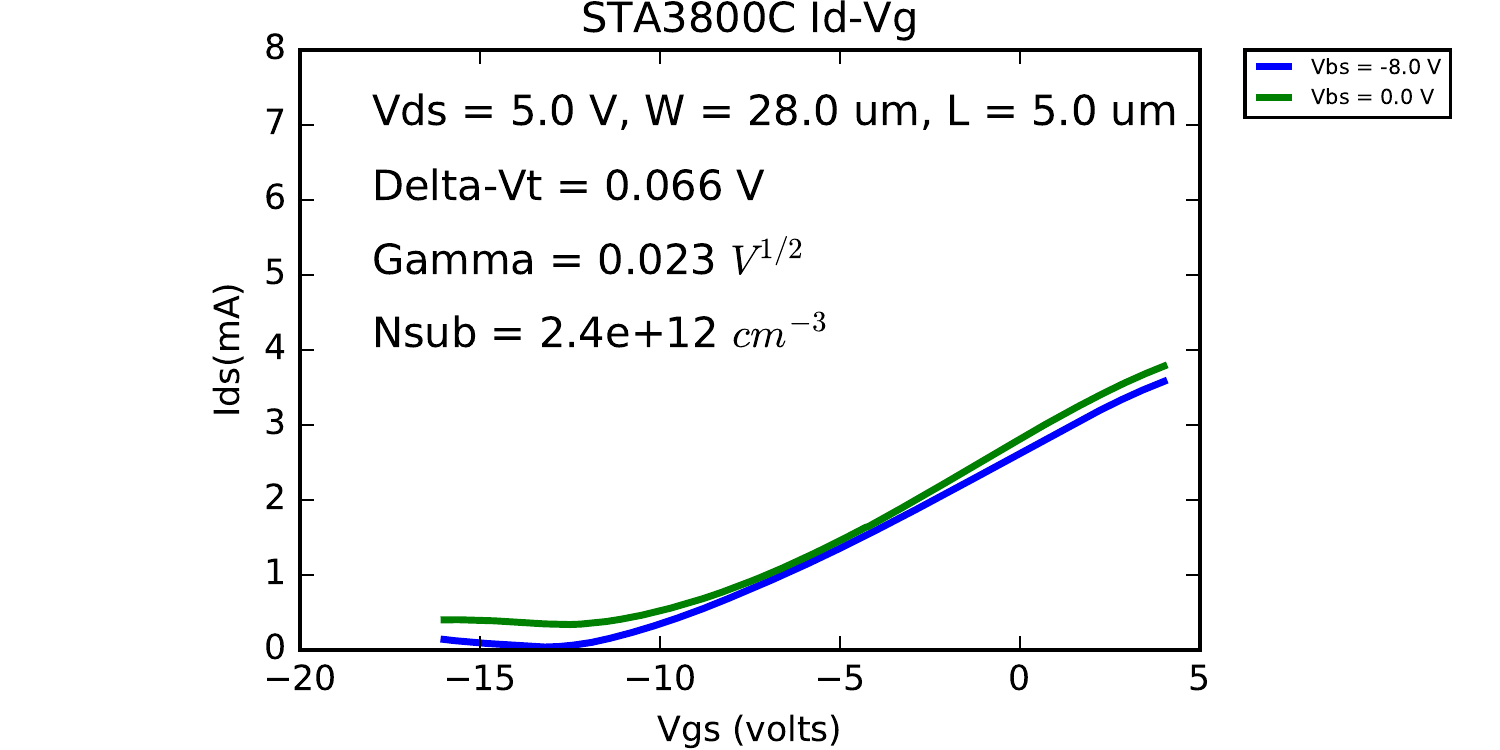}}
  \caption{Electrical measurements of the STA3800C output transistor.  The top panel shows the Id-Vd characteristics.  The center panel shows the device turn-on.  The bottom panel shows the variation in threshold voltage with back-bias, which allows us to calculate the substrate concentration.  The derived value of $\rm 2.4E12 cm^{-3}$ is close to expectations and was used to update the electrostatic model.}  
  \label{ITL_Output_Meas}
  % Trim is Left Bottom Right Top
\end{figure}

On the STA3800C, we were able to make electrical measurements of the output transistor.  The source (OS), drain (OD), and substrate (Gnd) of this device are connected to output pads.  The gate is not directly accessible, but we were able to bias the gate through the reset drain (RD) by properly biasing the reset gate (RG).  These electrical measurements are shown in Figure \ref{ITL_Output_Meas}.  In the next section, we discuss using these measurements to build a SPICE model of the output transistor.

\subsection{SPICE model development}
\label{SPICE_Model}
\begin{figure}[H]
  \begin{minipage}{0.49\textwidth}
    \begin{center}
      \includegraphics[trim=0.0in 0.0in 0.0in 0.0in,clip,width=0.95\textwidth]{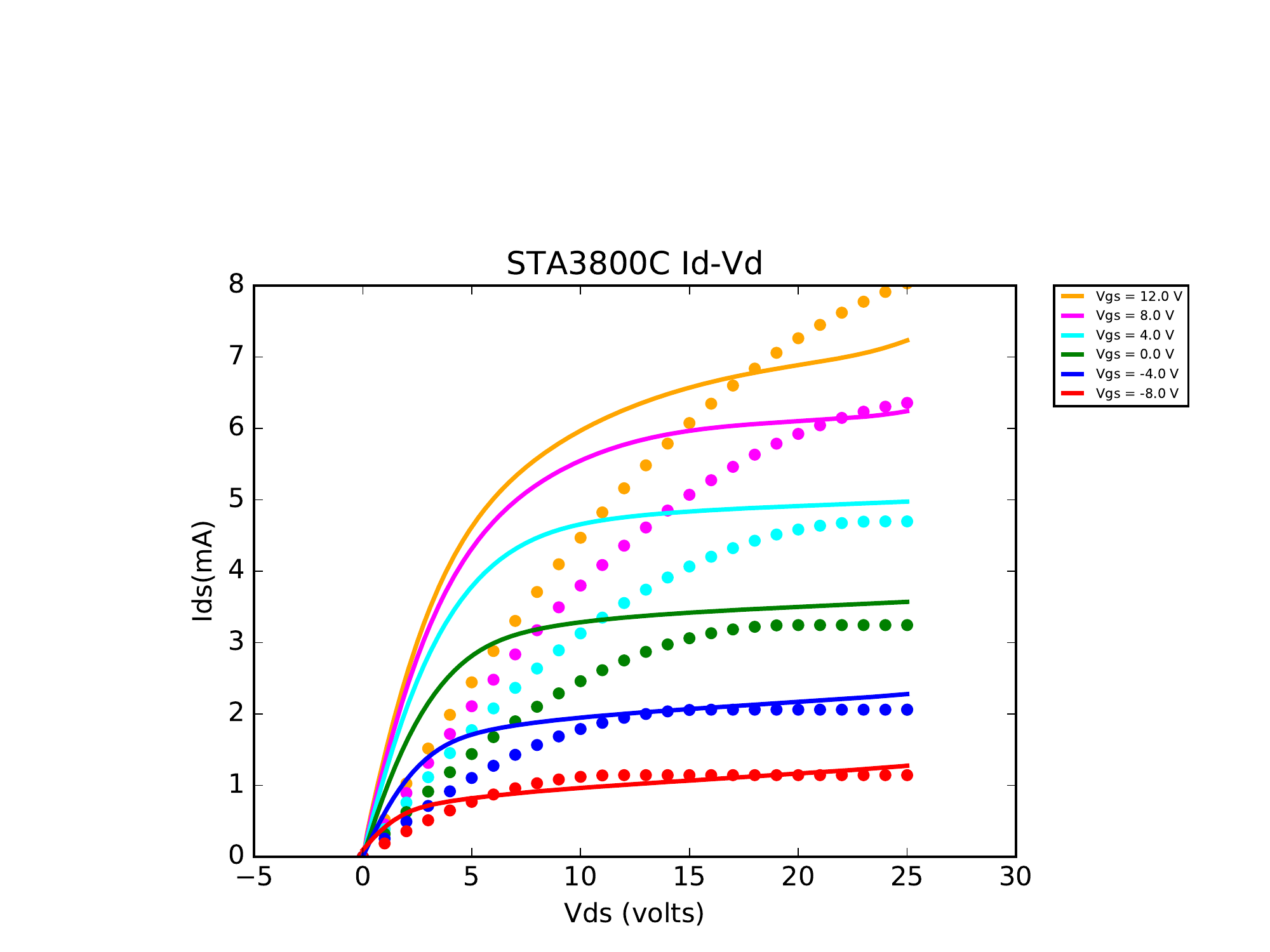}
    \end{center}
  \end{minipage}
  \begin{minipage}{0.49\textwidth}
    \begin{center}
      \includegraphics[trim=0.0in 0.0in 0.0in 0.0in,clip,width=0.95\textwidth]{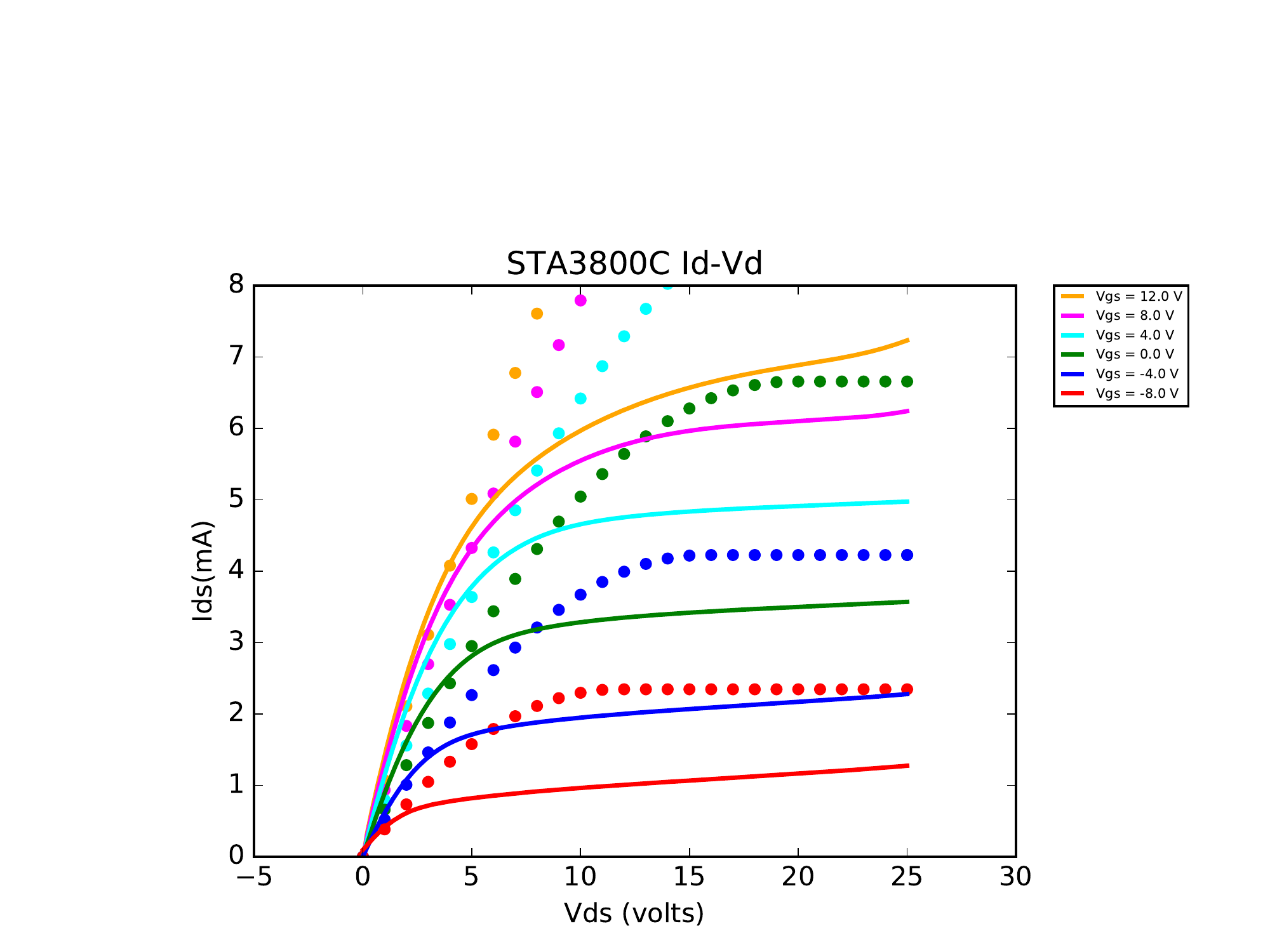}
    \end{center}
  \end{minipage}
  % Trim is Left Bottom Right Top
  \caption{Attempts to model the STA3800C output transistor as a single MOSFET.  On the left is one model set that attempted to fit the saturation current.  On the right is a second model set that attempted to fit the linear region.  Neither set fits the device well, and the models used had very non-physical parameters.  The solution is the composite device model shown in the Figure \ref{SPICE2}.}
  \label{SPICE1}
\end{figure}

\begin{figure}[H]
  \begin{minipage}{0.37\textwidth}
    \begin{center}
      \includegraphics[trim=1.0in 3.0in 0.5in 0.0in,clip,width=1.10\textwidth]{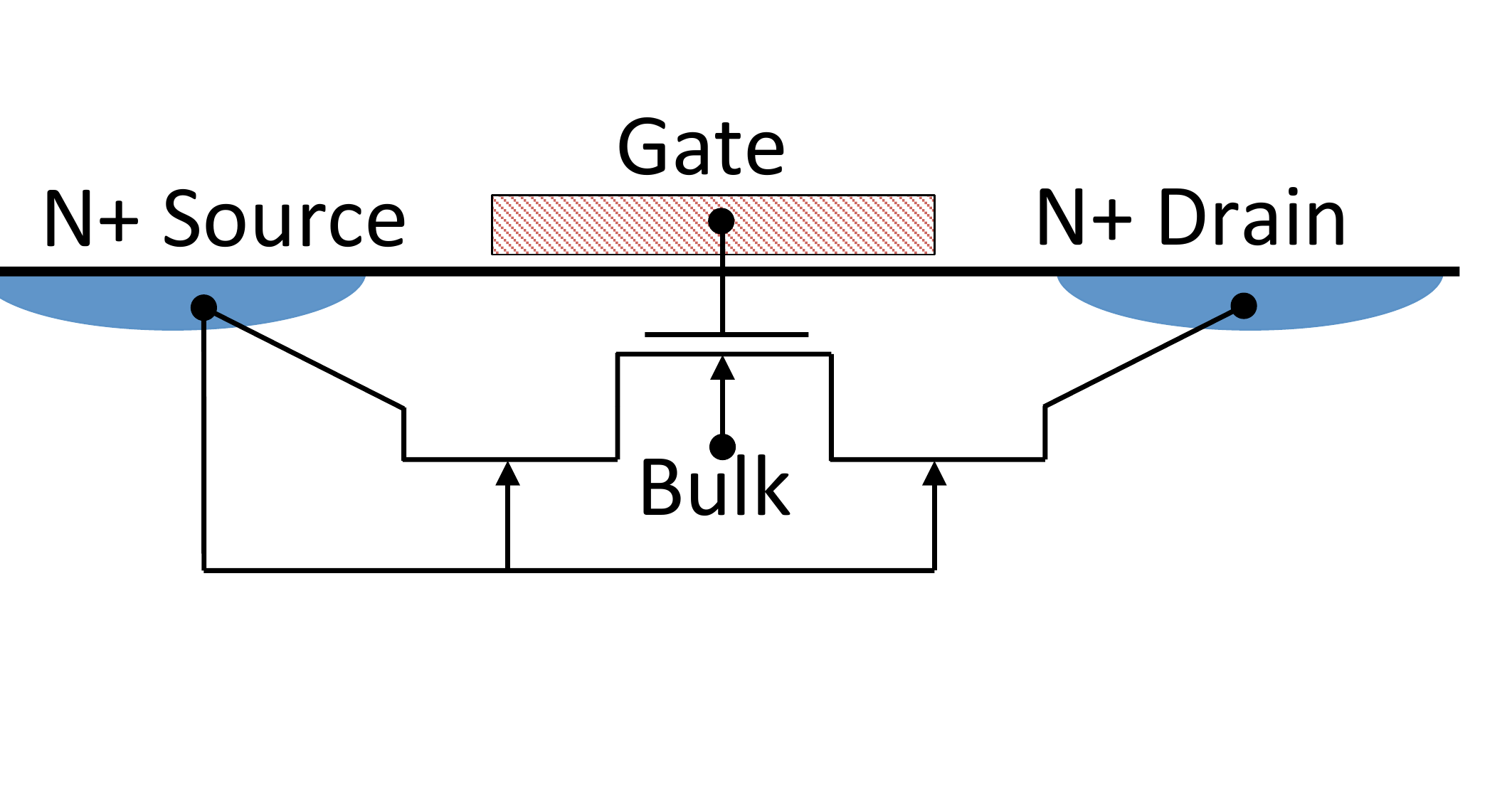}\\
    \end{center}
  \end{minipage}
  \begin{minipage}{0.32\textwidth}
    \begin{center}
      \includegraphics[trim=1.0in 0.0in 1.5in 1.4in,clip,width=0.99\textwidth]{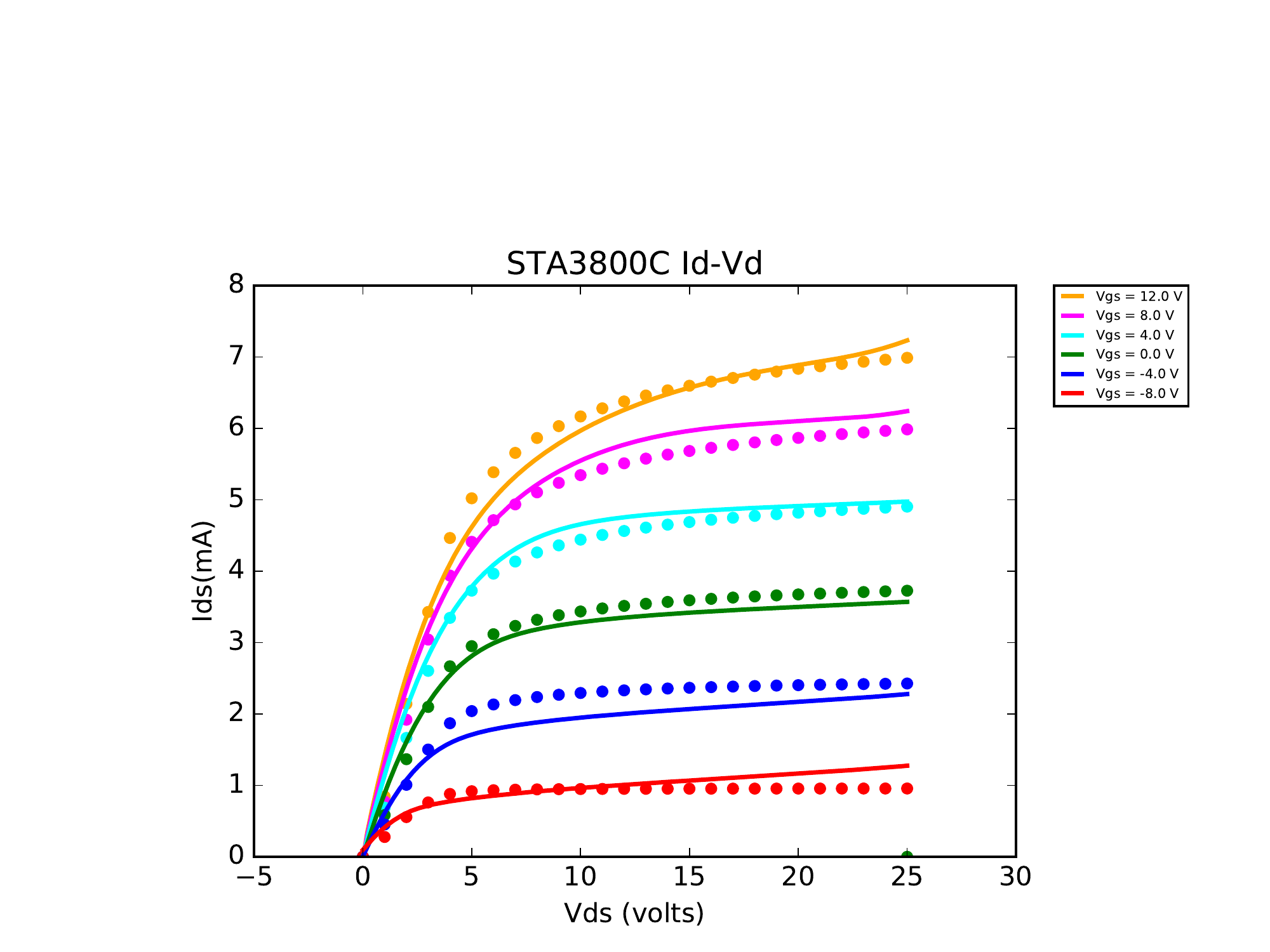}
    \end{center}
  \end{minipage}
  \begin{minipage}{0.32\textwidth}
    \begin{center}
      \includegraphics[trim=1.0in 0.0in 1.5in 1.4in,clip,width=0.99\textwidth]{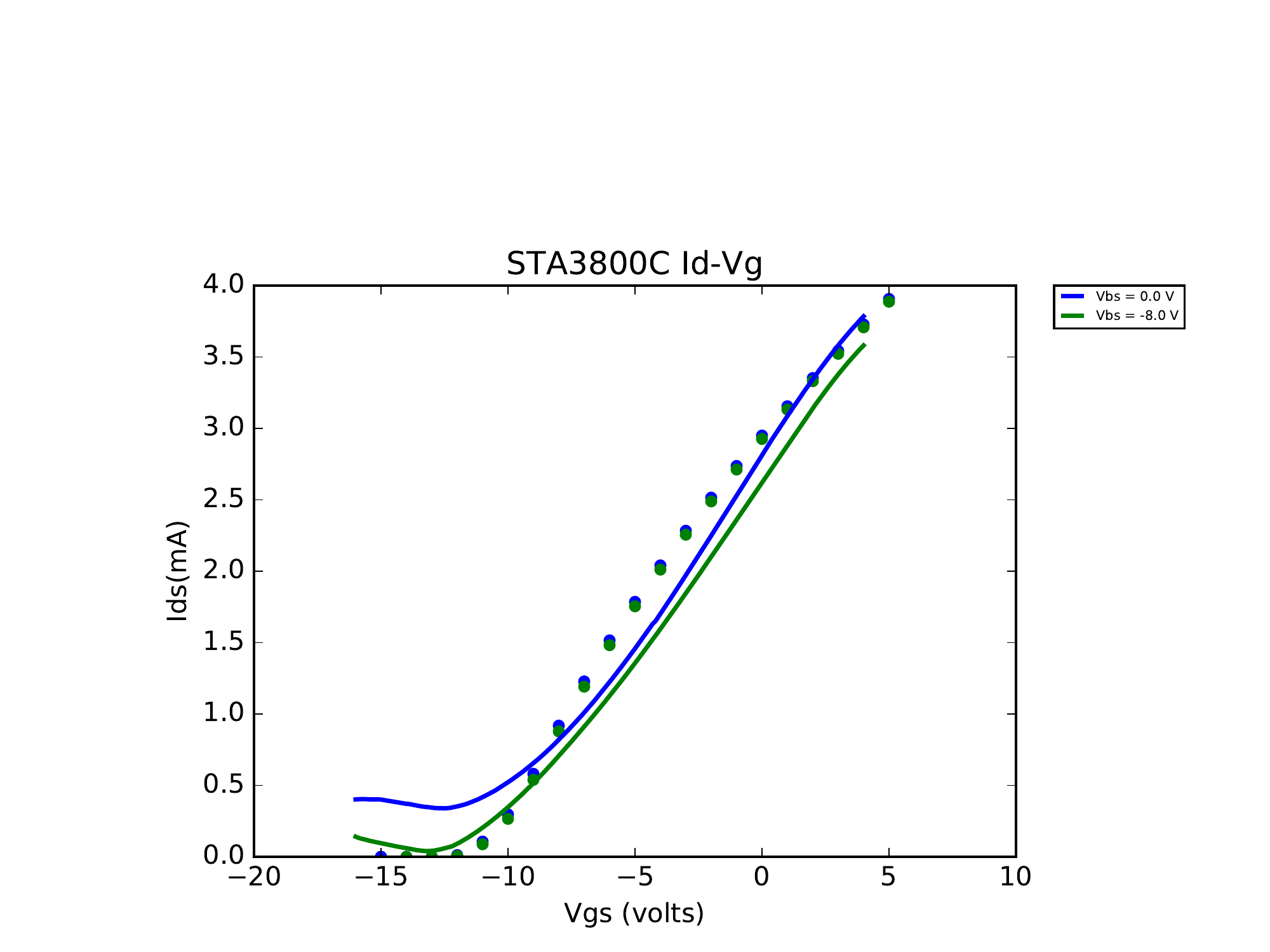}
    \end{center}
  \end{minipage}
  % Trim is Left Bottom Right Top
  \caption{DC measurements of the output characteristics of the output transistor on the STA3800C.  As shown on the left, the device is modeled as a MOSFET with two short-channel JFETs in series.  The fit of the measurements to the model is shown in the center (Id/Vd) and right (IdVg) panels, with the measurements as the solid lines and the SPICE simulations as the dots.  With this structure the device width(W), length(L), threshold voltage(Vt), and gate oxide thickness(Tox) all match the physical measurements, which was not the case when a MOSFET-only model was attempted. The details of the composite SPICE model are given in Appendix \ref{SPICE_Appendix}. }
  \label{SPICE2}
\end{figure}

To help understand the AC performance of the STA3800C device, we have also built a SPICE circuit-level simulation of the STA3800C output path. This has been calibrated to the above DC measurements and to several sets of AC measurements of the STA3800C.  We will first review the model used to fit the DC transistor measurements.  Figure \ref{SPICE1} shows that we were unable to model the device as a simple MOSFET, but that a composite device model as shown in Figure \ref{SPICE2} gave a reasonable fit to the measurements.

This SPICE model was used to analyze the output waveforms of the STA3800C CCD in two different controller environments.  The results of these analyses, including schematics and netlists, are given in Appendix \ref{SPICE_Appendix}.  This calibrated SPICE model is available for future analyses should the need arise.

\subsection{Analysis of glowing amplifiers}
\begin{figure}[H]
  \begin{center}
    \includegraphics[trim=0.0in 0.0in 0.0in 0.0in,clip,width=0.89\textwidth]{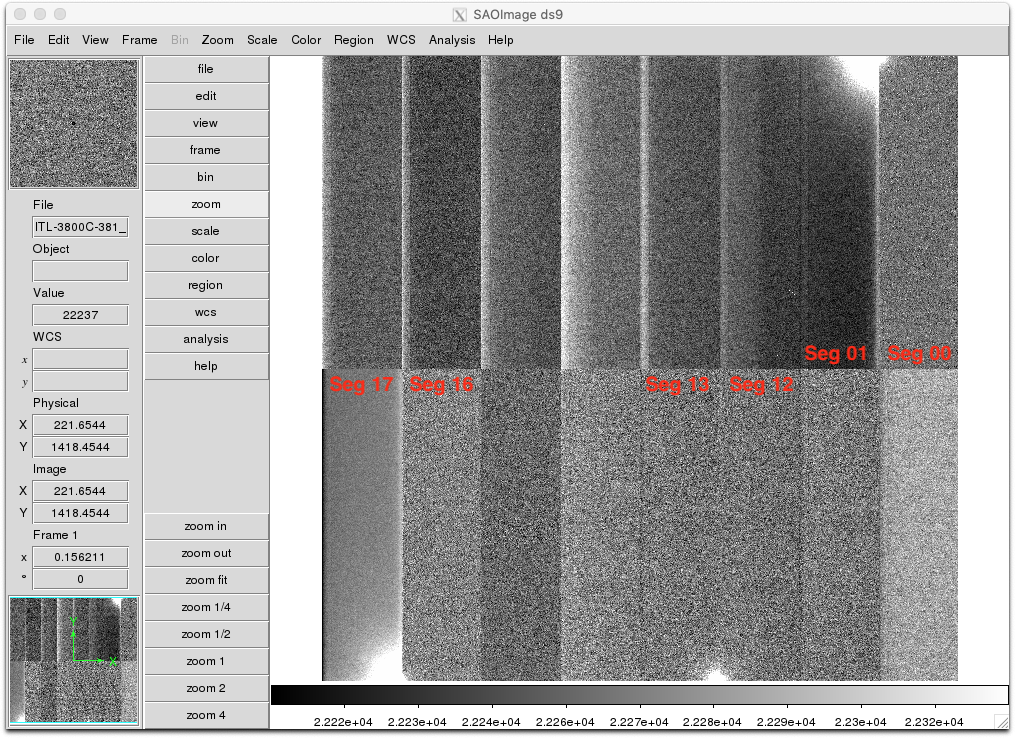}\\
    \includegraphics[trim=0.0in 0.0in 0.0in 0.0in,clip,width=0.49\textwidth]{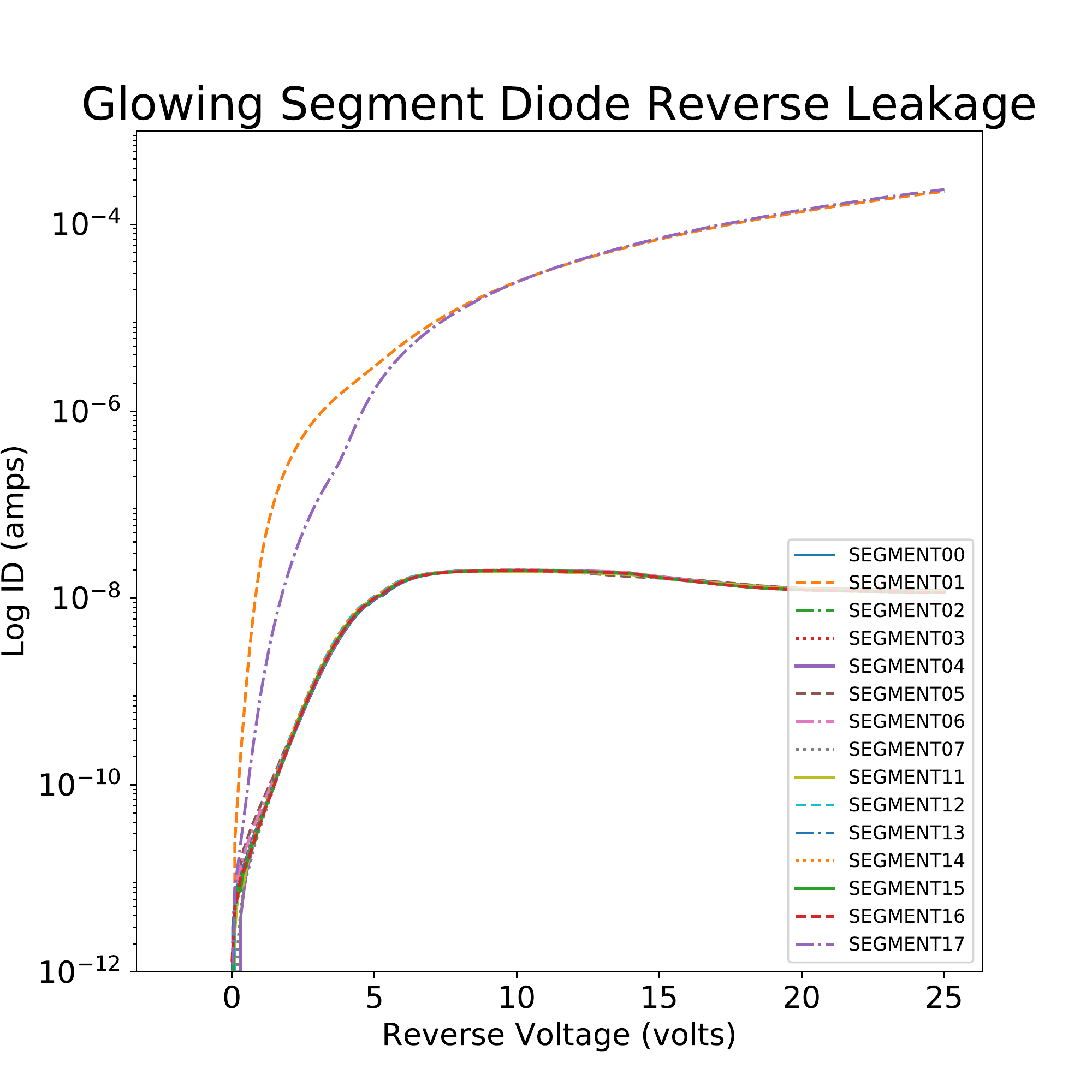}
    \includegraphics[trim=0.0in 0.0in 0.0in 0.0in,clip,width=0.49\textwidth]{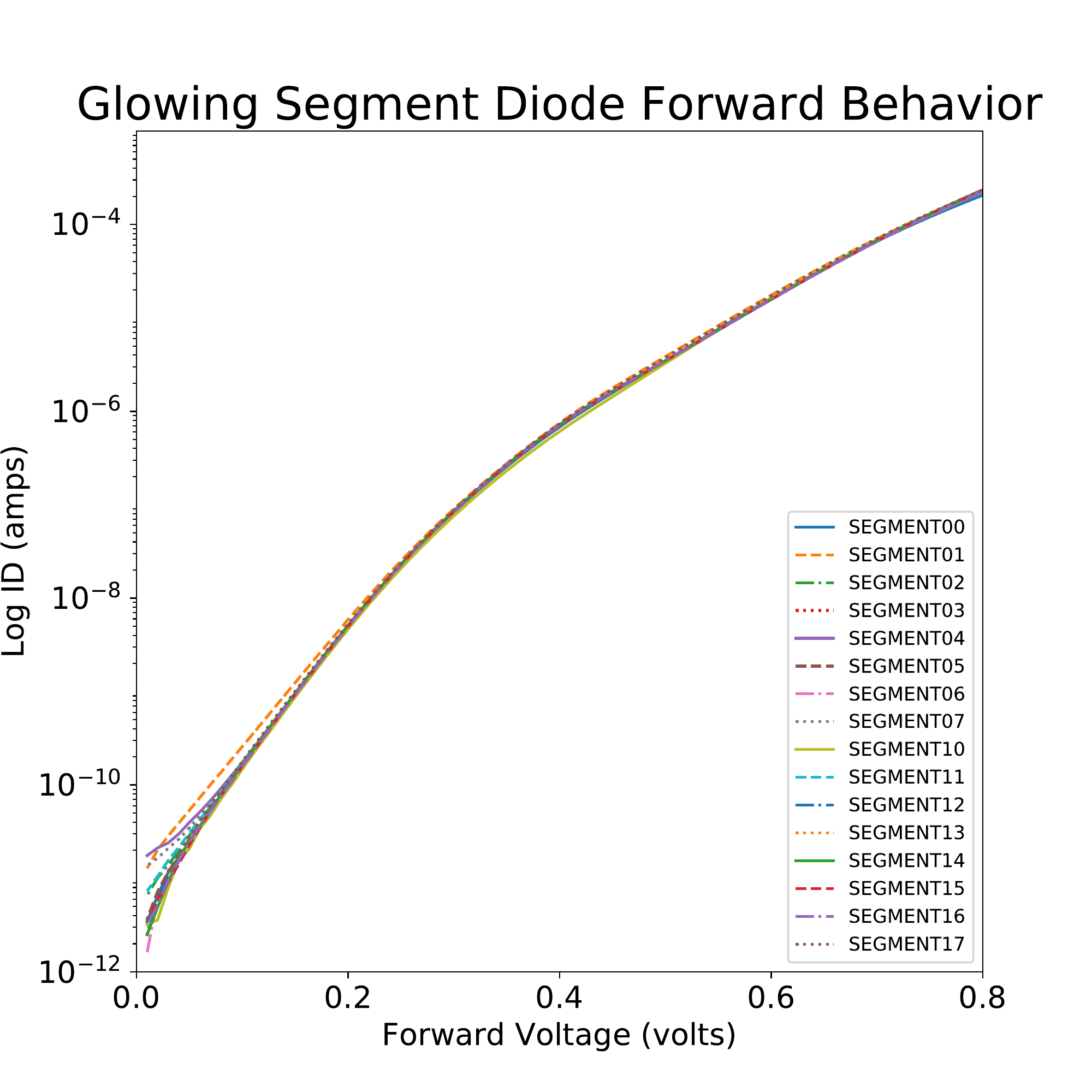}    
  \end{center}
  \caption{Diode leakage on good and glowing segments on ITL STA3800C-381.  The top image shows a dark image from this CCD, where segments 01 and 17 are exhibiting a strong glow, and segment 13 is exhibiting a weak glow.  The bottom two images show the forward and reverse diode behavior of all 16 amplifiers.  The two strongly glowing segments show reverse bias leakage current elevated by more than 10,000X, indicative of ESD damage.  The slightly glowing segment is indistinguishable from the good segments.}
  \label{Glow}
\end{figure}

During construction of the LSST focal plane, several amplifiers on STA3800C CCDs were seen to exhibit ``glow'', where a region of signal appearing to emanate from the amplifier region could be seen.  This was believed to be caused by damage to the amplifier transistor which causes the transistor to emit infrared light.  This infrared light then activates the light-sensitive CCD pixels.  One of these sensors, which exhibited three glowing segments, was obtained for electrical analysis.   The results are shown in Figure \ref{Glow}.  The two strongly glowing segments show reverse bias leakage current elevated by more than 10,000X, indicative of ESD damage.  The slightly glowing segment is indistinguishable from the good segments.  Other attempts to find anomalous behavior on the weakly glowing segment 13 were unsuccessful, so the cause of the glow from this segment is unknown.

Emission of light from damaged silicon diodes is well known.  See, for example, \cite{Rasras_2001}, especially Figure 10, where infrared light emission from damaged diodes has been measured, with the flux of light emission roughly proportional to the magnitude of the leakage current.  Because this light emission can interfere with sensitive astronomical measurements, sensors exhibiting this behavior should be avoided.

\section{E2V CCD250}

\subsection{Deprocessing}
\begin {figure}[H]
	\centering
	\subfigure[b][Schematic of CCD and support silicon]{\includegraphics[trim=3.0in 1.8in 3.0in 0.5in,clip,width=0.35\textwidth]{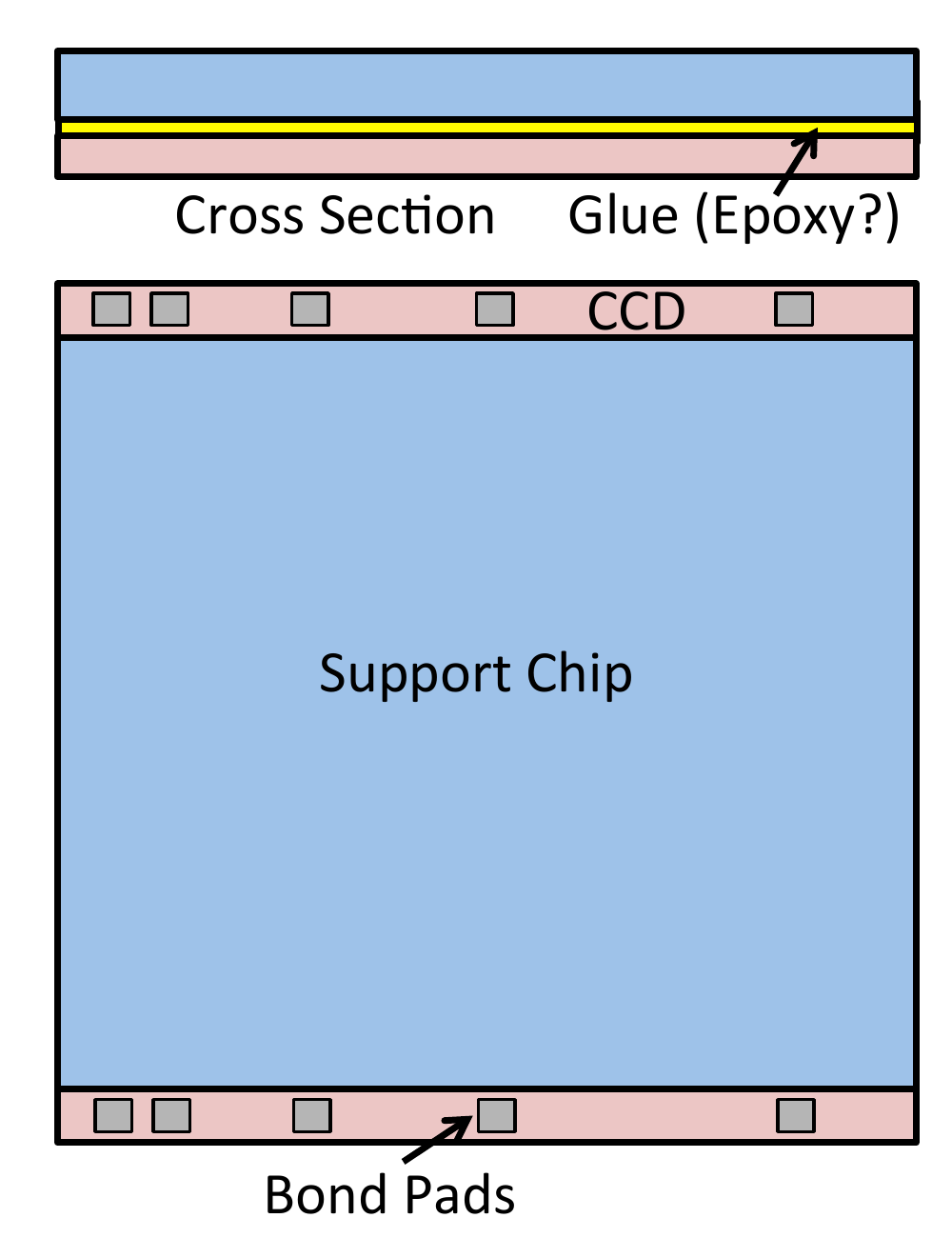}}
	\subfigure[b][Cross-section of CCD and support silicon]{\includegraphics[trim=2.0in 2.5in 2.0in 1.5in,clip,width=0.63\textwidth]{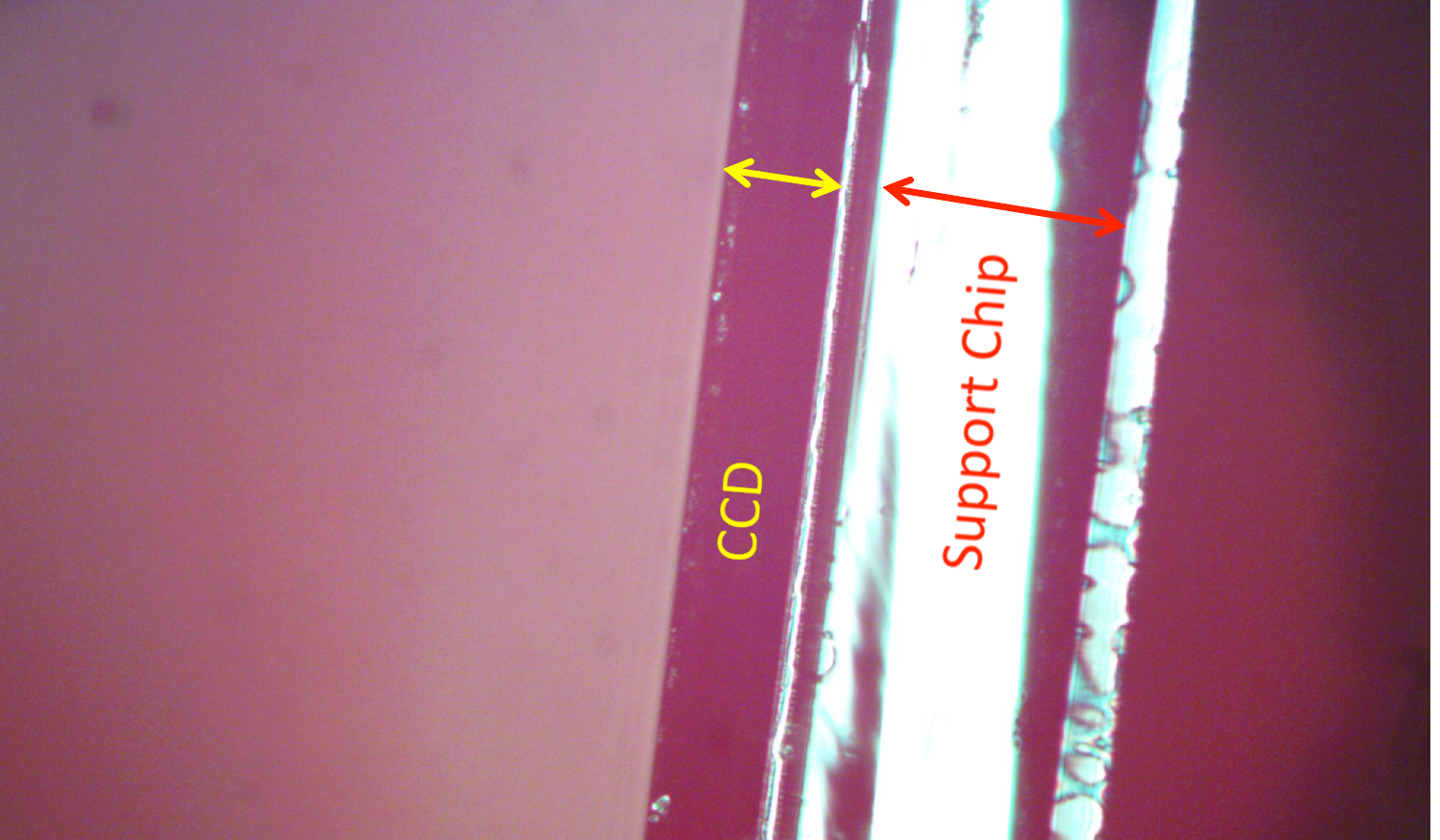}}
	\subfigure[b][Top view of CCD and support silicon]{\includegraphics[trim=0.0in 1.0in 0.0in 1.0in,clip,width=0.60\textwidth]{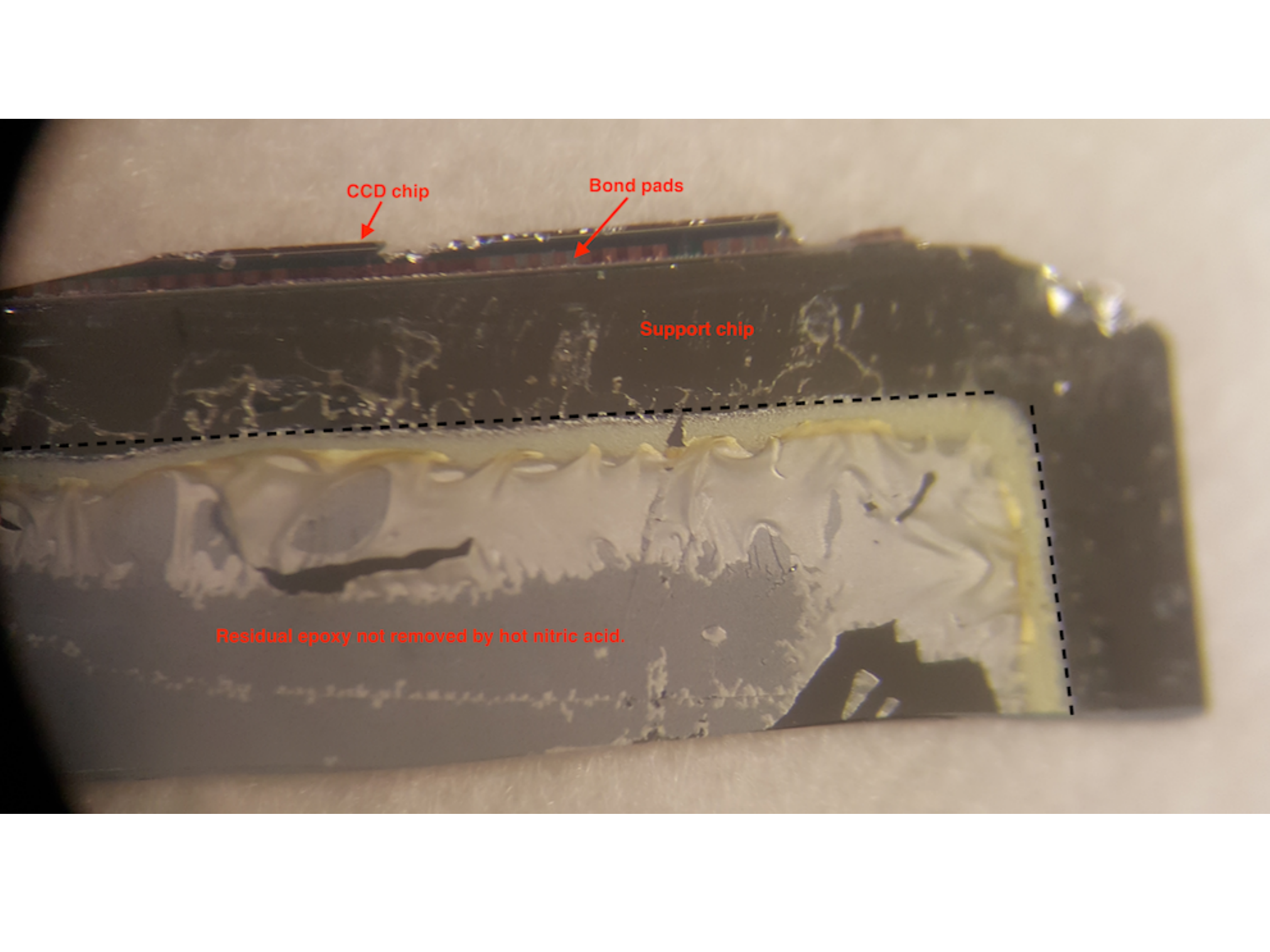}}
	\subfigure[b][Portion of CCD with support silicon attached]{\includegraphics[trim=0.0in 3.5in 0.0in 0.0in,clip,width=0.80\textwidth]{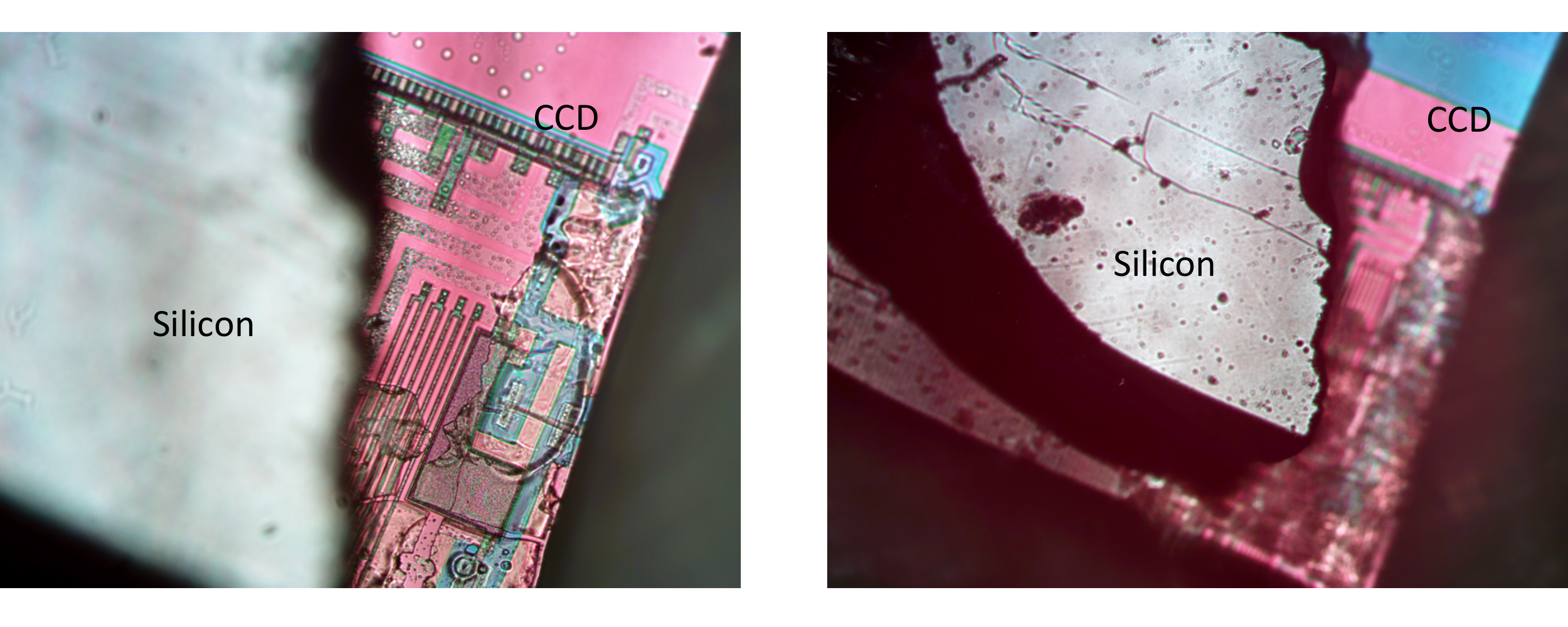}}        

  \caption{Various views of the E2V CCD with attached support silicon.  The support silicon is believed to have no electrical function, but merely give mechanical support.  The CCD is 100 microns thick and the support silicon is about 225 microns thick.}
  \label{Support}
  % Trim is Left Bottom Right Top
\end{figure}

In order to proceed with physical analysis of the E2V CCD250 chip, it was necessary to deprocess it and remove it from the package.  This proved to be a lengthy trial-and-error process, but eventually several pieces of the CCD were removed from the package, and these were sufficient to do the desired analysis.  A high-level overview of the deprocessing procedure is as follows:

\begin{itemize}
  \item The CCD was believed to be bonded to the package with epoxy adhesive.  IC deprocessing to remove epoxy adhesive is typically carried out with red fuming nitric acid, but this was not available.  Concentrated (70\%) nitric acid at 70C was chosen as an alternative.  After immersing the package in this solution for about two hours, the epoxy adhesive was undercut sufficiently to allow several pieces to be mechanically cleaved loose from the substrate with a razor blade.
  \item At this point it was apparent that the CCD was fastened to a piece of support silicon, apparently to improve the mechanical rigidity of the this CCD.  This ``sandwich'' was then fastened to the ceramic package.  Figure \ref{Support} shows several views of the CCD and support silicon.
  \item The support silicon was apparently also bonded to the CCD using epoxy adhesive.  Accordingly, the pieces containing portions of the CCD and support silicon were immersed in hot nitric acid for an additional 4 hours.  At this point, the epoxy between the CCD and the support silicon was again sufficiently undercut to allow them to be mechanically separated (again using a razor blade).
  \item This process then exposed the CCD surface for optical photographs.  There was still residual epoxy on the CCD surface, and the majority of this was removed with and additional 15 minutes in hot nitric acid.  At this point the CCD surface was relatively clean.  It is important to note that the metal layers were removed by the hot nitric acid, although a ``ghost'' image where the metal layers had been could still be seen.  Optical photographs were obtained and are detailed in the next section.
  \item To obtain SIMS analysis of dopant profiles, a larger portion of the CCD was needed, including a portion of the imaging array.  For this, it was not necessary to leave the oxide and metal intact.  In fact removing all layers down to bare silicon is desirable.  So for these samples, a portion of the CCD was immersed in 49\% HF for approximately 18 hours.  This undercut the oxide layers on the CCD and separated the CCD from the support silicon.  After examining these samples, it was seen that many pieces of polysilicon that had been floating in the HF solution had adhered to the silicon.  These were removed with short etches in Piranha ($\rm H_2SO_4 / H_2O_2$) followed by BOE (Buffered HF).  After this, the samples were fairly clean.  Optical micrographs of these samples were also obtained, and the samples were then sent for SIMS analysis.  These results are detailed in the next sections.
  \item An alternate method to dissolve the epoxy adhesive was also tried.  Properties of epoxy adhesives list methylene chloride ($\rm CH_2Cl_2$) as a solvent for epoxy.  However, even after immersing several samples in room temperature methylene chloride for two weeks, there was no apparent attack of the epoxy.
  
\end{itemize}

\subsection{Optical Micrographs}
\begin {figure}[H]
	\centering
	\subfigure[b][Output chain]{\includegraphics[trim=1.5in 2.0in 0.5in 2.0in,clip,angle=90,width=0.49\textwidth]{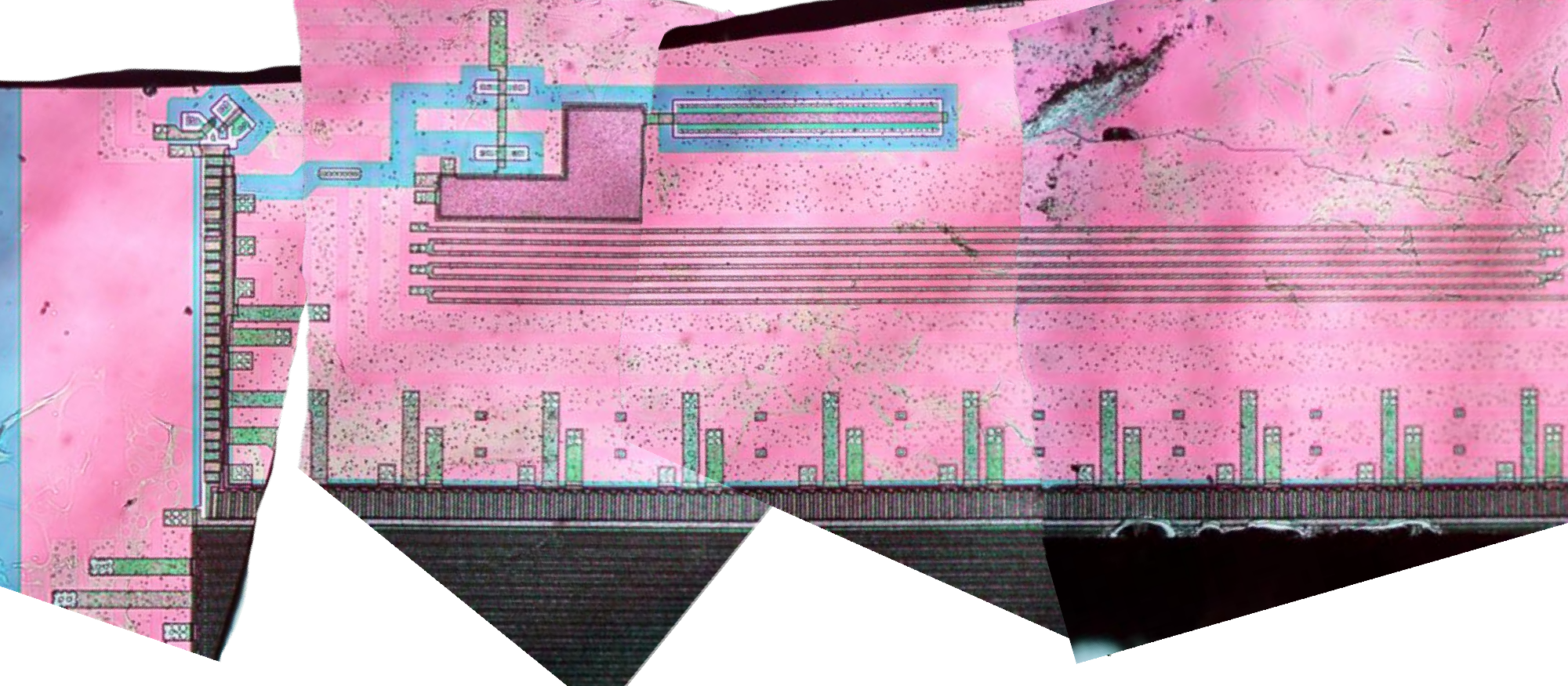}}
	\subfigure[b][Output chain with annotated metal lines]{\includegraphics[trim=1.5in 2.0in 0.5in 2.0in,clip,angle=90,width=0.49\textwidth]{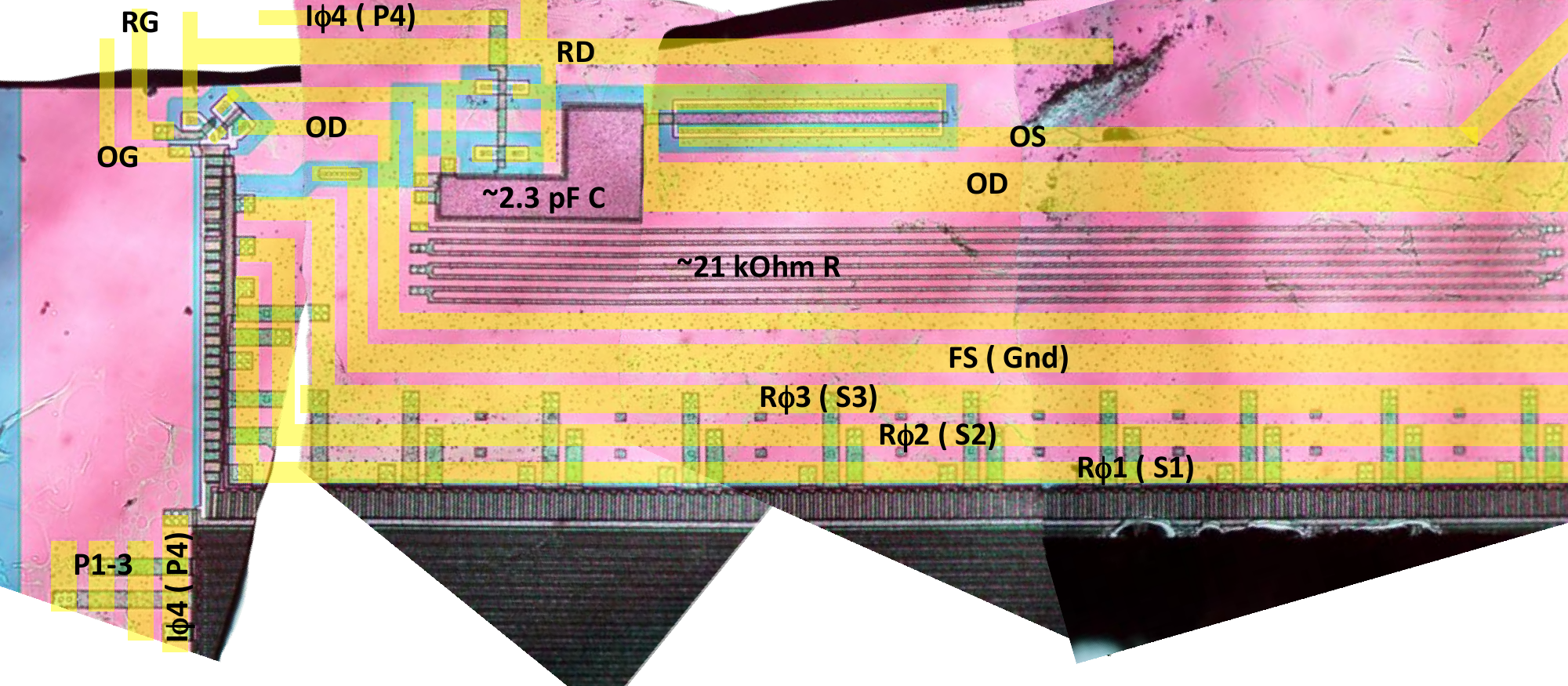}}

  \caption{Output chain on the E2V CCD250 device after deprocessing.  This is a composite of several pieces after deprocessing.  The deprocessing has removed the metal lines, but a ``ghost'' image of the metal lines can still be seen.  The image on the right has the metal lines drawn back in and labeled.  The component values given here are an estimate from measuring the photographs and should be considered approximate ($\pm 20 \%$ at best).}
  \label{E2V_Output}
  % Trim is Left Bottom Right Top
\end{figure}

\begin {figure}[H]
	\centering
	\subfigure[b][Higher mag view of output device and reset gate]{\includegraphics[trim=2.5in 3.0in 2.5in 1.0in,clip,width=0.69\textwidth]{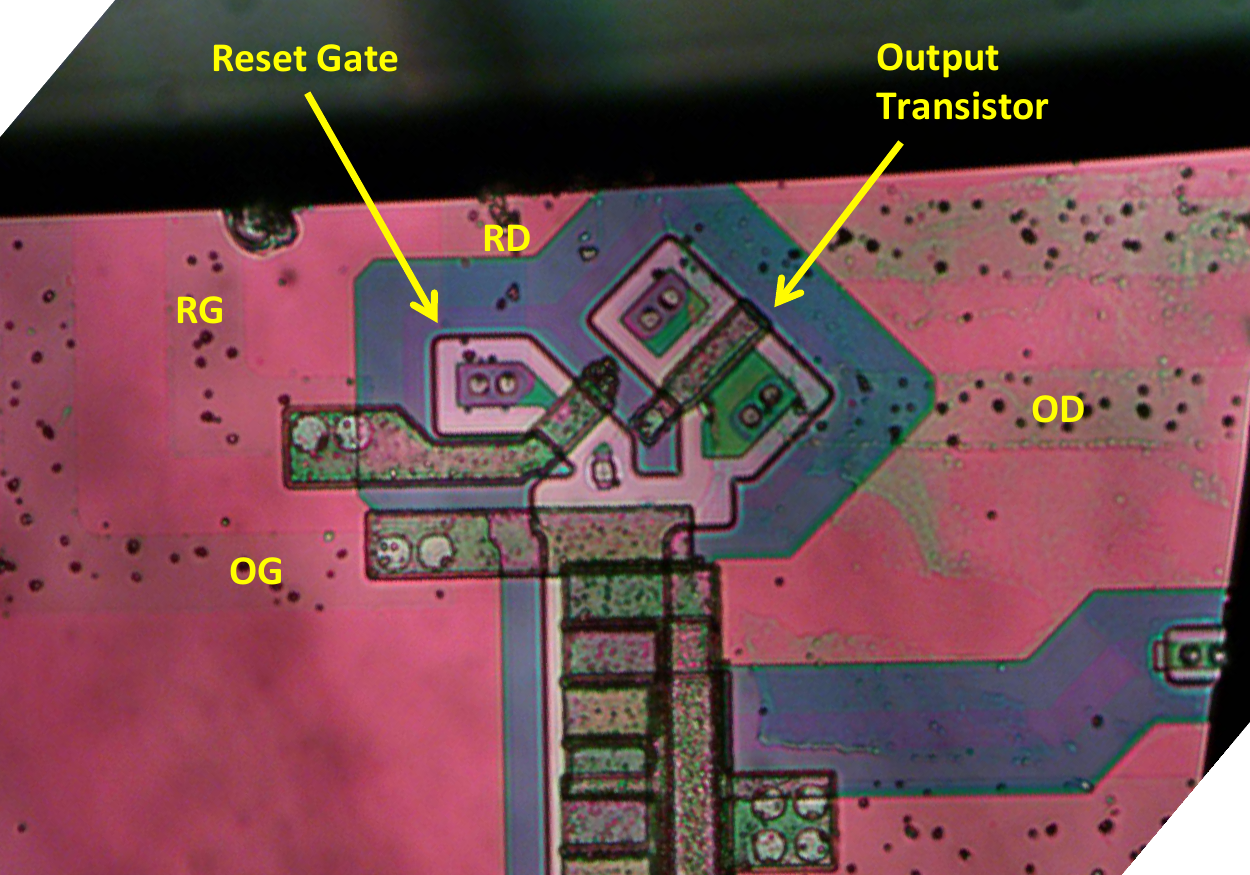}}
	\subfigure[b][Second stage clamp]{\includegraphics[trim=2.0in 2.0in 4.0in 2.0in,clip,width=0.69\textwidth]{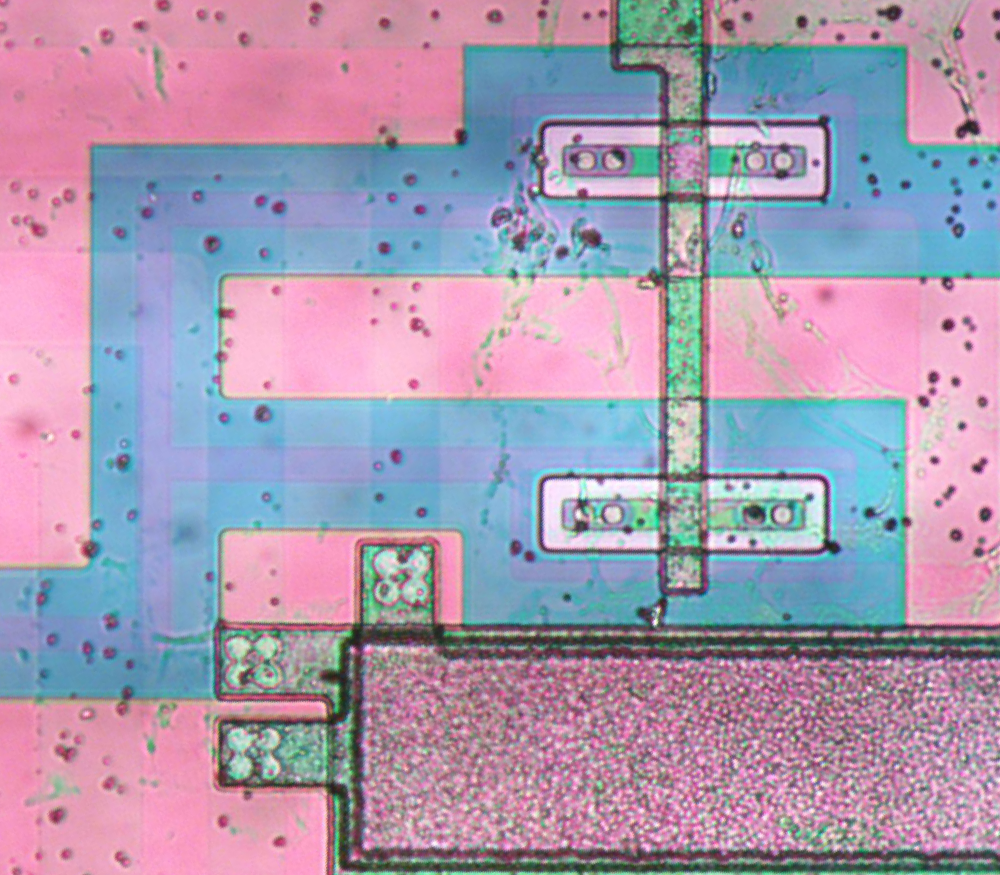}}
  \caption{Additional optical micrographs of the E2V CCD250 chip. }
  \label{Other_Optical_1}
  % Trim is Left Bottom Right Top
\end{figure}

\begin {figure}[H]
	\centering
	\subfigure[b][Bond pads and guard ring]{\includegraphics[trim=2.0in 2.0in 2.0in 2.0in,clip,width=0.69\textwidth]{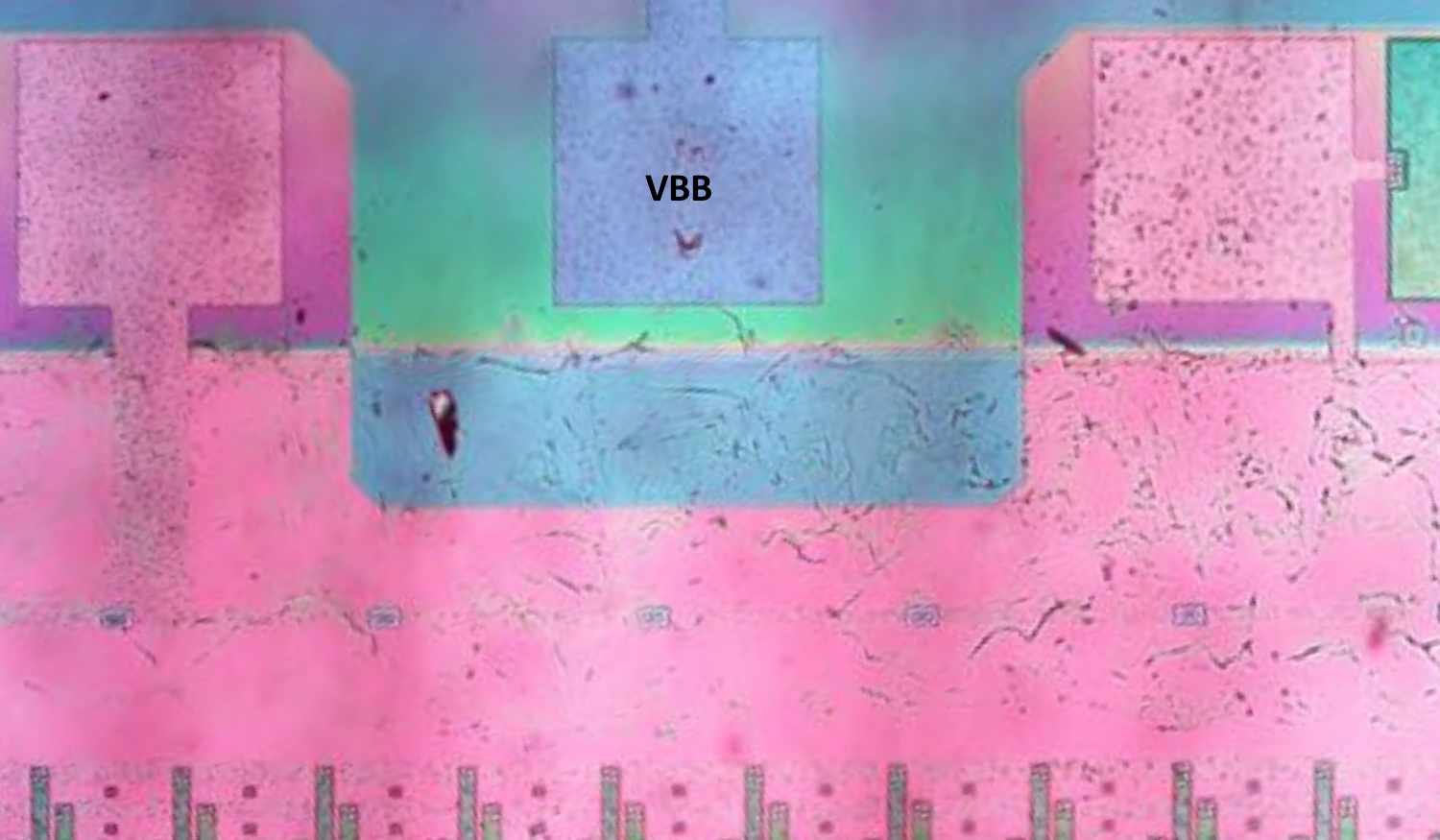}}
	\subfigure[b][Output chain stripped down to silicon]{\includegraphics[trim=2.0in 3.0in 3.0in 2.0in,clip,width=0.69\textwidth]{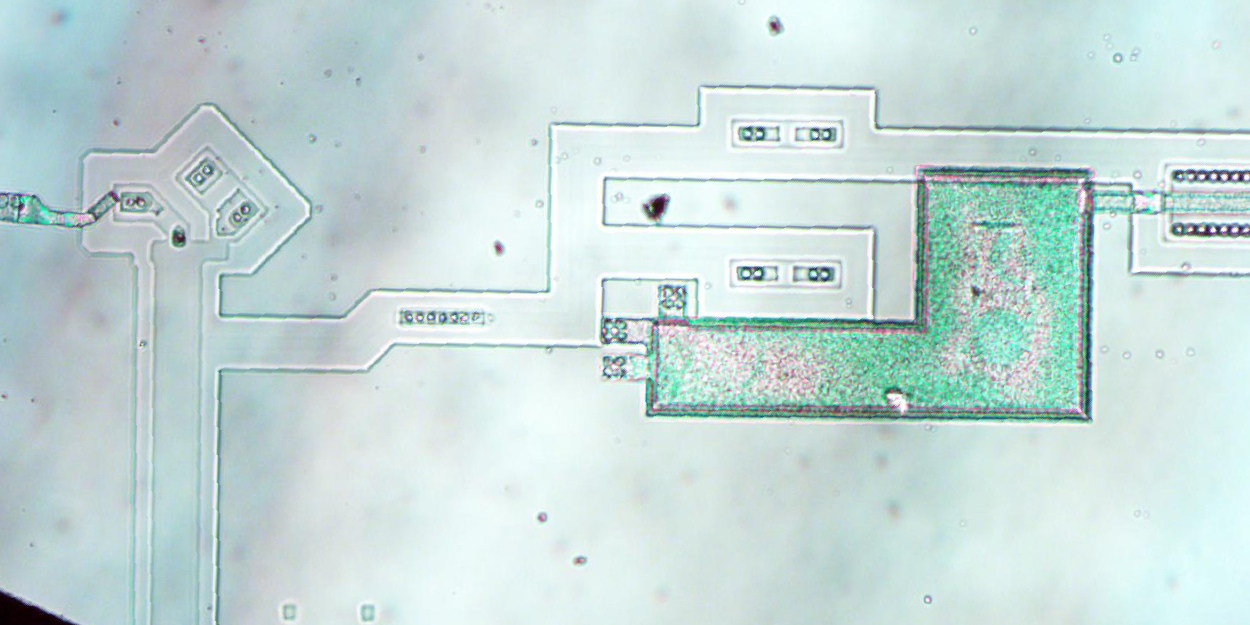}}
        
  \caption{Additional optical micrographs of the E2V CCD250 chip.}
  \label{Other_Optical_2}
  % Trim is Left Bottom Right Top
\end{figure}

\begin {figure}[H]
	\centering
	\includegraphics[trim=1.0in 2.0in 1.5in 1.5in,clip,width=0.49\textwidth]{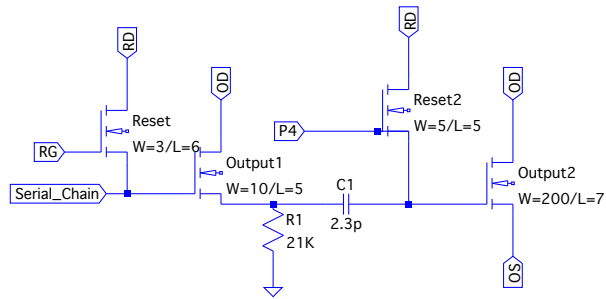}
  \caption{Approximate schematic of the E2V output chain.   The component values given here are an estimate from measuring the photographs and should be considered approximate ($\pm 20 \%$ at best).}
  \label{E2V_Output_Chain}
  % Trim is Left Bottom Right Top
\end{figure}

A series of optical micrographs of the E2V CCD250 chip are shown in Figures \ref{E2V_Output}, \ref{Other_Optical_1}, and \ref{Other_Optical_2}.  These illustrate the serial register with its right-angle bend, and the two stage output device circuitry.  A schematic of the output chain is shown in Figure \ref{E2V_Output_Chain}.  The component values given there are estimates from measuring the photographs and should be considered approximate ($\pm 20 \%$ at best).

\subsection{SIMS Analysis}
\label{E2V_SIMs_Section}

SIMS analysis (see Section \ref{ITL_SIMS_Section}) was also performed on the E2V CCD250 chip.  In the case of the ITL device, consultation with the vendor allowed us to identify large regions in the periphery of the chip where the dopant profiles in the array could be measured.  Not having this information on the E2V device required taking the SIMS profiles in the imaging array.  Since the SIMS measurement spot is much larger than one pixel, the profiles in the array include both channel and channel stop regions.  This requires some judgment to calculate the actual dopant profiles, so there is some uncertainty.  This is discussed in Figure \ref{E2V_SIMS}.  For the E2V chip a series of SIMS profiles were also taken outside the array, and these results are shown in Figure \ref{E2V_SIMS_2}.

\section{{\carlito Poisson\_CCD} Model Development}
\label{Poisson_Model}
With the physical analysis of the CCD in place, we were able to proceed with building calibrated electrostatic models of both CCDs.   The {\carlito Poisson\_CCD} code is designed to model astronomical CCDs and provide answers to questions of astronomical interest.  These include calculating the electric fields in the CCD, calculating the electron paths after photo-conversion, simulating the impact of diffusion on the PSFs, simulating the impact of lateral electric fields on phenomena such as the brighter-fatter effect (\cite{antilogus2014}, \cite{gruen2015}, \cite{Coulton_2018}) and edge roll-off(\cite{bradshaw2015}), and any other phenomena of interest.  The code is described in detail elsewhere (\cite{Lage_2017}, \cite{Poisson-CCD-paper}) and the code itself, with many examples, is available at \cite{Poisson-CCD-code}.

The intent here is not to go into detail of the {\carlito Poisson\_CCD} code, but to show an example of what has been accomplished with the physical characterization discussed above.  Perhaps the most important characteristic of the CCD which has been simulated and compared to experiment is the distortion of the pixel shapes due to the brighter-fatter effect.  As has been extensively discussed in the literature (\cite{antilogus2014}, \cite{gruen2015}, \cite{guyonnet2015}, \cite{Lage_2017}), as charge builds up in the central region of bright objects, the stored charge repels additional incoming charge and broadens the profile of these objects.  The impact of the stored charge on the pixel shapes can be measured by measuring the pixel-pixel correlations on a large number of flat images (\cite{antilogus2014}, \cite{Coulton_2018}).  These correlations are calculated from a large number of flat pairs of varying intensity (see \cite{Coulton_2018} for example) as:

\begin{equation}
C_{i,j} = \frac{\sum_{I,J} (f_{I,J} - \bar{f}) (f_{I+i,J+j} - \bar{f})}{\bar{f}^2(N_{pix} - 1)}
\end{equation}

where $\rm f_{i,j}$ is the difference in flux between the two flats at pixel i,j, and $\rm N_{pix}$ is the number of pixels summed over.  This calculation is implemented in the LSST image reduction pipeline \cite{2018arXiv181203248B}.

We show here that these correlations can be simulated using the {\carlito Poisson\_CCD} software with a model calibrated with the above-measured physical attributes of the CCDs.  Since the physical attributes of the CCD are either known silicon parameters or measured physical quantities, any adjustable parameters associated with the CCD structure have been removed.  The only remaining adjustable parameters are those associated with the details of the numerical solution to Poisson's equation.

First we show that the dopant profile model in the {\carlito Poisson\_CCD} simulator accurately reproduces the measured SIMS profiles, as shown in Figures \ref{ITL_SIMS_Fit} and \ref{E2V_SIMS_Fit}.  Next, Figures \ref{ITL_Pixel_Distortions} and \ref{E2V_Pixel_Distortions} show the location of the charge in three dimensions when one pixel contains 100,000 electrons and the surrounding pixels are empty.  These figures also show the pixel distortions which result in this case.  These pixel area distortions allow one to calculate the pixel-pixel correlations that result, and this is done and compared to measured correlations in Figures \ref{ITL_Correlation_Sims} and \ref{E2V_Correlation_Sims}.  Care was taken to make sure to match the conditions (namely the applied voltages and which parallel phases were high during image integration) between the flat measurements and the simulations.  The good agreement shows the value of obtaining good physical characterization to inform the simulations.

\begin {figure}[H]
	\centering
	\subfigure[b][Optical micrograph of SIMS ``pit'' after measurements]{\includegraphics[trim=0.0in 0.0in 0.0in 0.0in,clip,width=0.45\textwidth]{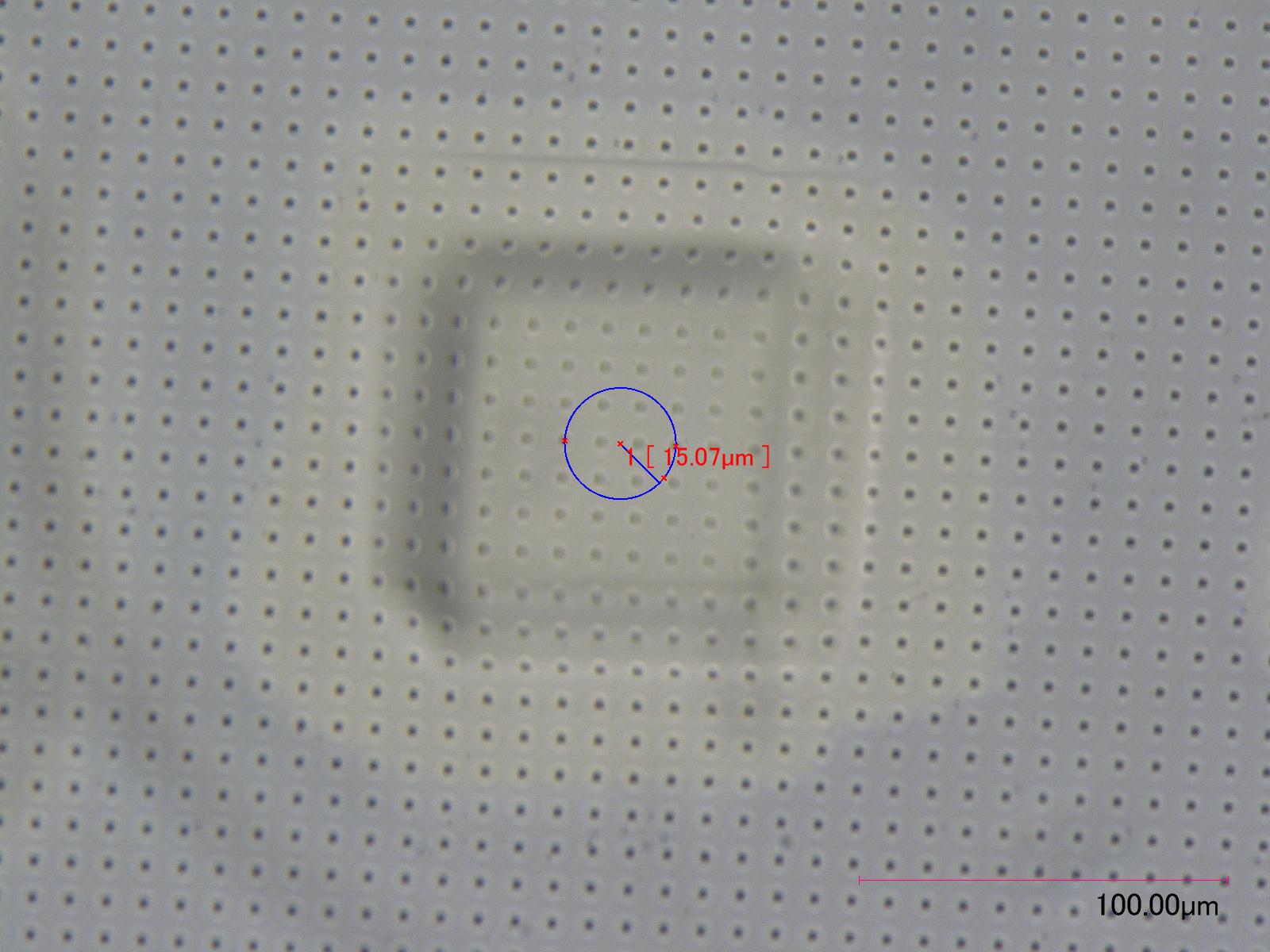}}
	\subfigure[b][Schematic of assumed structure in array.]{\includegraphics[trim=0.5in 0.5in 0.5in 1.5in,clip,width=0.54\textwidth]{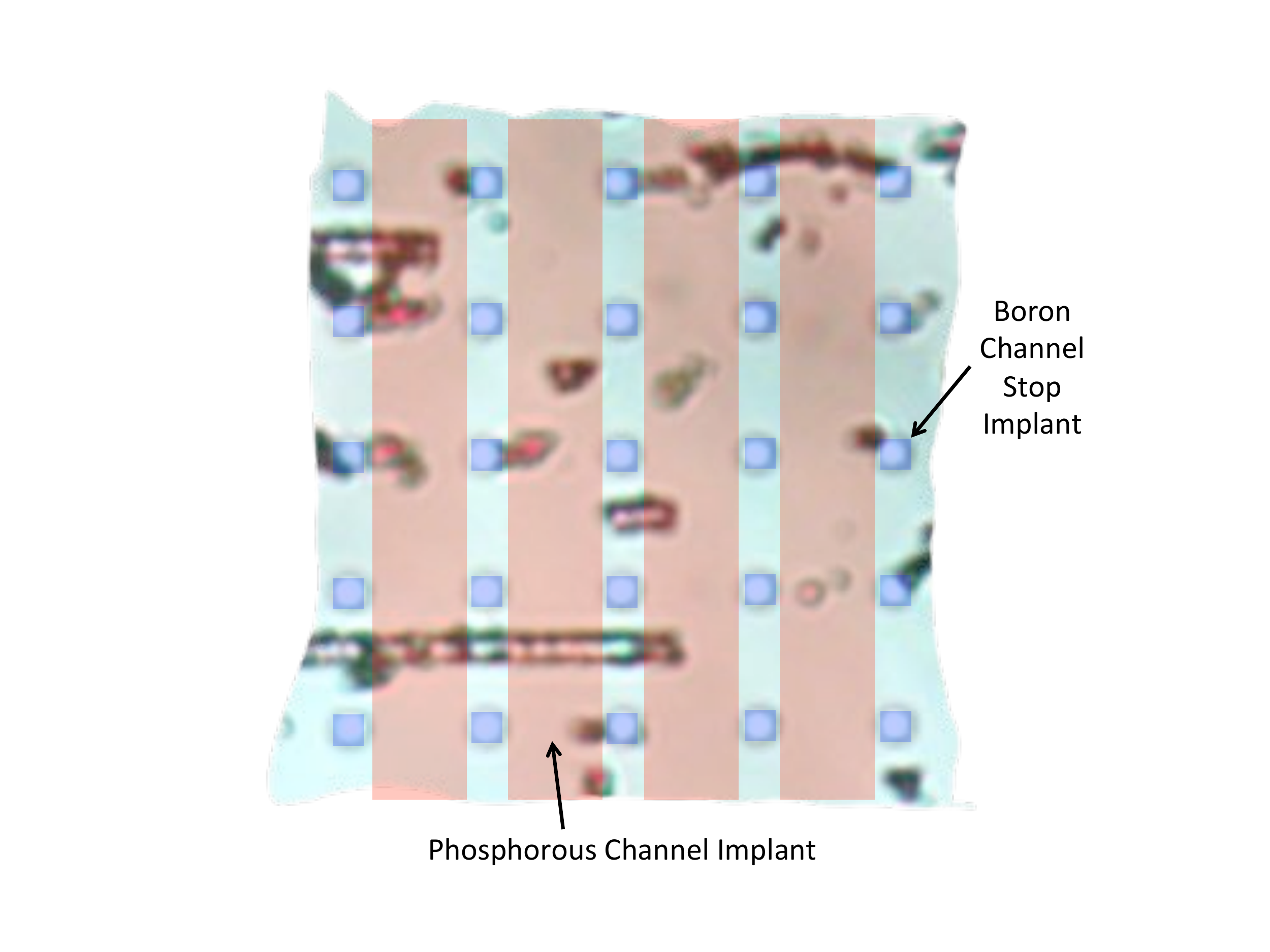}}
	\subfigure[b][Phosphorous channel profile.]{\includegraphics[trim=0.5in 0.5in 0.5in 0.5in,clip,width=0.49\textwidth]{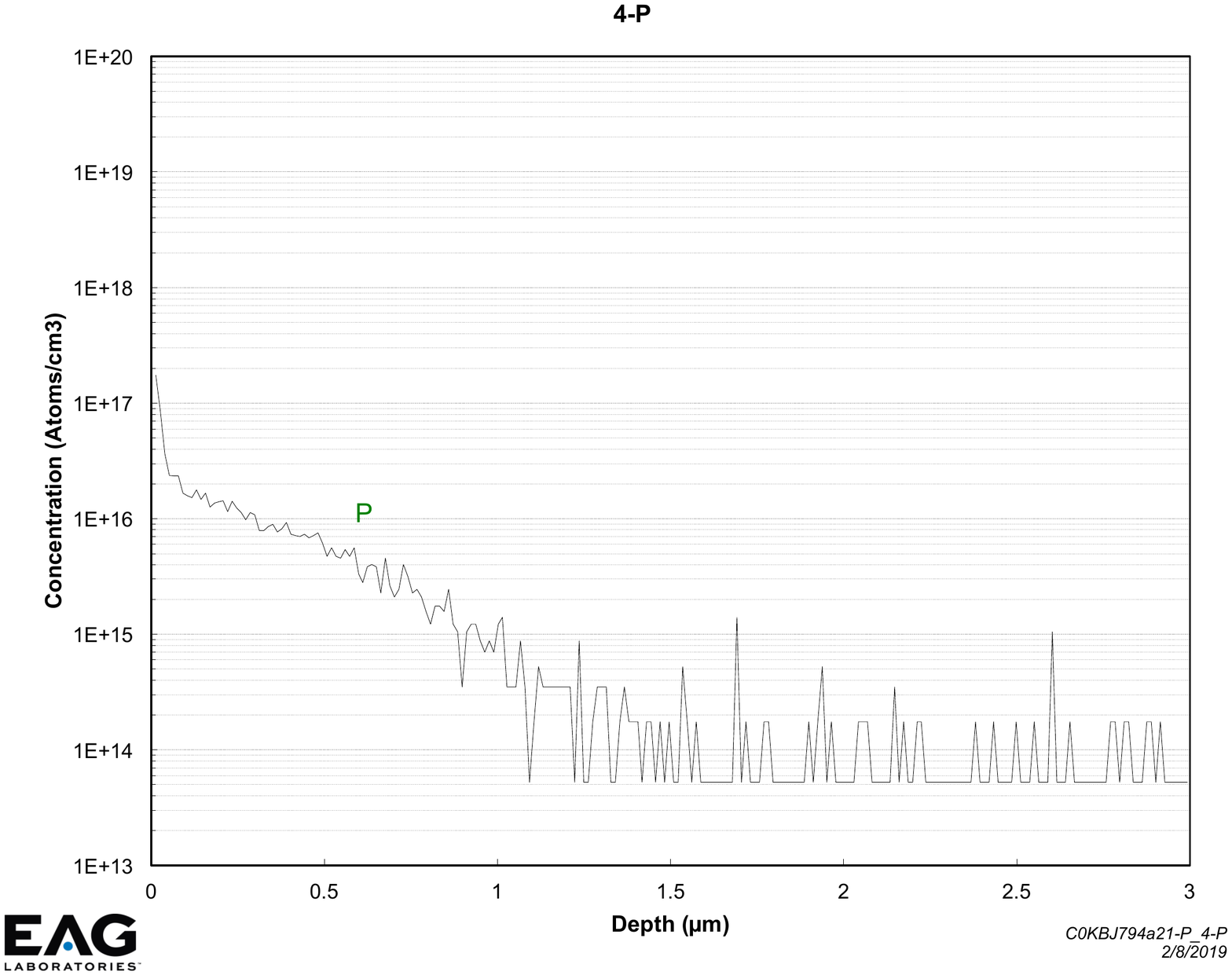}}
	\subfigure[b][Boron channel stop profile.]{\includegraphics[trim=0.5in 0.5in 0.5in 0.5in,clip,width=0.49\textwidth]{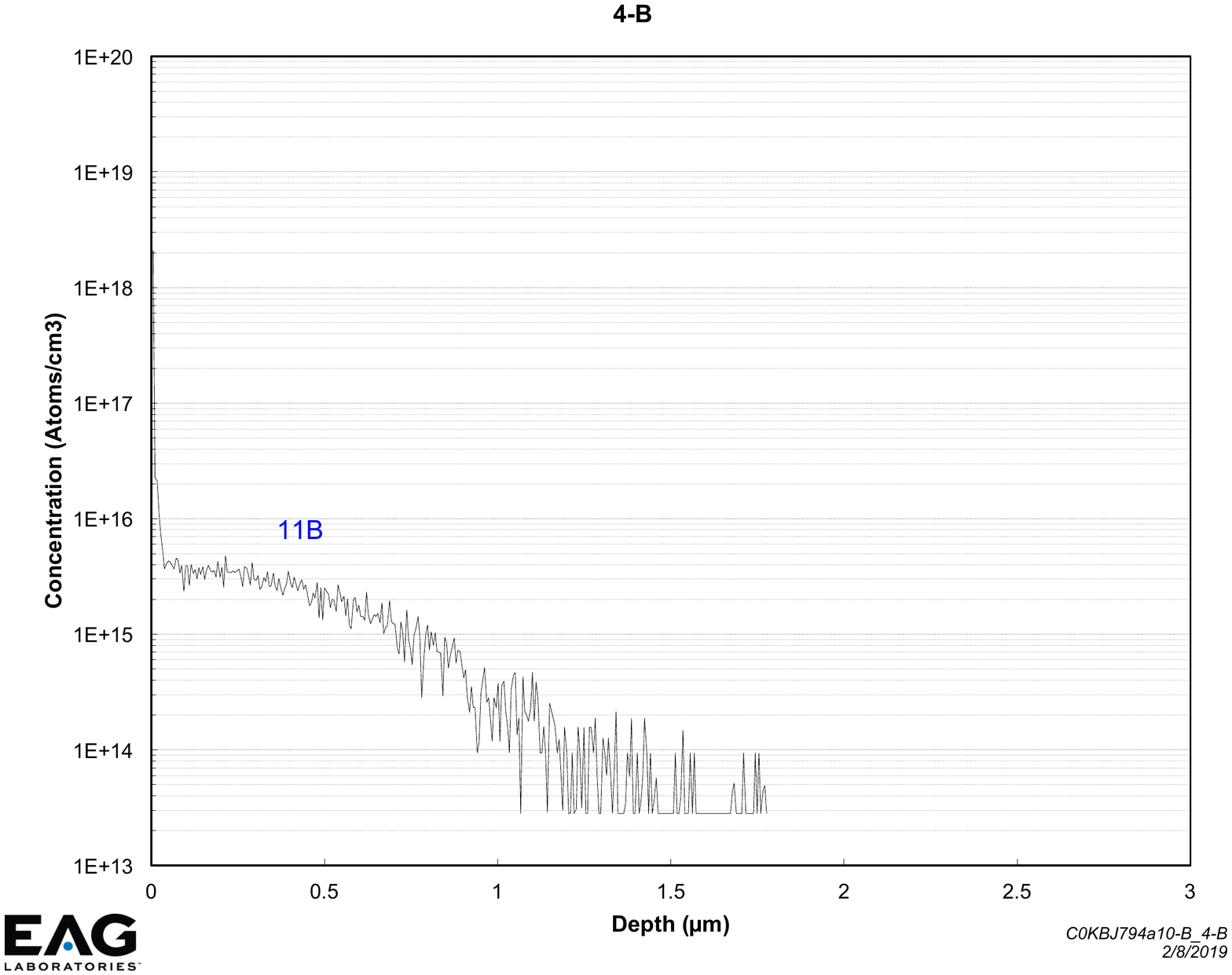}}

  \caption{SIMS dopant profiles of the E2V CCD250 device in the imaging array as measured by EAG laboratories \cite{EAG_website}.  The top left panel shows where in the array the measurements were made.  The top right panel shows the assumed device structure. The bottom two panels show the measured dopant profiles.  Based on the area factor, the phosphorous channel dopant values should be multiplied by 1.25, and the boron channel stop profiles should be multiplied by 25.0.}
  \label{E2V_SIMS}
  % Trim is Left Bottom Right Top
\end{figure}

\begin {figure}[H]
	\centering
	\subfigure[b][Diagram of the regions measured in the periphery.]{\includegraphics[trim=0.5in 0.5in 0.5in 1.5in,clip,width=0.49\textwidth]{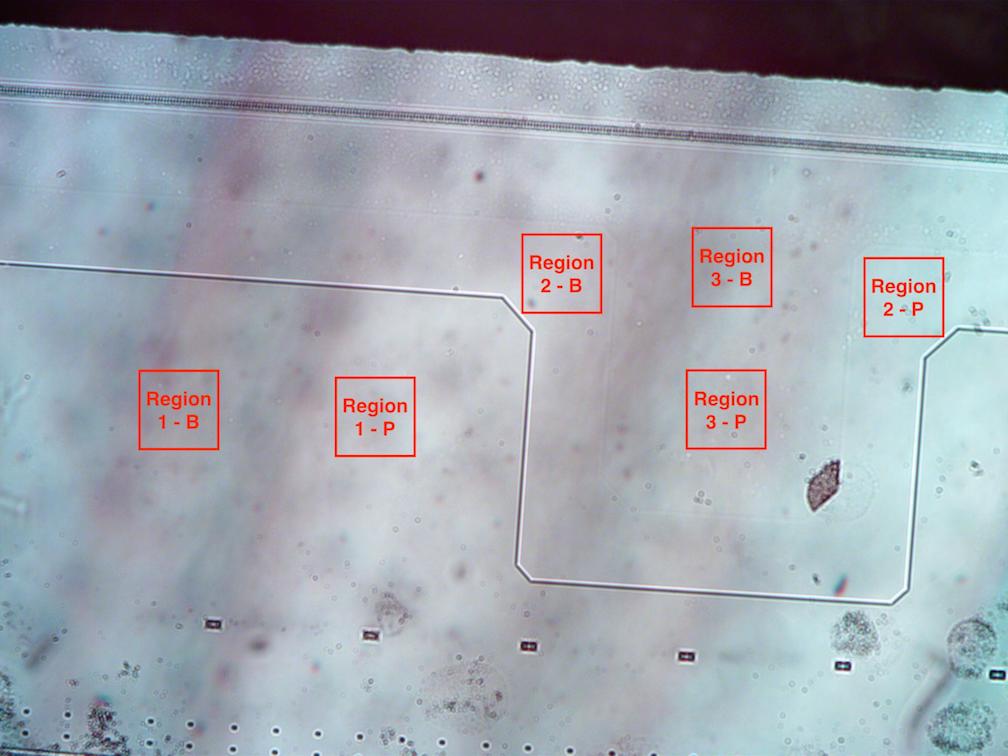}}
	\subfigure[b][Phosphorous channel profile.]{\includegraphics[trim=0.5in 0.5in 0.5in 0.5in,clip,width=0.49\textwidth]{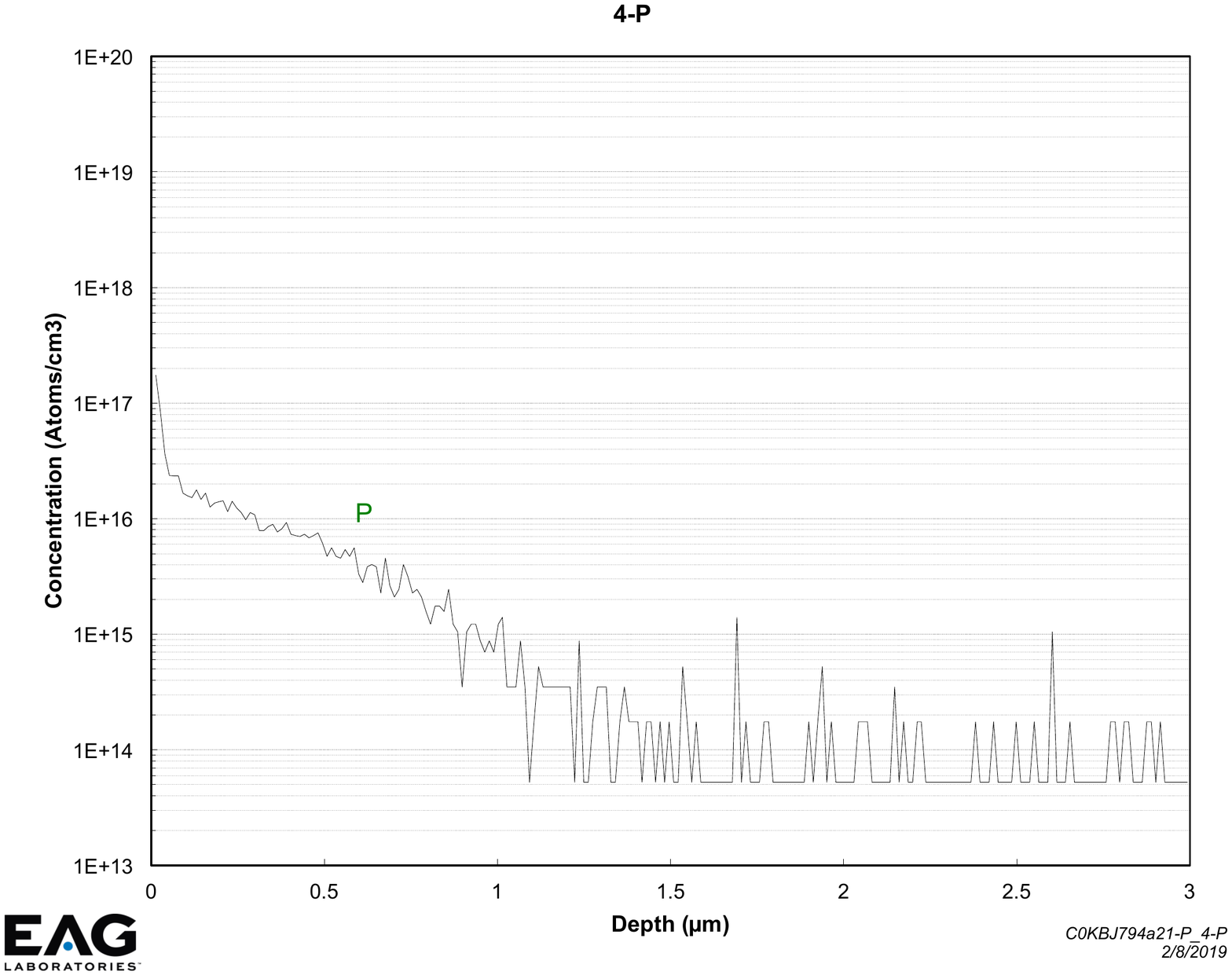}}
	\subfigure[b][Boron channel stop profile.]{\includegraphics[trim=0.5in 0.5in 0.5in 0.5in,clip,width=0.49\textwidth]{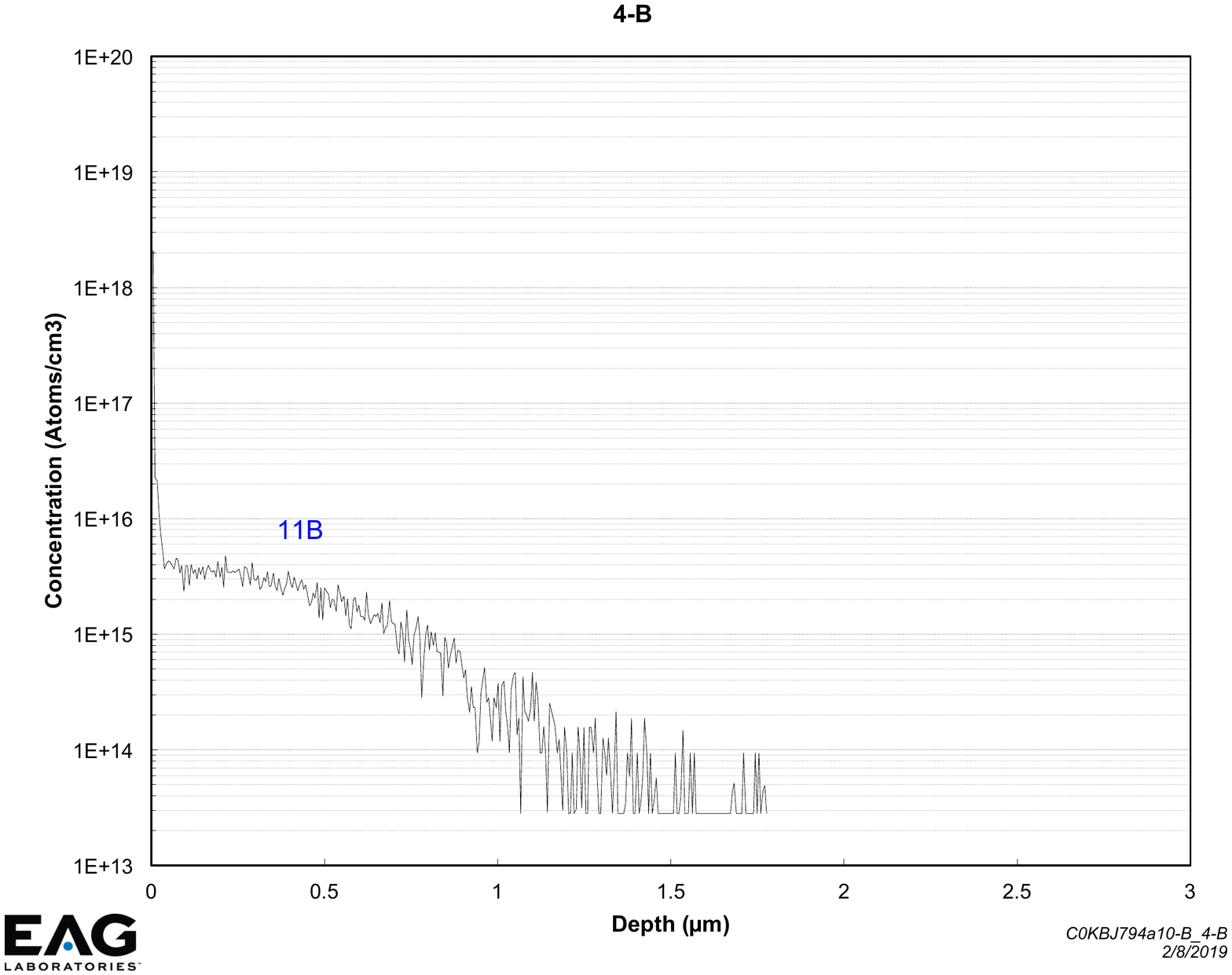}}

  \caption{SIMS dopant profiles of the E2V CCD250 device in the periphery as measured by EAG laboratories \cite{EAG_website}.  The top left panel shows where in the periphery the measurements were made.  The remaining panels show the measured profiles.}
  \label{E2V_SIMS_2}
  % Trim is Left Bottom Right Top
\end{figure}

\begin {figure}[H]
  \centering
      \includegraphics[trim=0.0in 0.0in 0.0in 0.0in,clip,width=0.65\textwidth]{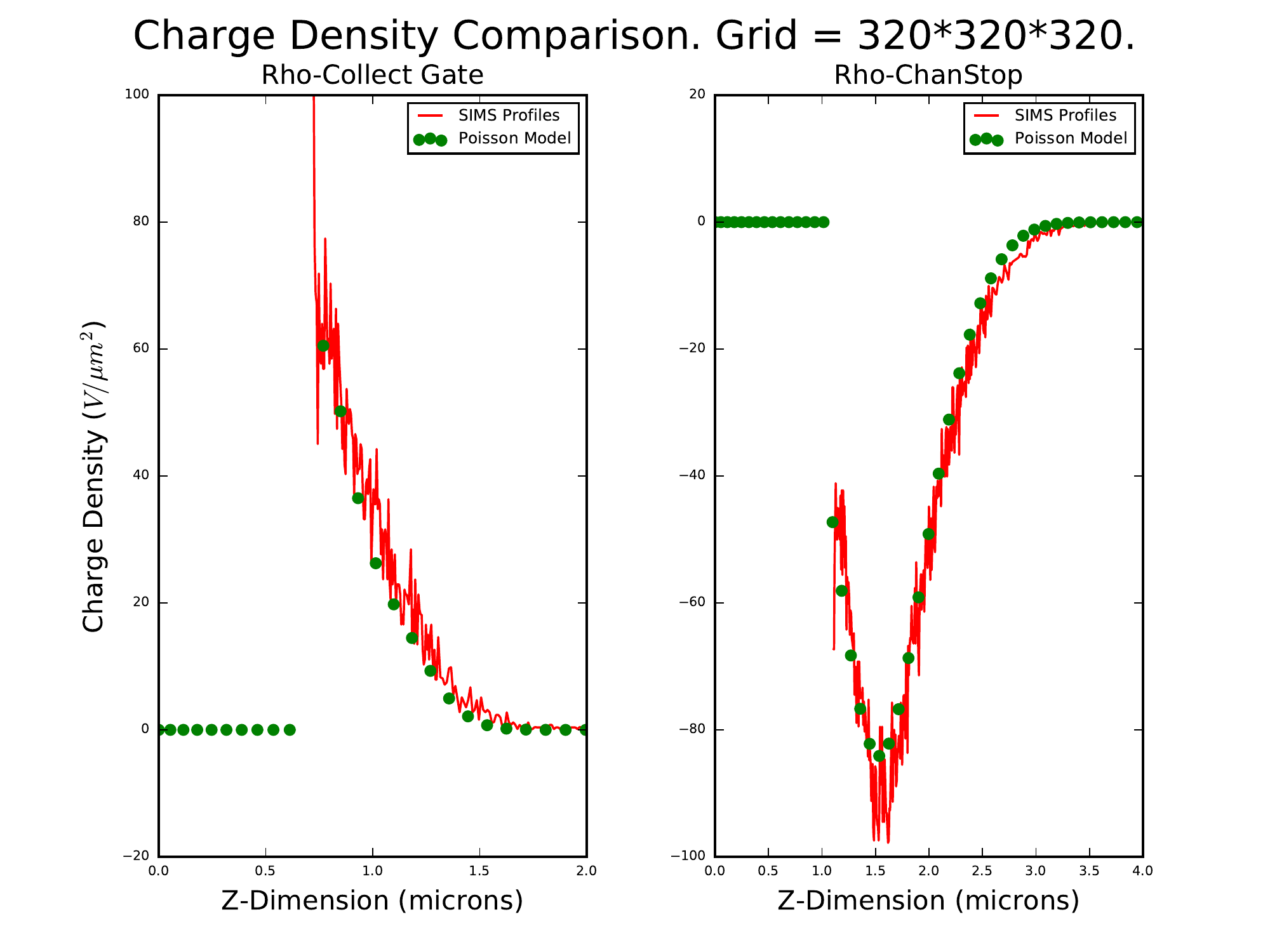}
  \caption{This figure shows that the dopant model for the ITL chip in the {\carlito Poisson\_CCD} simulator, which uses a combination of two Gaussians, accurately reproduces the measured SIMS profiles.  The parameters of the two Gaussians are given in the configuration file in Appendix \ref{ITL_Poisson_Appendix}.  The vertical scale is in ``code units'', which is the charge density divided by $\rm \epsilon_{Si}$.}
  \label{ITL_SIMS_Fit}
  % Trim is Left Bottom Right Top
\end{figure}

\begin {figure}[H]
  \centering
      \includegraphics[trim=0.0in 0.0in 0.0in 0.0in,clip,width=0.65\textwidth]{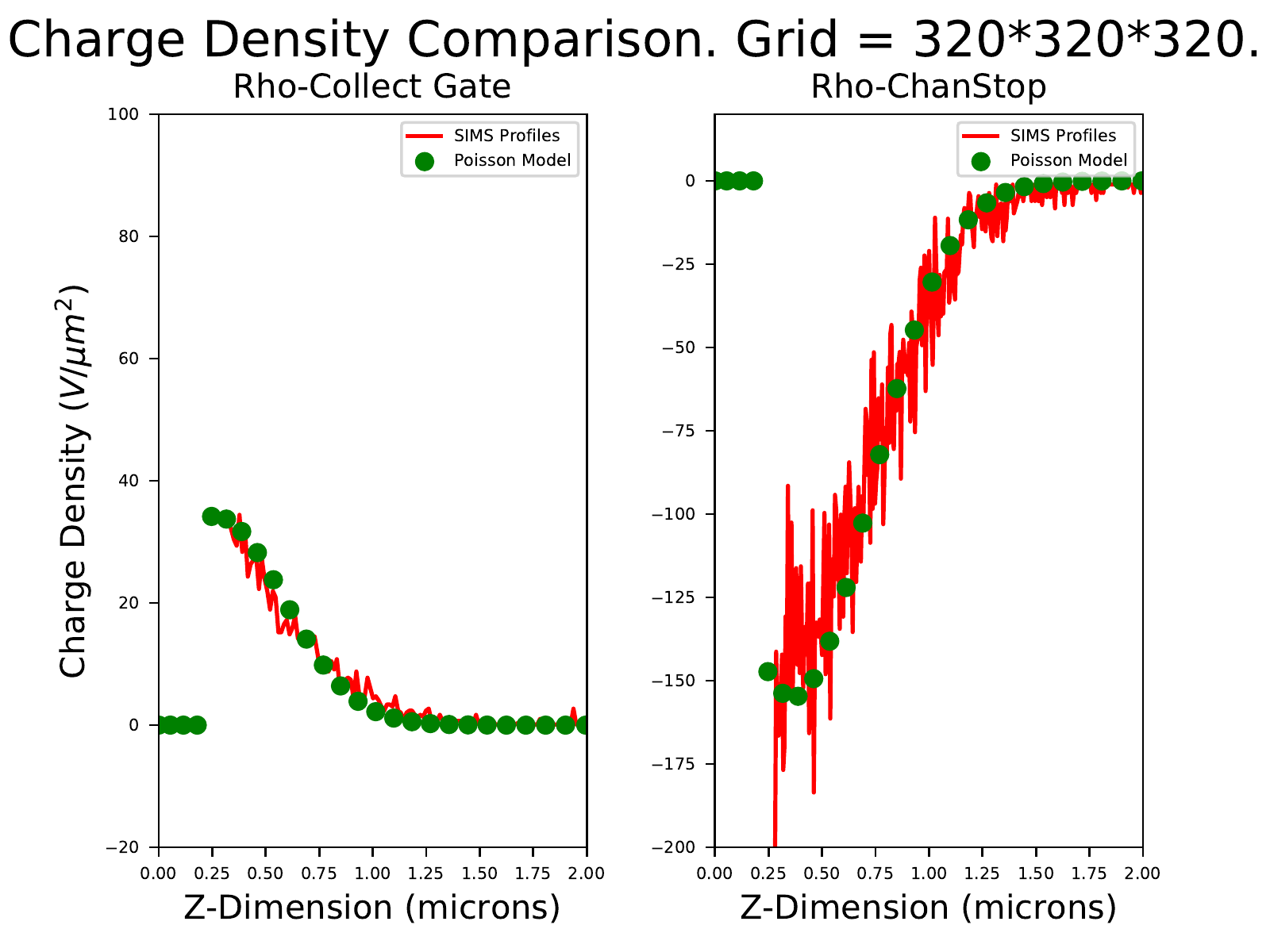}
  \caption{This figure shows that the dopant model for the E2V chip in the {\carlito Poisson\_CCD} simulator, which uses a single Gaussian profile for each region, accurately reproduces the measured SIMS profiles.  The parameters of the Gaussians are given in the configuration file in Appendix \ref{E2V_Poisson_Appendix}.  The vertical scale is in ``code units'', which is the charge density divided by $\rm \epsilon_{Si}$.}
  \label{E2V_SIMS_Fit}
  % Trim is Left Bottom Right Top
\end{figure}

\begin {figure}[H]
    \centering
	\subfigure[b][Charge packet with 100,000 electrons]{\includegraphics[trim=0.0in 0.0in 0.0in 0.0in,clip,width=0.40\textwidth]{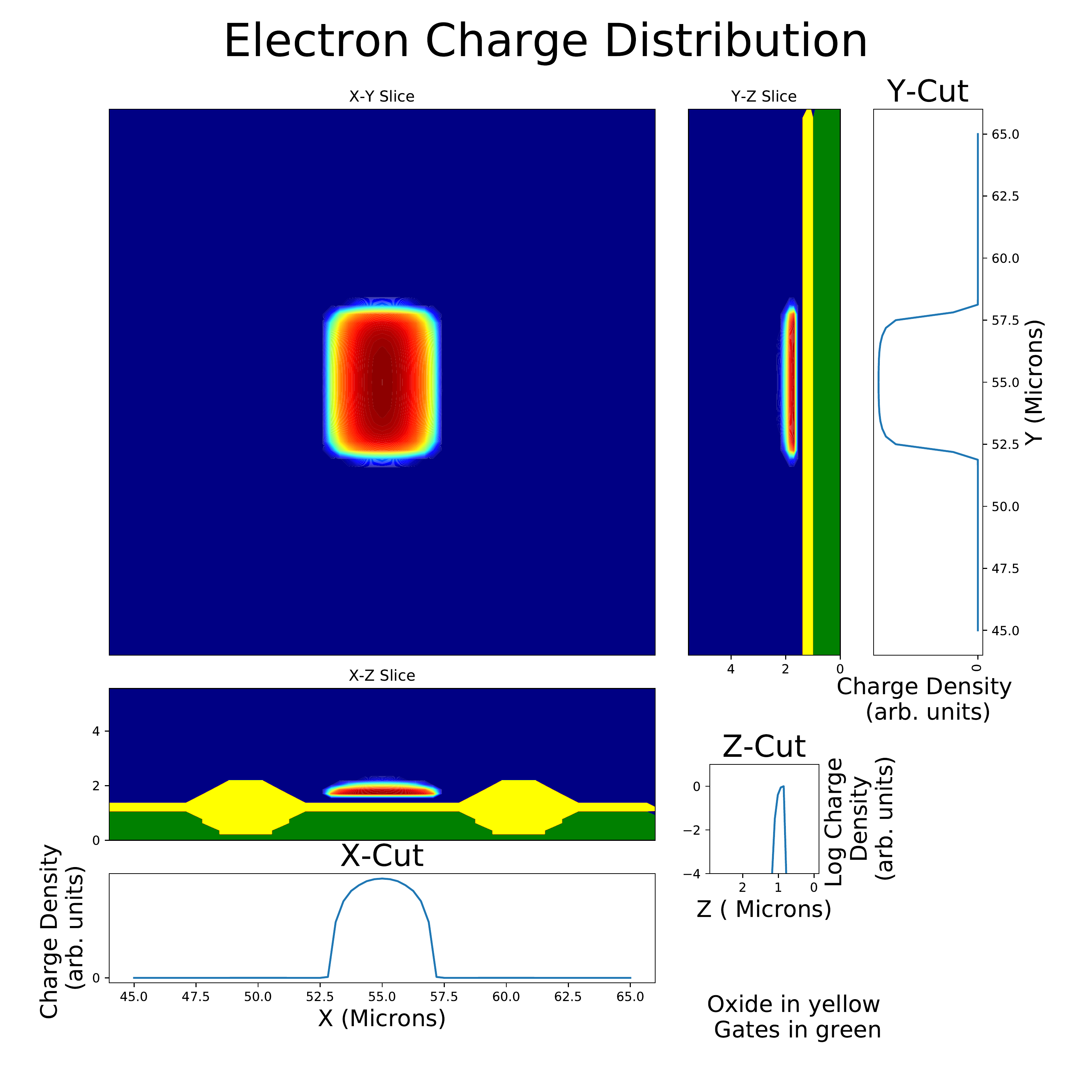}}    
	\subfigure[b][Pixel distortion from the central charge packet]{\includegraphics[trim=1.0in 0.0in 1.0in 0.0in,clip,width=0.65\textwidth]{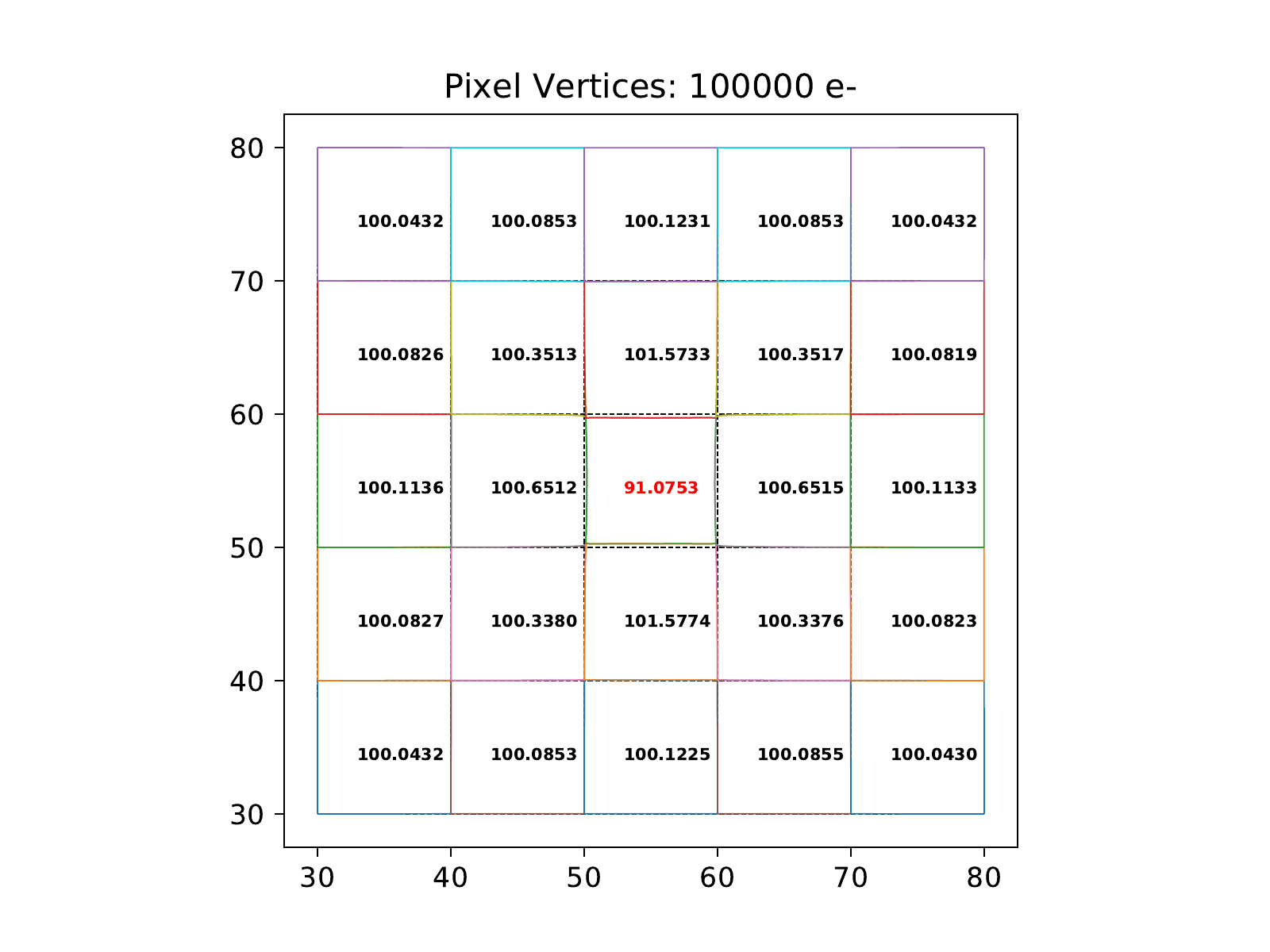}}
  \caption{Simulation of pixel distortions in an ITL chip when the central pixel contains 100,000 electrons and the surrounding pixels are empty.  These distortions are obtained by solving Poisson's equation for the potentials in the CCD, then tracking electrons down and using a binary search to determine the pixel boundaries.  As expected, the central pixel loses area and the surrounding pixels all gain area.  Note that the loss in area of the central pixel is greater than the sum of the area gains of the surrounding pixels because there are more distant pixels which are not plotted here and which also gain area.}
  \label{ITL_Pixel_Distortions}
  % Trim is Left Bottom Right Top
  \end{figure}

\begin {figure}[H]
    \centering
	\subfigure[b][Charge packet with 100,000 electrons]{\includegraphics[trim=0.0in 0.0in 0.0in 0.0in,clip,width=0.40\textwidth]{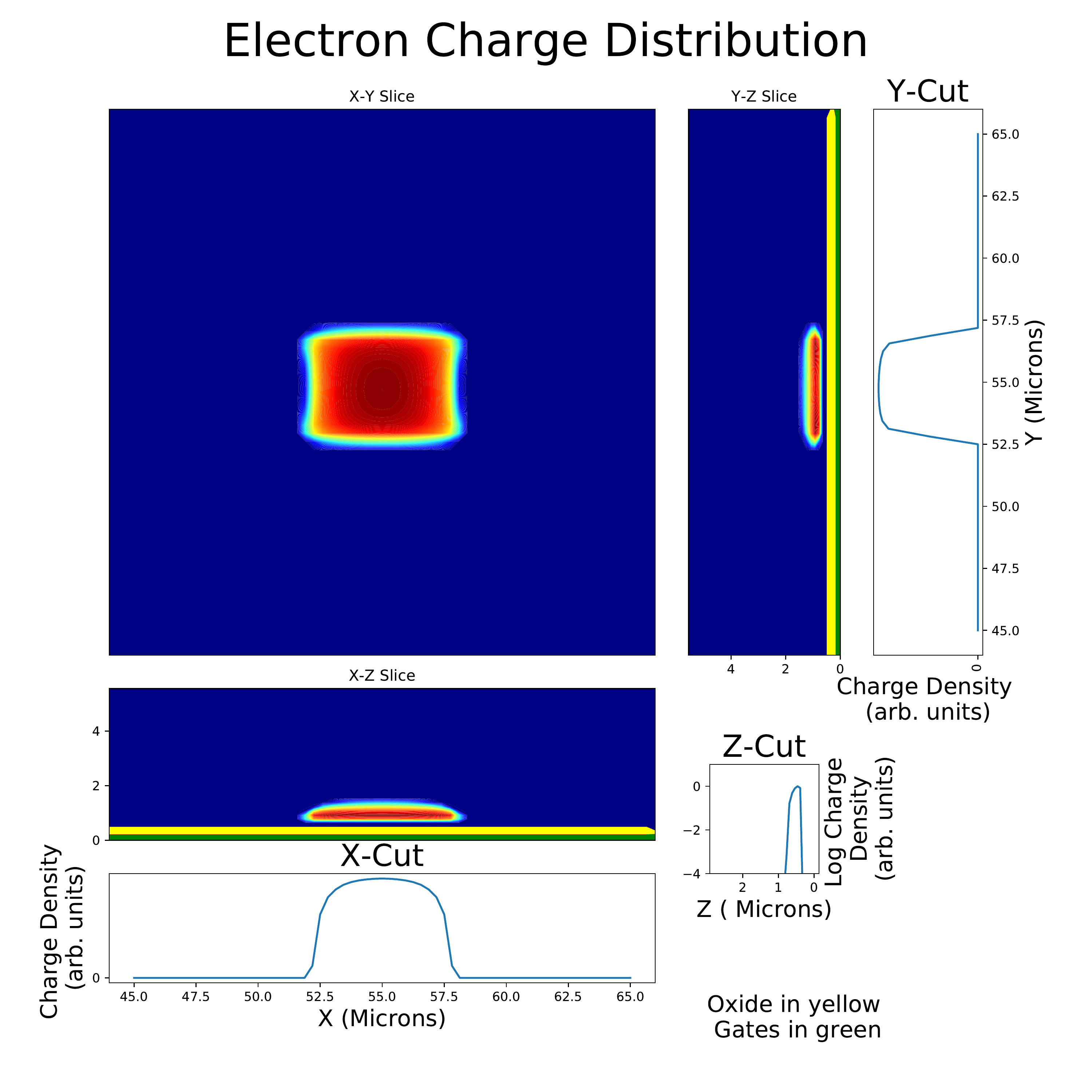}}    
	\subfigure[b][Pixel distortion from the central charge packet]{\includegraphics[trim=1.0in 0.0in 1.0in 0.0in,clip,width=0.65\textwidth]{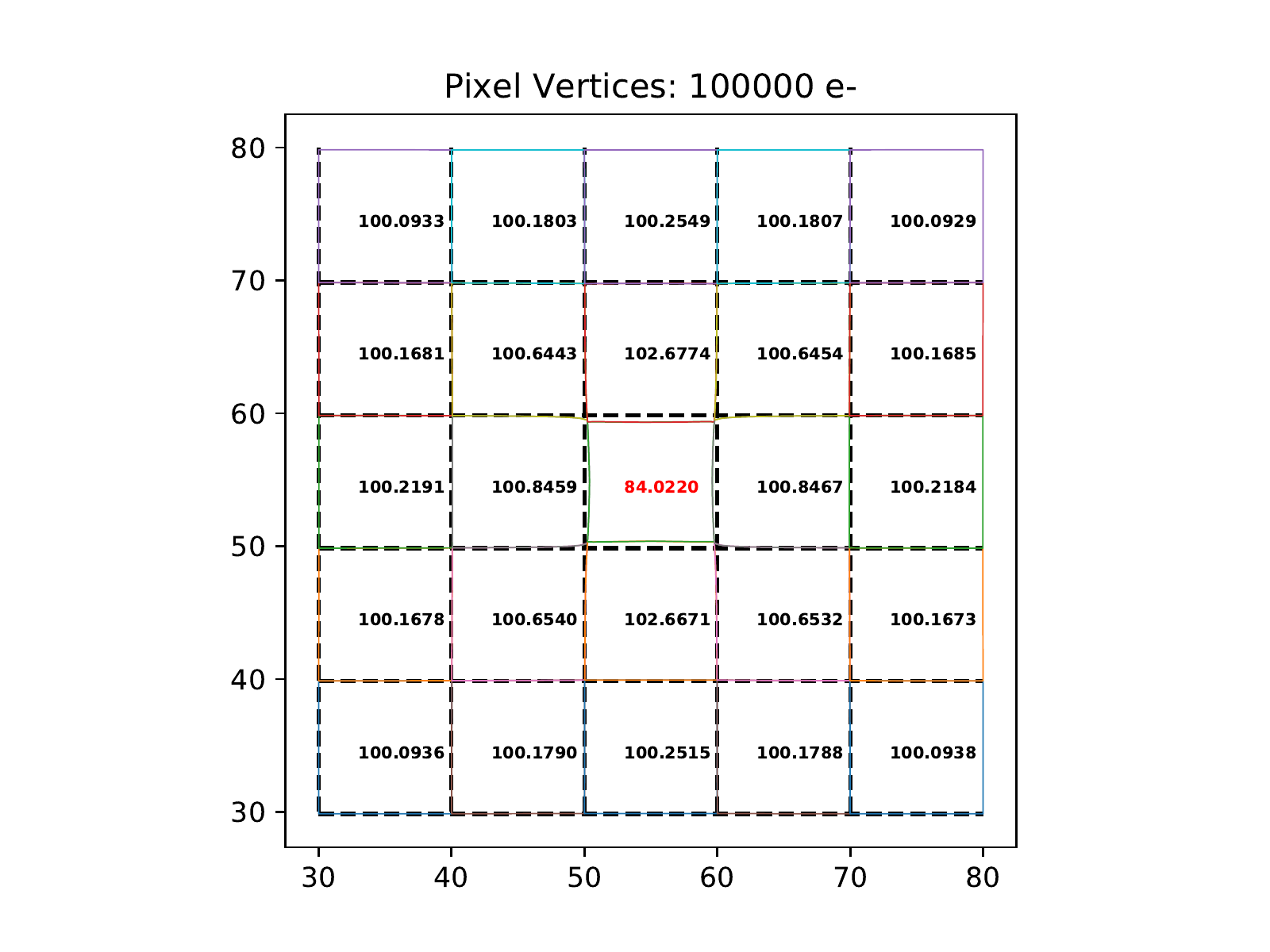}}
  \caption{This is the same as Figure \ref{ITL_Pixel_Distortions}, but for the E2V CCD250 chip.}
  \label{E2V_Pixel_Distortions}
  % Trim is Left Bottom Right Top
  \end{figure}

\begin {figure}[H]
	\centering
	\includegraphics[trim=0.5in 0.0in 1.0in 0.0in,clip,width=0.80\textwidth]{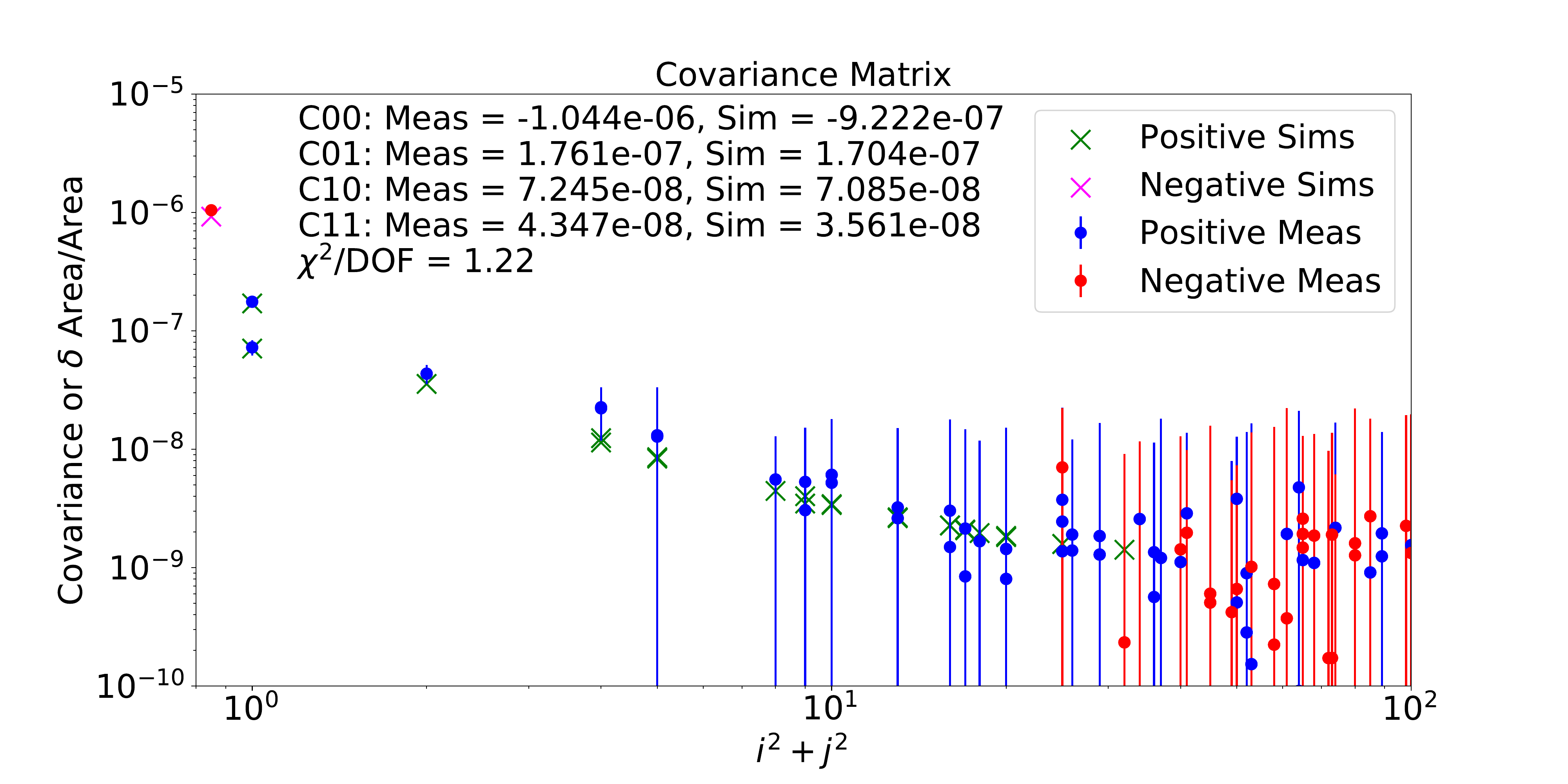}
  \caption{Correlation measurements and simulations for the ITL STA3800C chip.  The simulated pixel area distortions (see Figure \ref{ITL_Pixel_Distortions}) accurately determine the measured pixel-pixel correlations as measured on flat pairs.  The circles are the measured correlations, calculated as described in the text.  The crosses are the fractional area distortions as simulated by the {\carlito Poisson\_CCD} code and shown in  Figure \ref{ITL_Pixel_Distortions}.  The leftmost point (the central pixel) has been shifted to an X-axis value of 0.8 to allow plotting it on this log-log plot.  The simulations have been informed by physical analysis, including SIMS dopant profiling and measurements of physical dimensions, as discussed in the text.  Both the correlation measurements and the simulations have been normalized to the distortion caused by one electron.  The agreement is quite good.  The asymmetry of the nearest neighbor pixels is correctly modeled, and the simulated values agree with the measurements within the statistical errors.}
  \label{ITL_Correlation_Sims}
  % Trim is Left Bottom Right Top
  \end{figure}

\begin {figure}[H]
	\centering
        \includegraphics[trim = 0.5in 0.0in 1.0in 0.0in, clip, width=0.80\textwidth]{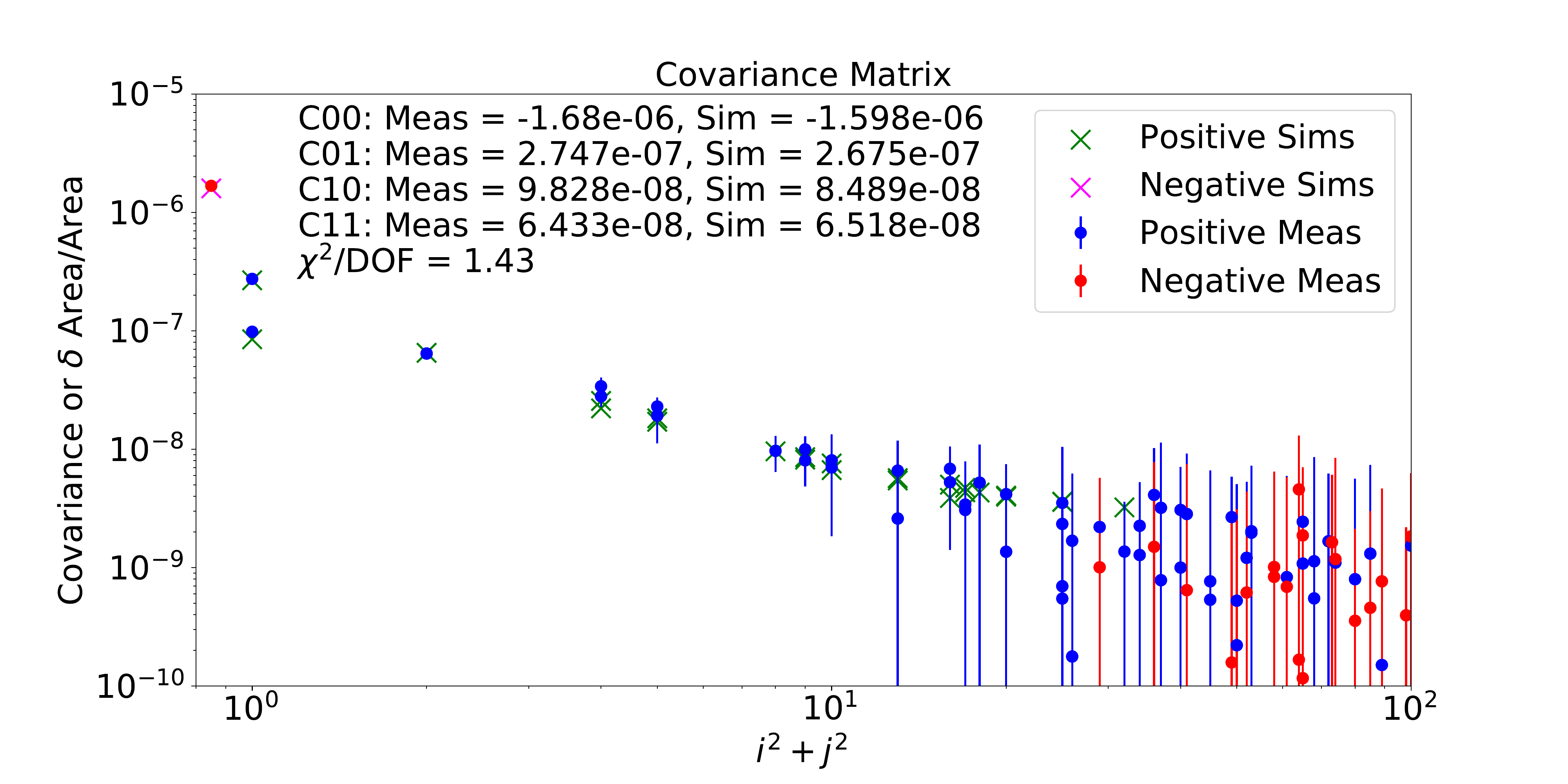}
  \caption{This is the same as Figure \ref{ITL_Correlation_Sims}, but for the E2V CCD250 chip, using the pixel distortions shown in Figure \ref{E2V_Pixel_Distortions}.  The agreement is not quite as good as in the ITL case, but is still good.}
  \label{E2V_Correlation_Sims}
  % Trim is Left Bottom Right Top
  \end{figure}

\begin {figure}[H]
	\centering
	\subfigure[b][``Dots'' in center]{\includegraphics[trim=0.0in 0.1in 0.0in 0.2in,clip,width=0.49\textwidth]{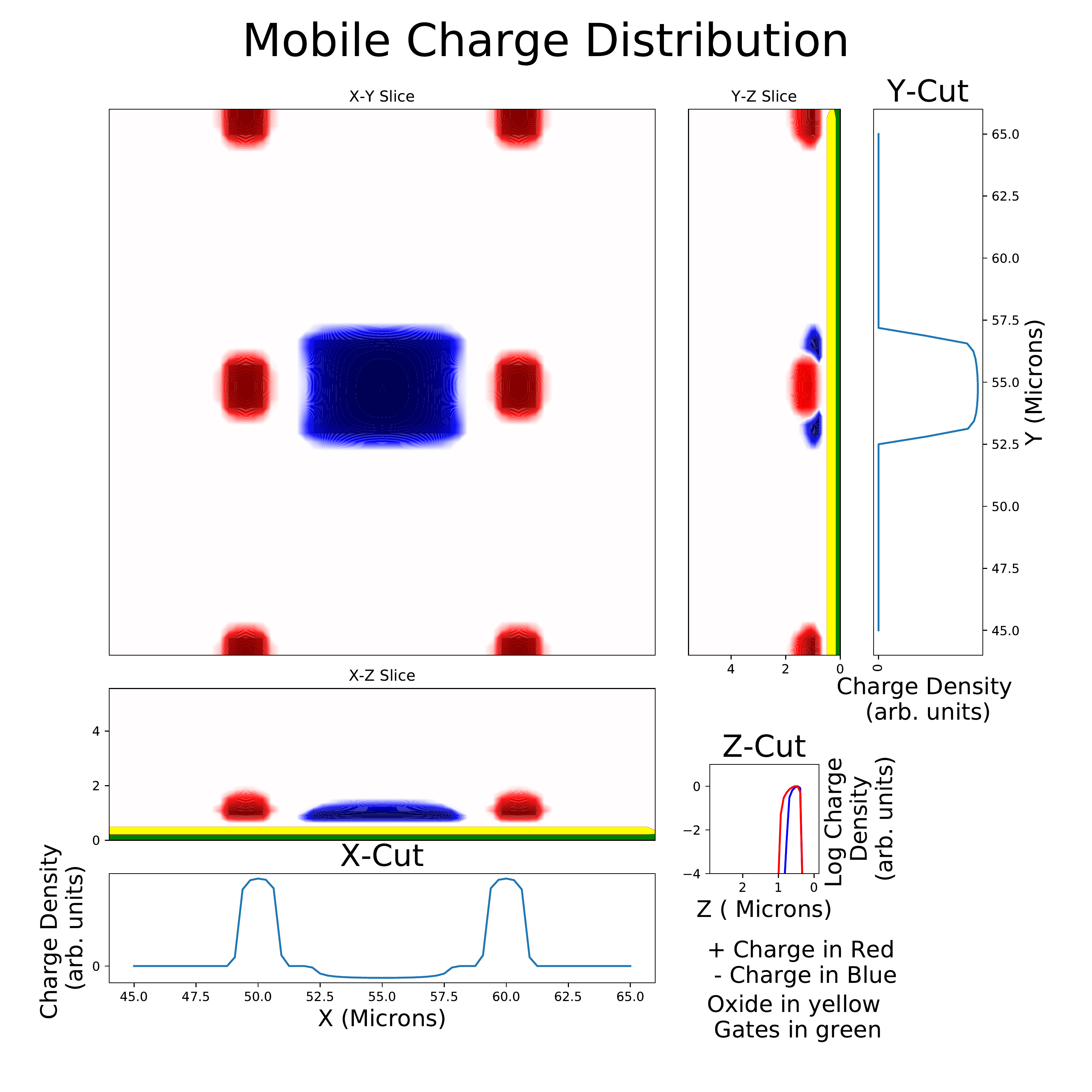}}
	\subfigure[b][``Dots'' in corners]{\includegraphics[trim=0.0in 0.1in 0.0in 0.2in,clip,width=0.49\textwidth]{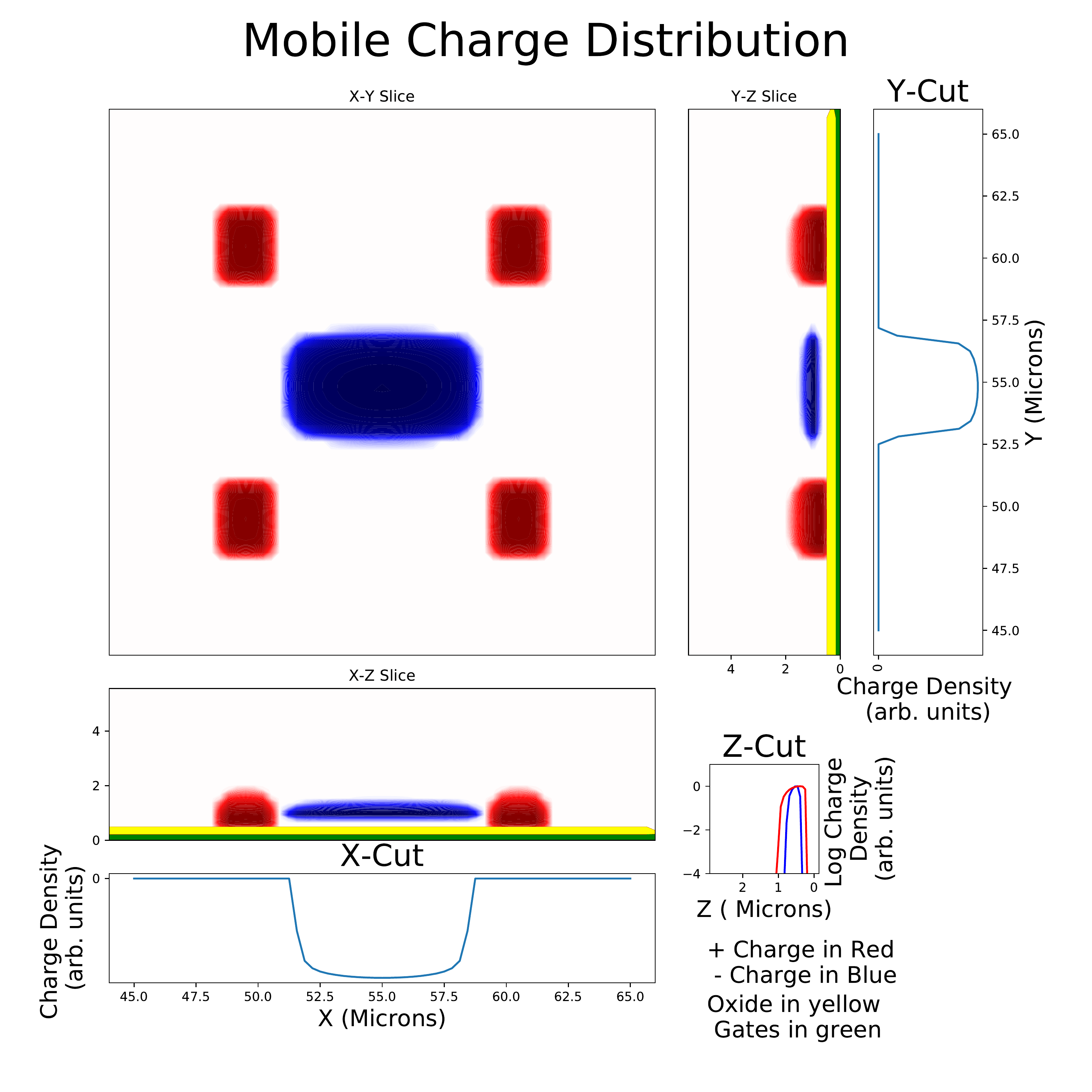}}
	\subfigure[b][``Dots'' in center]{\includegraphics[trim = 0.5in 0.1in 1.0in 0.5in, clip, width=0.65\textwidth]{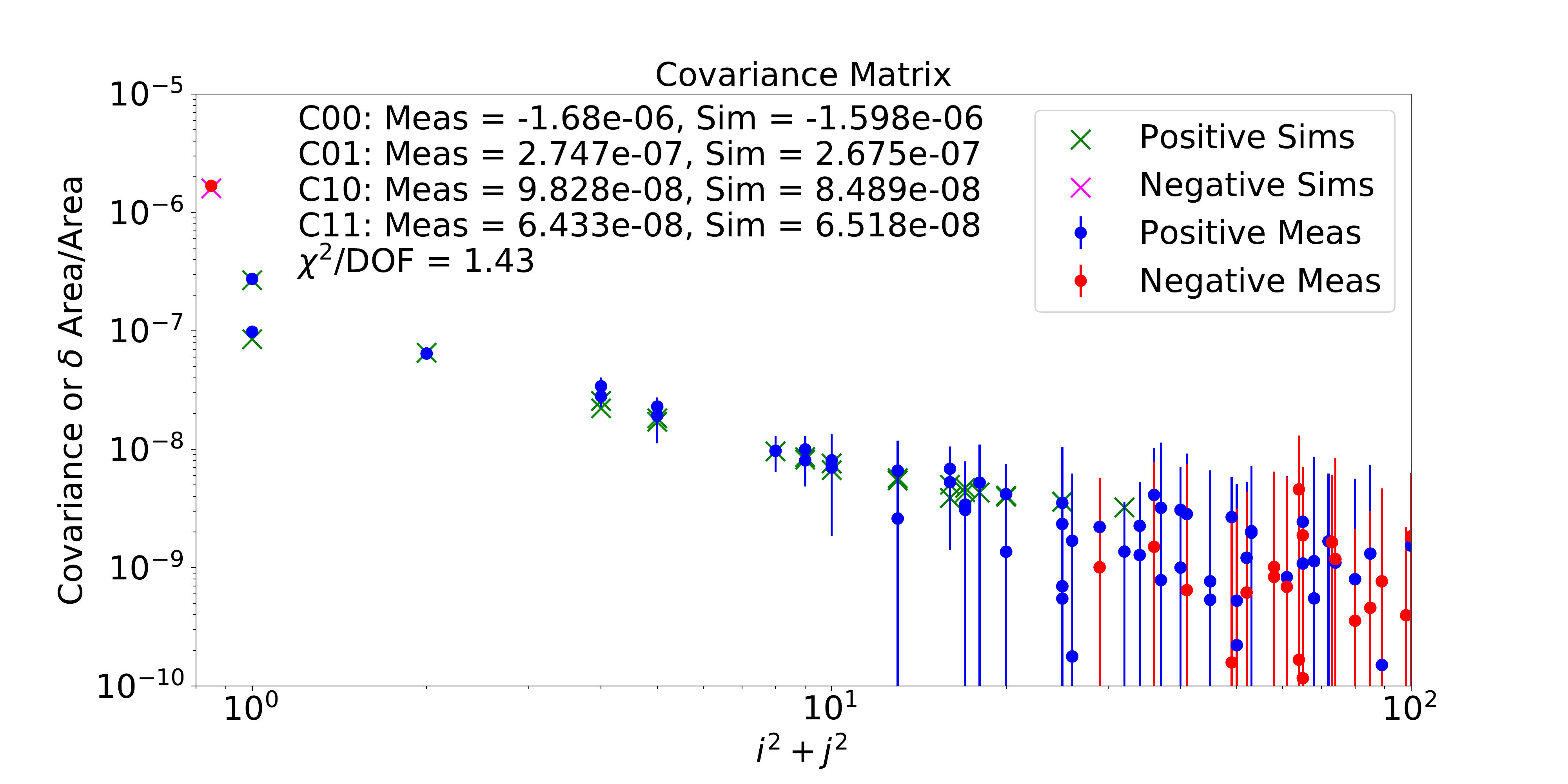}}
	\subfigure[b][``Dots'' in corners]{\includegraphics[trim = 0.5in 0.1in 1.0in 0.5in, clip, width=0.65\textwidth]{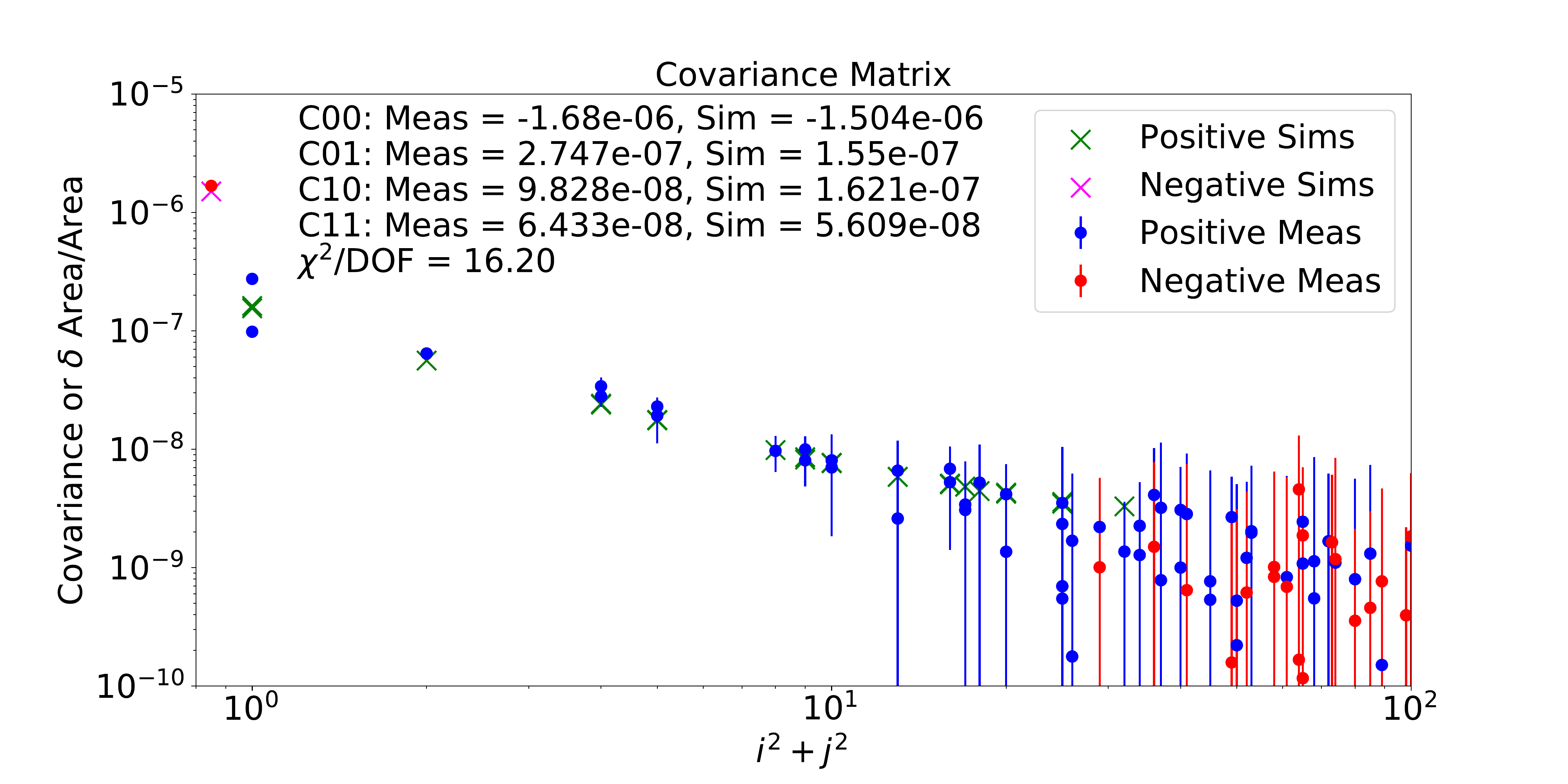}}

  \caption{Impact of the location of the ``dots'' in the E2V CCD array.  See Section \ref{Poisson_Model} for a fuller description of this issue.}
  \label{E2V_Dot_Location}
  % Trim is Left Bottom Right Top
\end{figure}

There is one caveat to be discussed in the case of the E2V sensor.  As discussed in Section \ref{E2V_SIMs_Section} and in Figure \ref{E2V_SIMS}, there is some uncertainty in the structure of the channel stops in the imaging array.  Here we assume that the channel stop implant is only present in the small rounded square ``dots'' which are visible in Figure \ref{E2V_SIMS}.  With this assumption, there is still some uncertainty as to the location of the ``dots'' with respect to the collecting gates when doing charge integration.  To try to answer this question, we ran the two simulations shown in Figure \ref{E2V_Dot_Location}.  The fit with the measured correlations is much better when assuming that the channel stop ``dots'' are centered on the collecting gates (the collecting gates are phases 2 and 3 in these measurements).  Therefore the simulations shown in Figure \ref {E2V_Pixel_Distortions} and Figure \ref{E2V_Correlation_Sims} are run with this assumption.

The configuration files used to perform the simulations shown in Figures \ref{ITL_Pixel_Distortions} and \ref{ITL_Correlation_Sims} are given in Appendix \ref{ITL_Poisson_Appendix}, and the files used for Figures \ref{E2V_Pixel_Distortions} and \ref{E2V_Correlation_Sims} are given in Appendix \ref{E2V_Poisson_Appendix}.

\section{Conclusions}
This work documents the physical and electrical measurements which have been done to characterize the CCDs from both vendors which are being used to build the LSST focal plane.  Based on these measurements, calibrated electrostatic simulations have been prepared to answer questions which may arise during LSST commissioning and operation.  A set of calibrated SPICE models for the ITL STA3800C CCD has also been built and is available should it prove useful. 

\section{Acknowledgments}
Andrew Bradshaw has been a key partner in building the UC Davis CCD lab, developing the software, and making the CCD measurements.  Kirk Gilmore's help in setting up the hardware and software for the CCDs from both vendors has also been invaluable.  Perry Gee has provided key support in software and networking.   Mike Lesser of ITL has provided helpful discussions and support.  Claire Juramy and Sven Herrmann contributed to the ITL waveform analysis in Appendix \ref{SPICE_Appendix}.   Of course, Tony Tyson's vision in building the CCD lab and constant support are much appreciated.  Financial support from DOE grant DE-SC0009999 and Heising-Simons Foundation grant 2015-106 are gratefully acknowledged.

\bibliographystyle{unsrt}
\bibliography{ccd}

\clearpage

%STILL NEEDS WORK

\clearpage
\appendix

\section{Appendix - STA3800C Output Driver SPICE Model}
\renewcommand{\thefigure}{A.\arabic{figure}}
\setcounter{figure}{0}
\label{SPICE_Appendix}

\begin{figure}[H]
  \begin{center}
    \hspace{-17.2mm}
    \includegraphics[trim=0.0in 5.3in 0.0in 1.3in,clip,width=12.85cm,height=1.0cm]{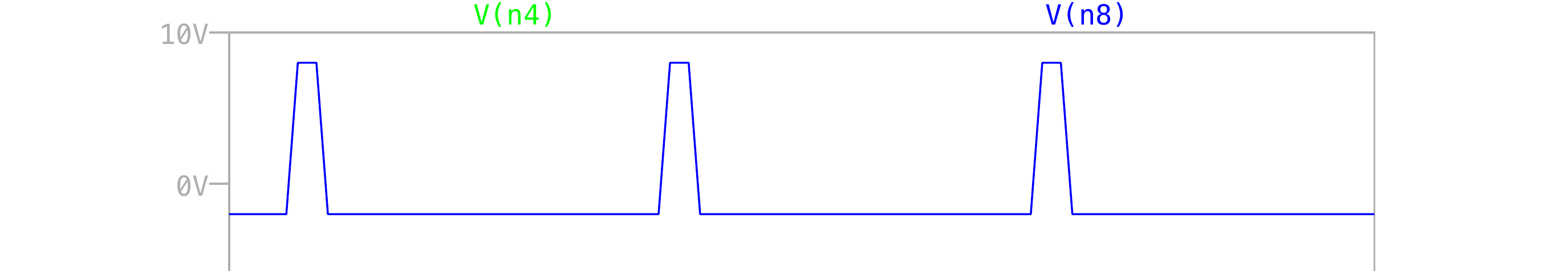} \\
    \line(1,0){300}\\
    \includegraphics[trim=0.0in 1.0in 0.0in 2.0in,clip,width=0.62\textwidth,height=4.3cm]{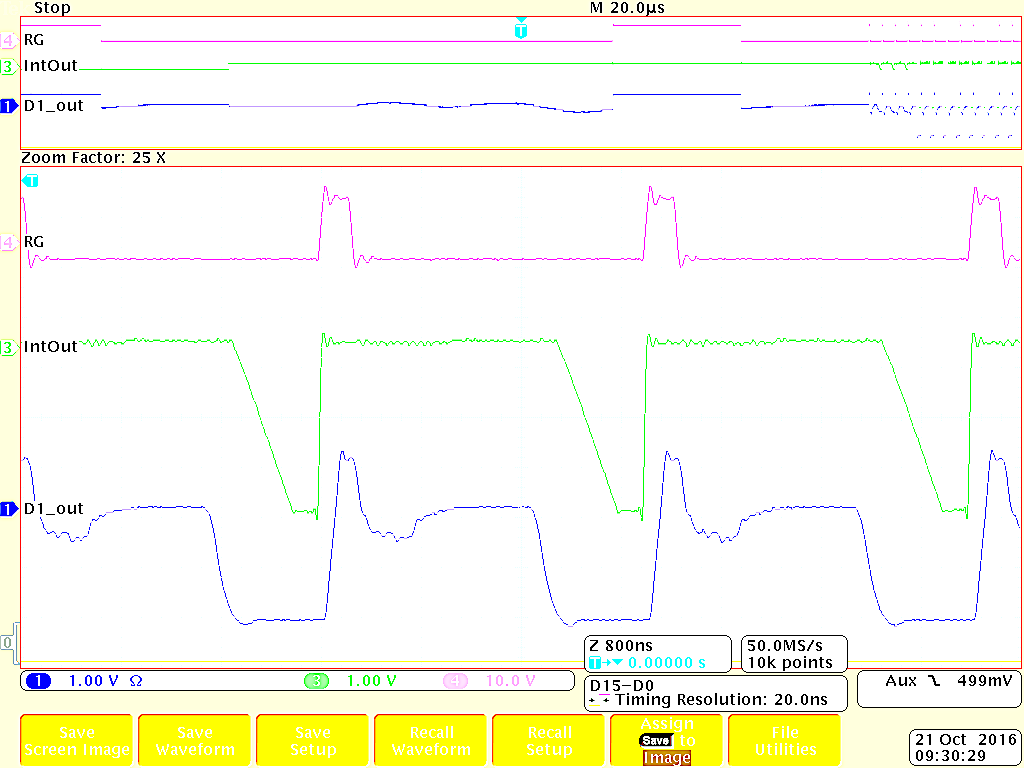}\\
    \line(1,0){300}\\
    \hspace{-14.4mm}
    \includegraphics[trim=0.0in 1.2in 0.5in 2.3in,clip,width=13.5cm,height=2.6cm]{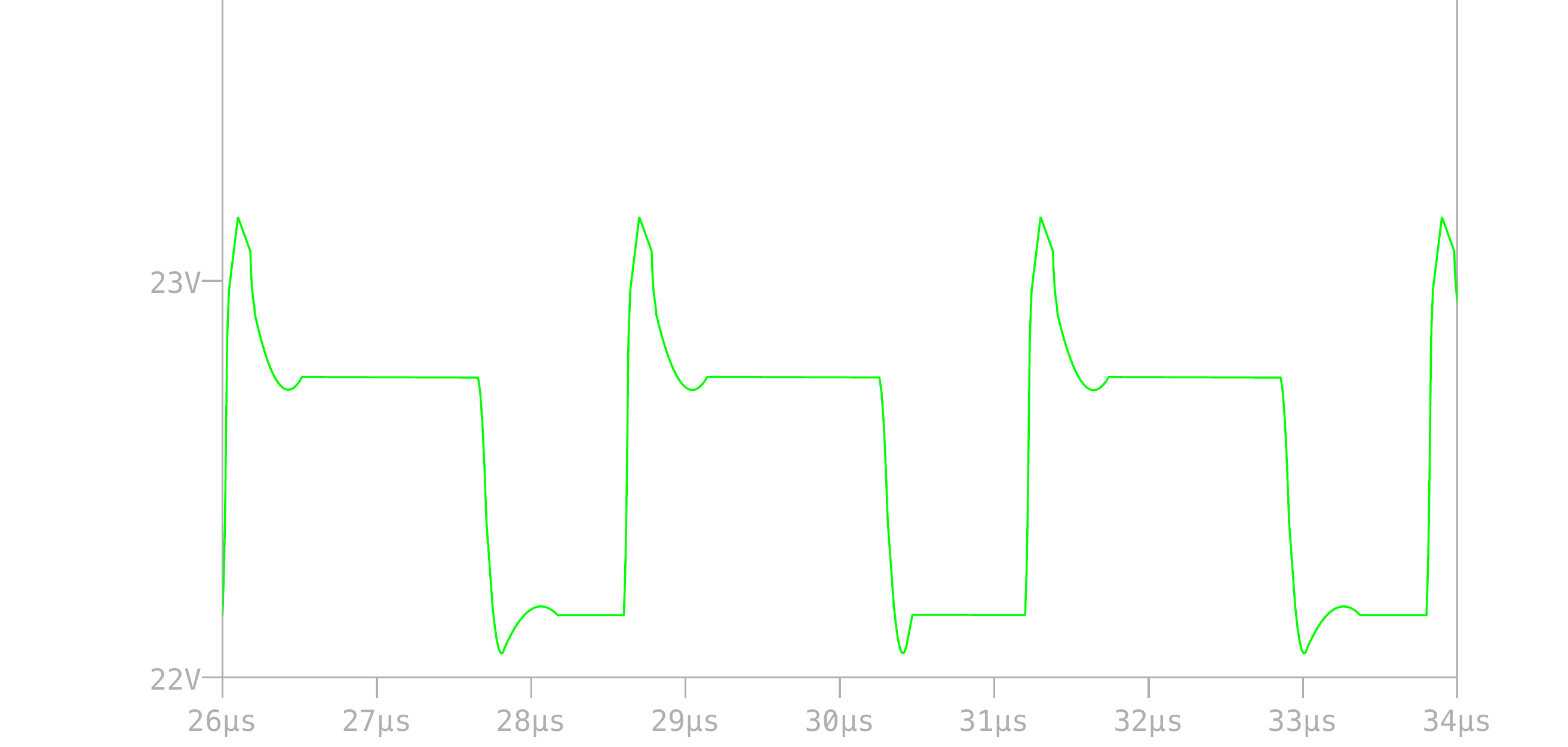}
  \end{center}
  \caption{Comparison of SPICE simulation to measured signals, as seen in the UC Davis LSST Optical Simulator. The top panel is the SPICE RG signal, which is an input to the simulation.  The center panel shows the measured waveforms, with RG in red, the dual-slope integrator output in green, and the CCD output in blue. The bottom panel is the SPICE simulation of the CCD output.  Scales have been adjusted to match.  The simulation of the CCD output matches the measured waveform relatively well.}
  \label{SPICE3}
\end{figure}

\begin{figure}[H]
  \centering
      \includegraphics[trim=0.0in 2.0in 0.0in 2.0in,clip,width=0.90\textwidth]{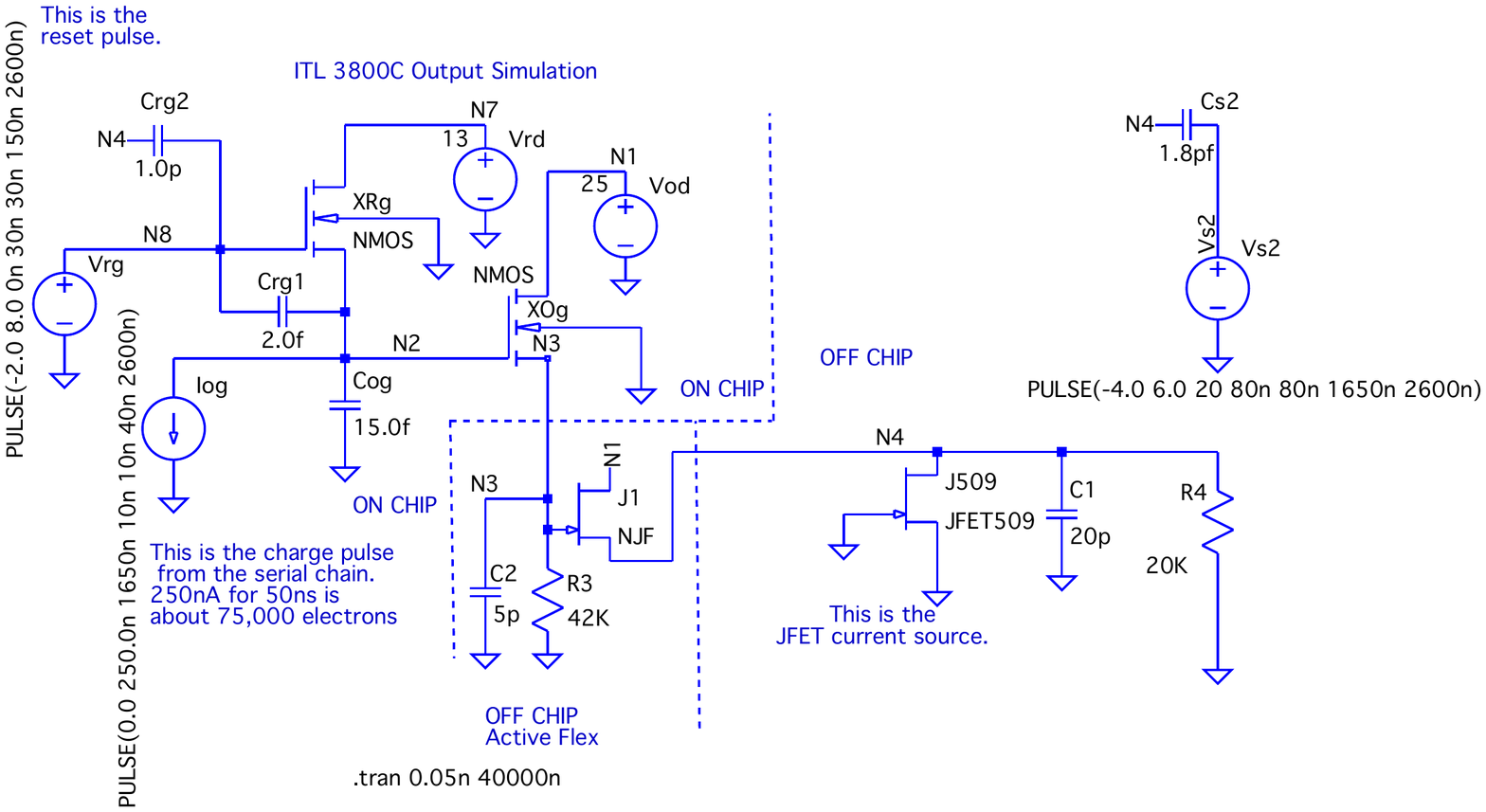}\\
  \begin{minipage}{0.49\textwidth}
    \begin{center}
      \includegraphics[trim=0.0in 0.0in 0.0in 0.0in,clip,width=0.99\textwidth]{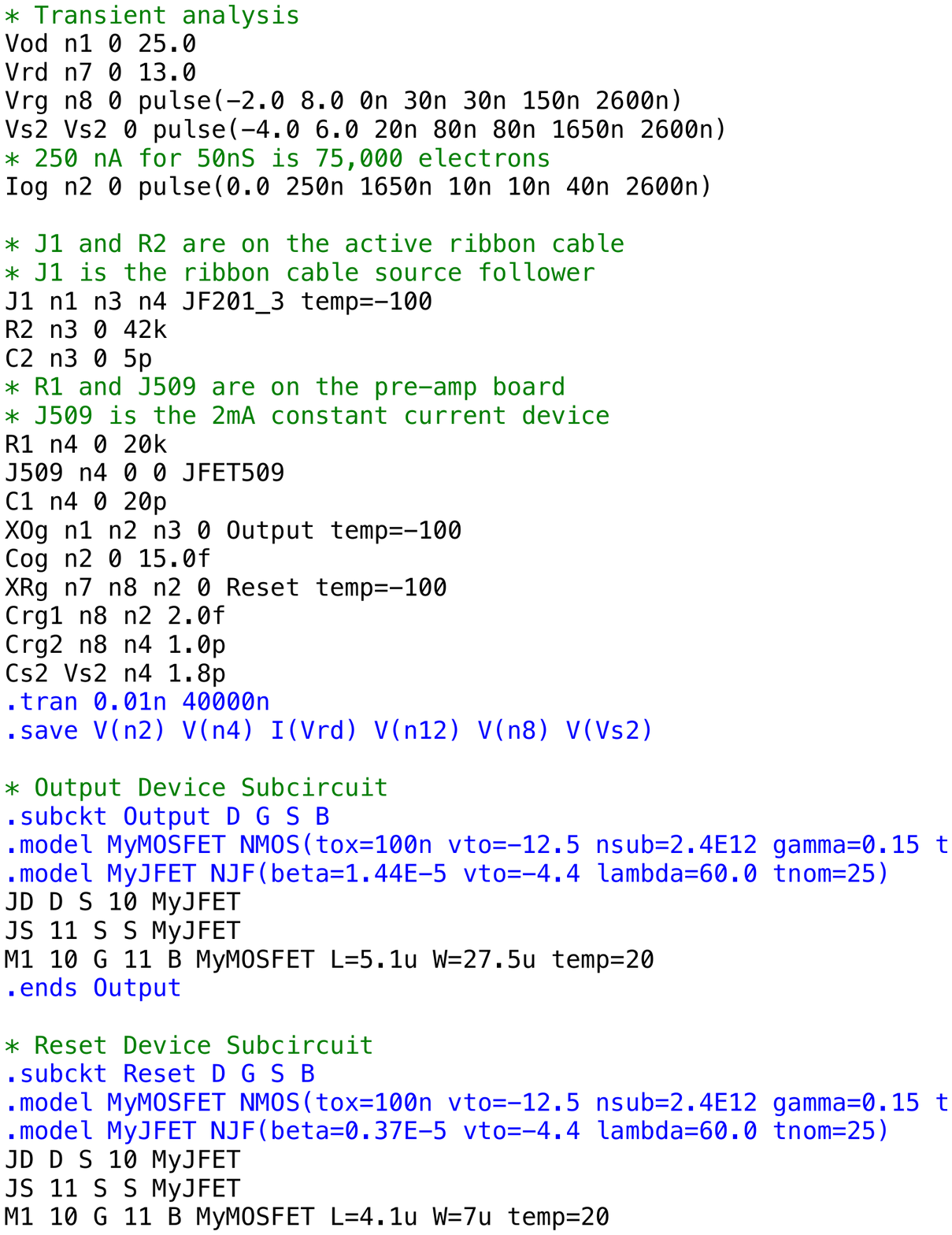}
    \end{center}
  \end{minipage}
  \begin{minipage}{0.49\textwidth}
    \begin{center}
      \includegraphics[trim=0.0in 0.0in 0.0in 0.0in,clip,width=0.99\textwidth]{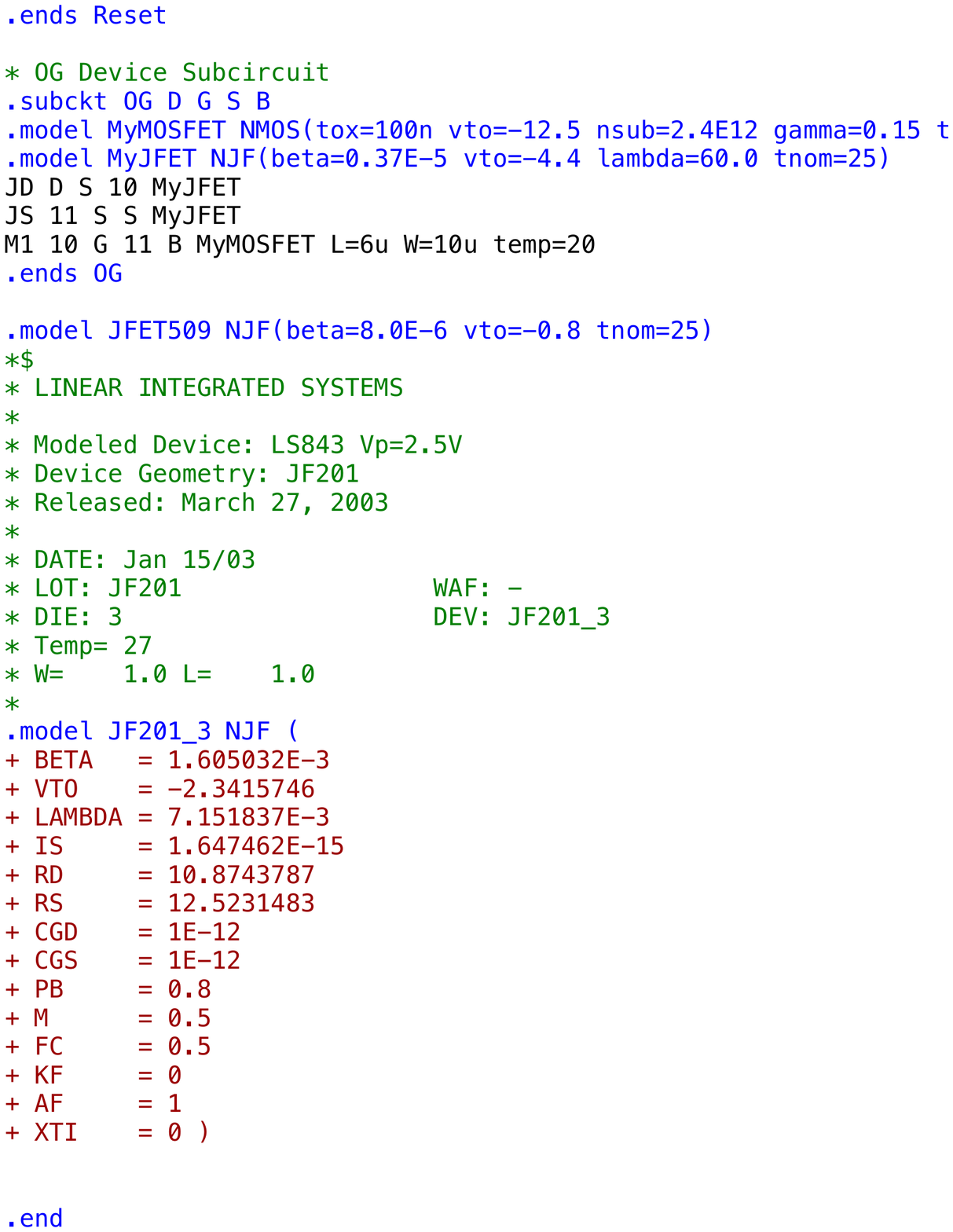}
    \end{center}
  \end{minipage}
  % Trim is Left Bottom Right Top
  \caption{Schematic and netlist used in the simulation shown in Figure \ref{SPICE3}.}
  \label{Schem_Net1}
\end{figure}

The model of the STA3800C output transistor that was discussed in Section \ref{SPICE_Model} was incorporated into more detailed SPICE simulations and used to model the output of the STA3800C CCD in two different controller environments.  The first used an SAO controller that was used in the LSST Optical Simulator at UC Davis (\cite{tyson2014}, \cite{bradshaw2015}, \cite{Lage_2017} ).  Figure \ref{SPICE3} shows the fit obtained between the measured output waveforms and the simulation.  The schematic and netlist used for this simulation are given in Figure \ref{Schem_Net1}.

\begin {figure}[H]
  \centering
  \subfigure[b][Waveforms of 9 ITL CCDs, as measured using the ASPIC Transparent Mode.]{\includegraphics[trim=0.4in 0.0in 0.7in 0.0in,clip,width=0.46\textwidth]{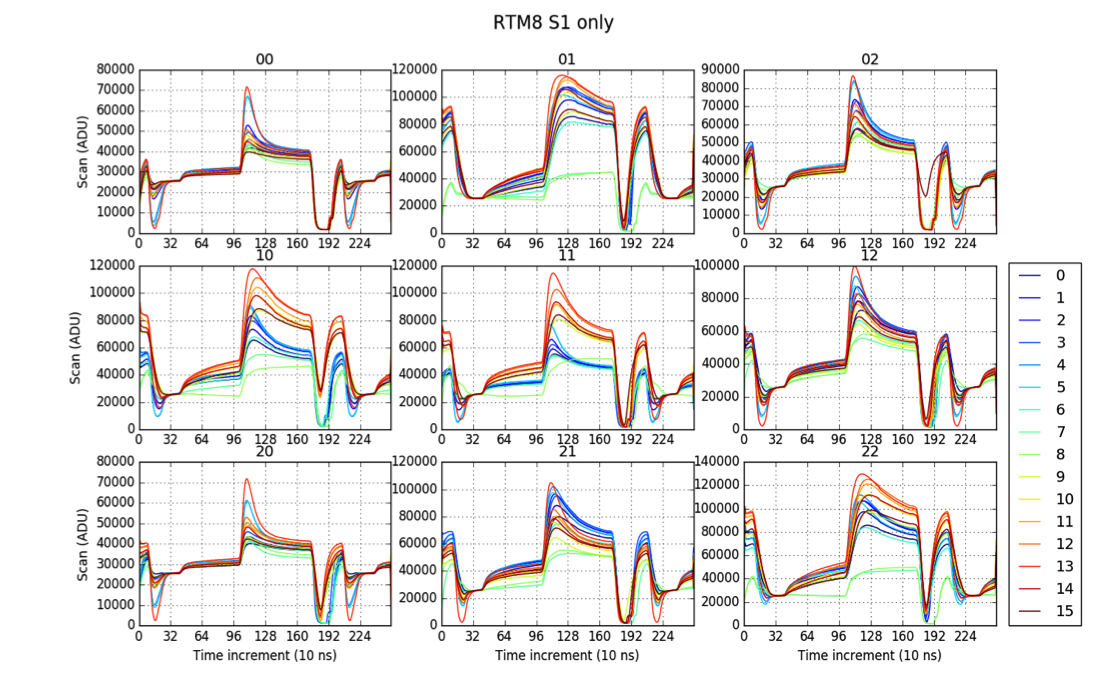}}
  \hspace{5mm}
  \subfigure[b][Simulation with varying Rsub. Rsub = 100(green), 1000(blue), 2000(red), 4000(cyan) Ohms.]{\includegraphics[trim=0.5in 0.0in 0.5in 1.0in,clip,width=0.46\textwidth]{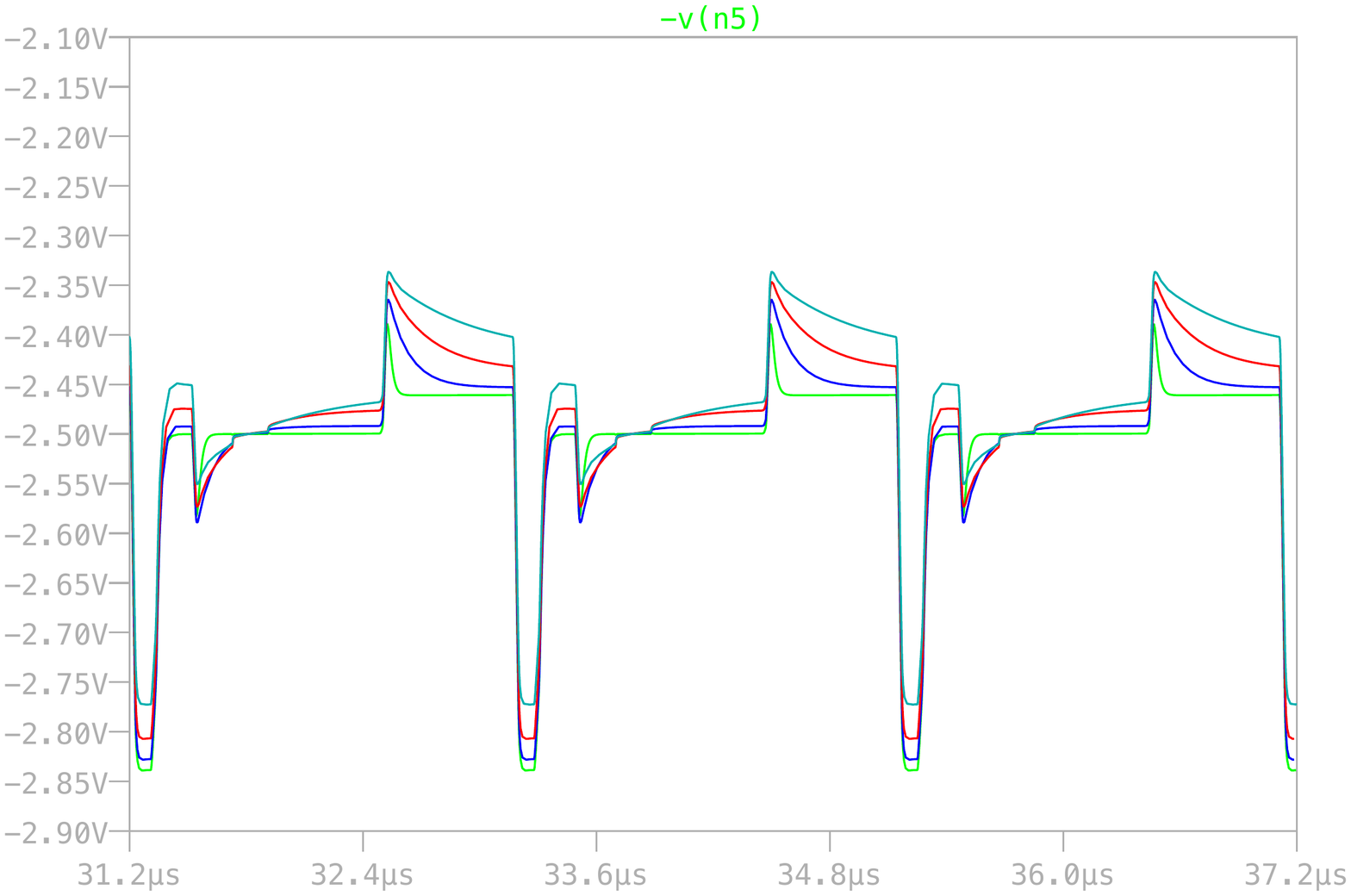}}
  \caption{Output waveforms with slow recovery are seen on some STA3800C chips.  The left panel shows these measured waveforms, and the right panel shows SPICE simulations of the output waveforms when substrate resistance is added to the output transistor.  The added substrate resistance causes the output waveform to mimic what is seen in the measurements.}
  \label{SPICE_RTM8}
  % Trim is Left Bottom Right Top
\end{figure}

\begin{figure}[H]
  \centering
      \includegraphics[trim=0.0in 1.0in 0.0in 2.0in,clip,width=0.90\textwidth]{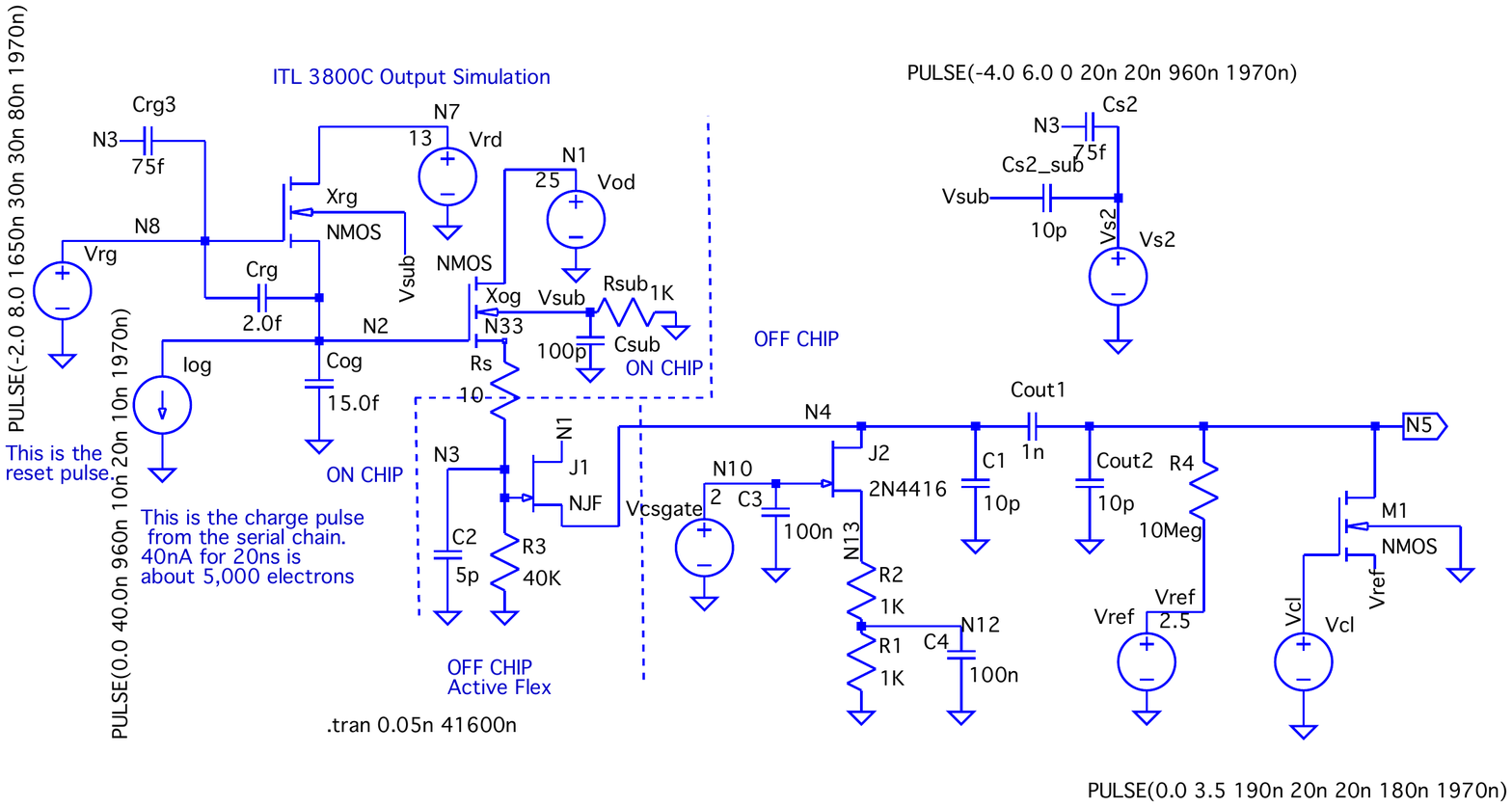}\\
  \begin{minipage}{0.49\textwidth}
    \begin{center}
      \includegraphics[trim=0.0in 0.0in 0.0in 0.0in,clip,width=0.99\textwidth]{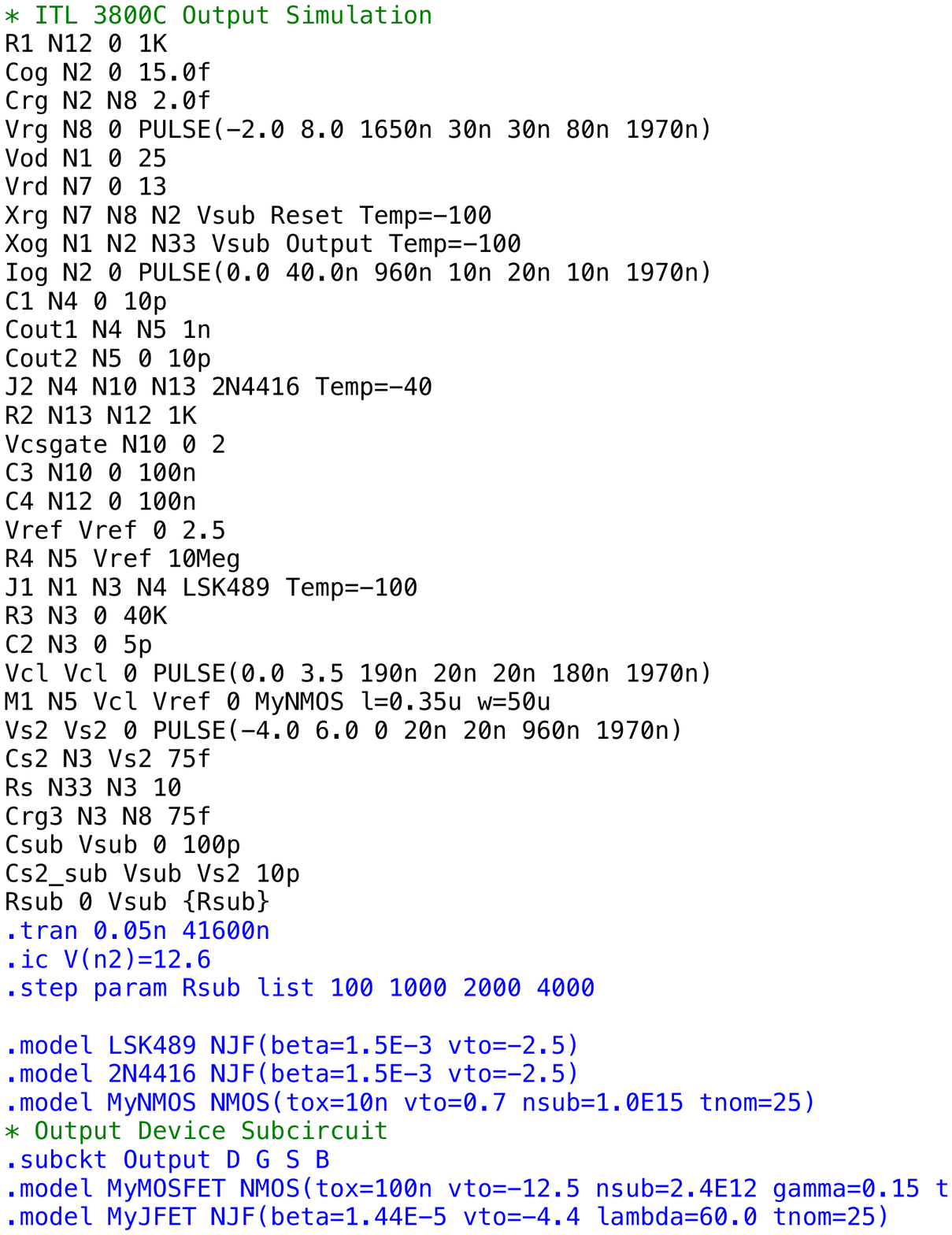}
    \end{center}
  \end{minipage}
  \begin{minipage}{0.49\textwidth}
    \begin{center}
      \includegraphics[trim=0.0in 0.0in 0.0in 0.0in,clip,width=0.99\textwidth]{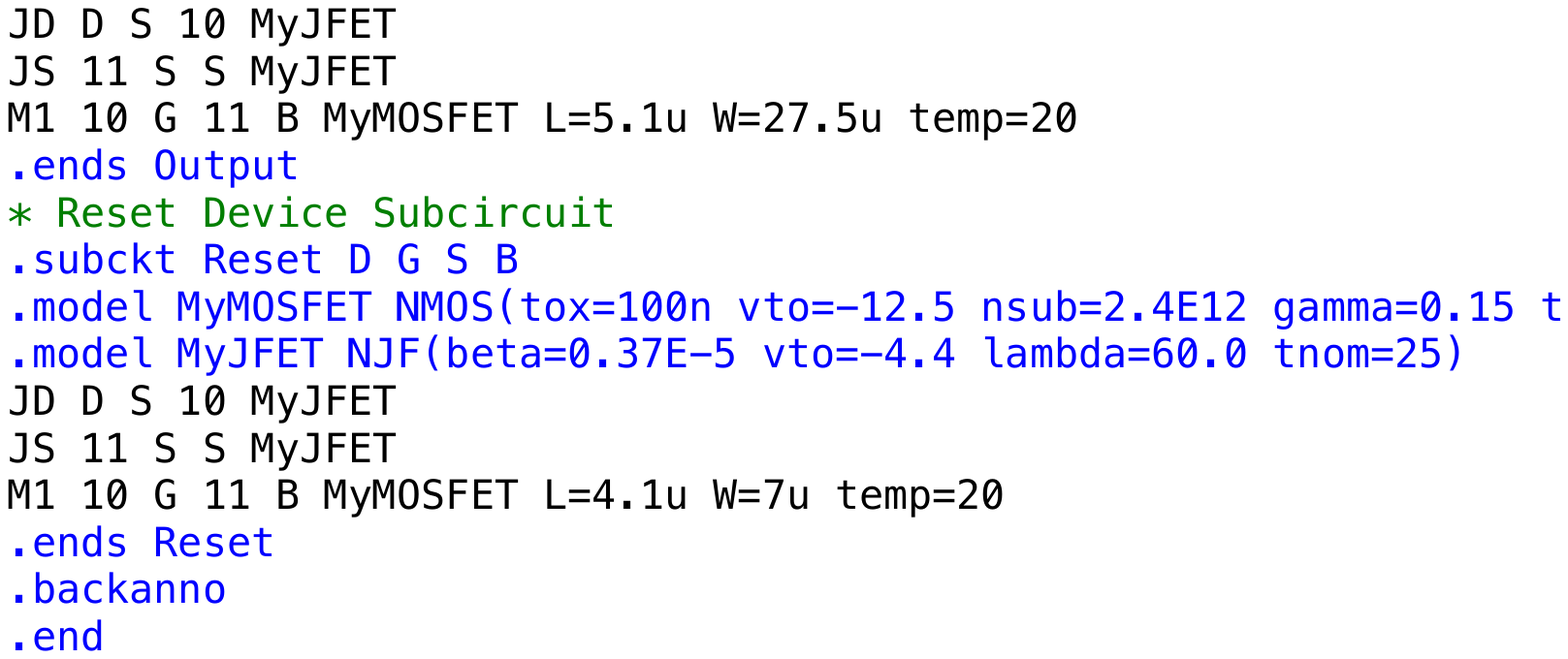}
    \end{center}
  \end{minipage}
  % Trim is Left Bottom Right Top
  \caption{Schematic and netlist used in the simulation shown in Figure \ref{SPICE_RTM8}.}
  \label{Schem_Net2}
\end{figure}

A more detailed simulation was carried out to understand anomalous output characteristics seen in the video chain used in a portion of the LSST focal plane.  Some amplifiers on some sensors display much slower output recovery than others, as seen in Figure \ref{SPICE_RTM8}.  Using the schematic and netlist shown in Figure \ref{Schem_Net2}, we were able to reproduce this behavior by adding substrate resistance to the output transistor.  While it is not certain that this is the cause of the problem, we believe that the very lightly doped CCD substrate can lead to high and variable substrate resistance of the output transistor and lead to the observed slow response.

\section{Appendix - STA3800C {\carlito Poisson\_CCD} configuration file}
\renewcommand{\thefigure}{A.\arabic{figure}}
\setcounter{figure}{0}
\label{ITL_Poisson_Appendix}
\begin{minipage}{0.49\textwidth}
{\tiny
\begin{verbatim}
#  ------------------------------------------------------------------------------
#  Author: Craig Lage, UC Davis
#  Date: Sep 3, 2015
#
#  Standalone cpp Poisson solver
#
#
# Poisson Solver configuration file
VerboseLevel = 1 
# Poisson solver constants
# These control the numerics of the Poisson solver
# They should not need to be changed unless you test for convergence

w = 1.8 			# Successive Over-Relaxation factor
ncycle = 128			# Number of SOR cycles at finest grid
iterations = 1			# Number of VCycles
#  ------------------------------------------------------------------------------
# Overall setup - these control the size and scale of the simulated volume
ScaleFactor = 2    	   	# Power of 2 that sets the grid size
# ScaleFactor = 1 means grid size is 0.625 micron, 160 grids in the z-direction
# ScaleFactor = 2 cuts grid size by a factor of 2
# ScaleFactor = 4 cuts grid size by a factor of 4, etc.
SensorThickness = 100.0         # Sensor thickness in microns
PixelSizeX = 10.0  	     	# Pixel size in microns in x 
PixelSizeY = 10.0  	     	# Pixel size in microns in y 
GridsPerPixelX = 16		# Number of grids per pixel in x at ScaleFactor = 1
GridsPerPixelY = 16		# Number of grids per pixel in y at ScaleFactor = 1
Nx = 160      			# Number of grids in x at ScaleFactor = 1 (Must be a multiple of 32)
Ny = 160      			# Number of grids in y at ScaleFactor = 1 (Must be a multiple of 32)
Nz = 160			# Number of grids in z at ScaleFactor = 1 (Must be a multiple of 32)
Nzelec = 24			# Number of grids in electron and hole arrays
NZExp = 10.0                    # Non-linear Z-axis slope at z=0
      				# A value of 1.0 makes the z-axis linear
				# A value of 10.0 gives a 10X magnification at z=0
				# A value of 10.0 is recommended.

XBCType = 1			# Set X direction boundary conditions: 0 - Free (Eperp = 0), 1 - Periodic
YBCType = 1			# Set Y direction boundary conditions: 0 - Free (Eperp = 0), 1 - Periodic
SimulationRegionLowerLeft = 5.0 5.0	  # Allows adjustment of X, Y coordinates
#  ------------------------------------------------------------------------------
# Fixed charges and oxides in the silicon
GateOxide = 0.10                 # Gate Oxide thickness in microns
ChannelStopWidth = 1.8 		 # Width of ChannelStop region in microns
FieldOxide = 1.0                 # Field Oxide thickness in microns
FieldOxideTaper = 1.1            # Field Oxide taper width in microns
BackgroundDoping = -2.4E12 	 # Background doping in cm^-3

# Channel Doping: Use the syntax below for a square profile
#ChannelProfile = 0		# 0 = Square profile, N = N Gaussian profiles
#ChannelDoping = 1.0E12		# Doping in cm^-2
#ChannelDepth = 1.0		# Depth in microns

ChannelProfile = 2		# 0 = Square profile, N = N Gaussian profiles
ChannelDose_0 = 9.6E11		# Doping in cm^-2
ChannelPeak_0 = 0.07		# Location of peak below silicon surface in microns
ChannelSigma_0 = 0.15		# Sigma in microns
ChannelDose_1 = 5.4E11		# Doping in cm^-2
ChannelPeak_1 = 0.40		# Location of peak below silicon surface in microns
ChannelSigma_1 = 0.19		# Sigma in microns
ChannelSurfaceCharge = 1.1E12	# Surface charge density in cm^-2

# Channel Stop doping: Use the syntax below for a square profile
#ChannelStopProfile = 0		# 0 = Square profile, N = N Gaussian profiles
#ChannelStopDoping = -2.0E12	# Doping in cm^-2
#ChannelStopDepth = 2.0		# Depth in microns

ChannelStopProfile = 2		# 0 = Square profile, N = N Gaussian profiles
ChannelStopDose_0 = -4.5E12	# Doping in cm^-2
ChannelStopPeak_0 = 0.45	# Location of peak below silicon surface in microns
ChannelStopSigma_0 = 0.39	# Sigma in microns
ChannelStopDose_1 = -0.8E12	# Doping in cm^-2
ChannelStopPeak_1 = 1.1		# Location of peak below silicon surface in microns
ChannelStopSigma_1 = 0.37	# Sigma in microns
ChannelStopSurfaceCharge = 0.0	# Surface charge density in cm^-2
ChannelStopSideDiff = 1.1       # Side diffusion in microns
#  ------------------------------------------------------------------------------
# Mobile charge calculation control parameters
ElectronMethod = 2	    	 # Controls electron calculation
	       	 		 # 0 - Leave electrons where they land from tracking
				 # 1 - Set QFe (QFe is always used in Fixed Regions)
				 # 2 - Electron conservation and constant QFe
				 # If 1 is specified, you must provide a *_QFe.dat file, either by
				 # Setting BuildQFeLookup = 1 or by copying a file into the data directory.
#BuildQFeLookup = 1
#NQFe = 81			 # If building QFe lookup, you need to provide at
       				 # least NQFe pixels in the PixelRegion
#QFemin = 10.0
#QFemax = 18.0
qfh = 0.0			 # Controls hole calculation.
      				 # Currently this applies to the whole volume,
				 # unless over-ridden in Fixed Regions
\end{verbatim}
}
\end{minipage}
\begin{minipage}{0.49\textwidth}
{\tiny
\begin{verbatim}
#  ------------------------------------------------------------------------------
# Voltages - these should be self-explanatory
Vbb = -60.0			# Back bias
Vparallel_lo = -8.0		# Parallel gate low voltage
Vparallel_hi = 4.0		# Parallel gate high voltage
NumPhases = 3	  		# Number of clock phases (typically either 3 or 4)
CollectingPhases = 2            # Number of Parallel gates high in collecting region
#  ------------------------------------------------------------------------------
# Pixel Regions
# These allow one to set up one or more regions of regularly spaced pixels.
# Each pixel region will need its extents defined
# Within each pixel region, one can fill multiple collecting wells with arbitrary amounts of charge
NumberofPixelRegions = 1	  	  # 
PixelRegionLowerLeft_0 = 0.0 0.0	  #
PixelRegionUpperRight_0 = 110.0 110.0	  #
NumberofFilledWells_0 = 1		  #
CollectedCharge_0_0 = 	100000		  # Collected charge in e-
FilledPixelCoords_0_0 = 55.0 55.0	  # (x,y) coords of pixel center
#  ------------------------------------------------------------------------------
# Constant Voltage Regions - this allows a number of regions of fixed surface potential
# Each Constant Voltage region will need its extents defined
# Example syntax below
NumberofFixedRegions = 0
#FixedRegionLowerLeft_0 = 0.0 367.0	  # 
#FixedRegionUpperRight_0 = 110.0 430.0	  #
#FixedRegionVoltage_0 = -60.0		  #
#FixedRegionDoping_0 = 0		  # Doping - 0-None; 1-Channel; 2-ChanStop 
#FixedRegionOxide_0 = 2			  # Oxide - 0-None; 1-Channel; 2-ChanStop
#FixedRegionQFe_0 = 100.0		  #
#FixedRegionQFh_0 = -58.0		  #
#FixedRegionBCType_0 = 0		  # Boundary conditions - 0-Fixed voltage; 1-Free (Eperp = 0)
#  ------------------------------------------------------------------------------
# Pixel Boundary Tests - This allows tracing the pixel boundaries and electron paths
PixelBoundaryLowerLeft = 10.0 10.0
PixelBoundaryUpperRight = 100.0 100.0
PixelBoundaryNx = 9	   	      	   # Number of pixels in postage stamp
PixelBoundaryNy = 9	   	      	   # Number of pixels in postage stamp

PixelBoundaryTestType = 1		   # 0 - Run a grid of equally spaced electrons,
		      			   # 1 - Run a random set of electrons with a Gaussian pattern
					   # 2 - Run a random set of electrons inside PixelBoundary
#PixelBoundaryStepSize = 0.2 0.2	   # Needed if PixelBoundaryTestType = 0

# The following parameters are used if PixelBoundaryTestType = 2
Sigmax = 10.0					 # Sigma of incoming light profile
Sigmay = 10.0					 # Sigma of incoming light profile
Xoffset = 0.0					 # Center offset of incoming light profile
Yoffset = 0.0					 # Center offset of incoming light profile
NumSteps = 1					 # Number of steps, each one adding NumElec electrons

NumElec = 0					 # Number of electrons to be traced between field recalculation

CalculateZ0 = 0				   # 0 - don't calculate - Use ElectronZ0
	      				   # 1 - calculate from filter and SED.
#FilterBand = r				   # Filter band from LSST used to calculate Z0
#FilterFile = notebooks/gclef_pdf.dat	   # SED used to calculate Z0
ElectronZ0Fill = 95.0       	      	   # Starting z value of electron for tracking.
ElectronZ0Area = 95.0       	      	   # Starting z value of electron for Area/Vertex finding.

LogEField = 1	 	       	      	   # 0 - don't calculate E-Field, 1 - Calculate and store E-Field
LogPixelPaths = 0			   # 0 - only the final (z~0) point is logged, 1 - Entire path is logged
PixelAreas = 1				   # -1 - Don't calculate areas, N - calculate areas every nth step
NumVertices = 32 			   # Number of vertices per side for the pixel area calculation.
	      				   # Since there are also 4 corners, there will be:
					   # (4 * NumVertices + 4) vertices in each pixel
#  ------------------------------------------------------------------------------
# Electron tracking parameters

CCDTemperature = 173.0			   # Temp in Degrees K.  Used to calculate diffusion steps.

DiffMultiplier = 1.0			   # Used to adjust the amount of diffusion.
       	 				   # A value of 1.0 gives the theoretical amount of diffusion
					   # A value of 0.0 turns off diffusion completely
EquilibrateSteps = 100			   # Number of diffusion steps each electron takes after reaching the bottom,
					   # and before beginning to log the charge.
BottomSteps = 1000			   # Number of diffusion steps each electron takes 
                                           # while logging final charge location				   
NumDiffSteps = 1			   # A speed/accuracy trade-off. A value of 1 uses the theoretical diffusion
	       				   # step.  A higher value takes larger steps. I have done a few tests
					   # but I recommend using a value of 1 unless you test larger values.
SaturationModel = 0			   # Saturation Model 1=On, 0=Off; Experimental!
#  ------------------------------------------------------------------------------
# These control the location and naming of the output
outputfiledir = data/run100K
outputfilebase 	= Pixel
SaveData = 1 				# 0 - Save only Pts data, N - Save all data every Nth step
SaveElec = 1 				# 0 - Save only Pts data, N - Save Elec data every Nth step
SaveMultiGrids = 0			
#  ------------------------------------------------------------------------------
# These control the continuation if you want to save a simuation before it is complete

Continuation = 0			# Use this to continue an existing simulation and read in where you left off
	       				# 0 - No continuation
					# 1 Continue at step LastContinuationStep
LastContinuationStep = 0
\end{verbatim}
}
\end{minipage}

\section{Appendix - E2V CCD250 {\carlito Poisson\_CCD} configuration file}
\renewcommand{\thefigure}{A.\arabic{figure}}
\setcounter{figure}{0}
\label{E2V_Poisson_Appendix}
\begin{minipage}{0.49\textwidth}
{\tiny
\begin{verbatim}
#
#  ------------------------------------------------------------------------------
#  Author: Craig Lage, UC Davis
#  Date: Sep 3, 2015
#
#  Standalone cpp Poisson solver
#
#
# Poisson Solver configuration file

VerboseLevel = 1

# Poisson solver constants
# These control the numerics of the Poisson solver
# They should not need to be changed unless you test for convergence

w = 1.8 			# Successive Over-Relaxation factor
ncycle = 512			# Number of SOR cycles at finest grid
iterations = 1			# Number of VCycles
#  ------------------------------------------------------------------------------
# Overall setup - these control the size and scale of the simulated volume
ScaleFactor = 2
# Power of 2 that sets the grid size
# ScaleFactor = 1 means grid size is 0.625 micron, 160 grids in the z-direction
# ScaleFactor = 2 cuts grid size by a factor of 2
# ScaleFactor = 4 cuts grid size by a factor of 4, etc.
SensorThickness = 100.0         # Sensor thickness in microns
PixelSizeX = 10.0  	     	# Pixel size in microns in x 
PixelSizeY = 10.0  	     	# Pixel size in microns in y 
GridsPerPixelX = 16		# Number of grids per pixel in x at ScaleFactor = 1
GridsPerPixelY = 16		# Number of grids per pixel in y at ScaleFactor = 1
Nx = 160      			# Number of grids in x at ScaleFactor = 1 (Must be a multiple of 32)
Ny = 160      			# Number of grids in y at ScaleFactor = 1 (Must be a multiple of 32)
Nz = 160			# Number of grids in z at ScaleFactor = 1 (Must be a multiple of 32)
Nzelec = 24			# Number of grids in electron and hole arrays
NZExp = 10.0                    # Non-linear Z-axis slope at z=0
      				# A value of 1.0 makes the z-axis linear
				# A value of 10.0 gives a 10X magnification at z=0
				# A value of 10.0 is recommended.

XBCType = 1			# Set X direction boundary conditions: 0 - Free (Eperp = 0), 1 - Periodic
YBCType = 1			# Set Y direction boundary conditions: 0 - Free (Eperp = 0), 1 - Periodic
SimulationRegionLowerLeft = 5.0 5.0	  # Allows adjustment of X, Y coordinates
#  ------------------------------------------------------------------------------
# Fixed charges and oxides in the silicon

GateOxide = 0.134                 # Gate Oxide thickness in microns
ChannelStopWidth = 2.0 		 # Width of ChannelStop region in microns
ChannelStopHeight = 2.0
FieldOxide = 0.134                 # Field Oxide thickness in microns
FieldOxideTaper = 0.0            # Field Oxide taper width in microns
BackgroundDoping = -2.4E12 	 # Background doping in cm^-3

# Channel Doping: Use the syntax below for a square profile
#ChannelProfile = 0		# 0 = Square profile, N = N Gaussian profiles
#ChannelDoping = 1.0E12		# Doping in cm^-2
#ChannelDepth = 1.0		# Depth in microns

ChannelProfile = 1		# 0 = Square profile, N = N Gaussian profiles
ChannelDose_0 = 1.0E12		# Doping in cm^-2
ChannelPeak_0 = 0.05		# Location of peak below silicon surface in microns
ChannelSigma_0 = 0.32		# Sigma in microns

# Channel Stop doping: Use the syntax below for a square profile
#ChannelStopProfile = 0		# 0 = Square profile, N = N Gaussian profiles
#ChannelStopDoping = 0.0		# Doping in cm^-2
#ChannelStopDepth = 0.0		# Depth in microns

ChannelStopProfile = 1		# 0 = Square profile, N = N Gaussian profiles
ChannelStopDose_0 = -6.0E12	# Doping in cm^-2
ChannelStopPeak_0 = 0.15	# Location of peak below silicon surface in microns
ChannelStopSigma_0 = 0.36	# Sigma in microns
ChannelStopSurfaceCharge = 0.0	# Surface charge density in cm^-2
#  ------------------------------------------------------------------------------
# Mobile charge calculation control parameters
ElectronMethod = 2	    	 # Controls electron calculation
	       	 		 # 0 - Leave electrons where they land from tracking
				 # 1 - Set QFe (QFe is always used in Fixed Regions)
				 # 2 - Electron conservation and constant QFe
				 # If 1 is specified, you must provide a *_QFe.dat file, either by
				 # Setting BuildQFeLookup = 1 or by copying a file into the data directory.
BuildQFeLookup = 0
NQFe = 81			 # If building QFe lookup, you need to provide at
       				 # least NQFe pixels in the PixelRegion
QFemin = 19.0
QFemax = 27.0

qfh = -10.0			 # Controls hole calculation.
      				 # Currently this applies to the whole volume,
				 # unless over-ridden in Fixed Regions
\end{verbatim}
}
\end{minipage}
\begin{minipage}{0.49\textwidth}
{\tiny
\begin{verbatim}
#  ------------------------------------------------------------------------------
# Voltages - these should be self-explanatory
Vbb = -70.0			# Back bias
Vparallel_lo = -7.2		# Parallel gate low voltage
Vparallel_hi = 3.5		# Parallel gate high voltage
NumPhases = 4	  		# Number of clock phases (typically either 3 or 4)
CollectingPhases = 2            # Number of Parallel gates high in collecting region
#  ------------------------------------------------------------------------------
# Pixel Regions
# These allow one to set up one or more regions of regularly spaced pixels.
# Each pixel region will need its extents defined
# Within each pixel region, one can fill multiple collecting wells with arbitrary amounts of charge
NumberofPixelRegions = 1	  	  # 1
PixelRegionLowerLeft_0 = 0.0 0.0	  #
PixelRegionUpperRight_0 = 110.0 110.0	  #
NumberofFilledWells_0 = 1		  #
CollectedCharge_0_0 = 	100000		  # Collected charge in e-
FilledPixelCoords_0_0 = 55.0 55.0	  # (x,y) coords of pixel center
#  ------------------------------------------------------------------------------
# Constant Voltage Regions - this allows a number of regions of fixed surface potential
# Each Constant Voltage region will need its extents defined
# Example syntax below
NumberofFixedRegions = 0
#FixedRegionLowerLeft_0 = 0.0 367.0	  # 
#FixedRegionUpperRight_0 = 110.0 430.0	  #
#FixedRegionVoltage_0 = -60.0		  #
#FixedRegionDoping_0 = 0		  # Doping - 0-None; 1-Channel; 2-ChanStop 
#FixedRegionOxide_0 = 2			  # Oxide - 0-None; 1-Channel; 2-ChanStop
#FixedRegionQFe_0 = 100.0		  #
#FixedRegionQFh_0 = -58.0		  #
#FixedRegionBCType_0 = 0		  # Boundary conditions - 0-Fixed voltage; 1-Free (Eperp = 0)
#  ------------------------------------------------------------------------------
# Pixel Boundary Tests - This allows tracing the pixel boundaries and electron paths

PixelBoundaryLowerLeft = 10.0 10.0
PixelBoundaryUpperRight = 100.0 100.0
PixelBoundaryNx = 9	   	      	   # Number of pixels in postage stamp
PixelBoundaryNy = 9	   	      	   # Number of pixels in postage stamp

PixelBoundaryTestType = 1		   # 0 - Run a grid of equally spaced electrons,
		      			   # 1 - Run a random set of electrons with a Gaussian pattern
					   # 2 - Run a random set of electrons inside PixelBoundary
#PixelBoundaryStepSize = 0.2 0.2	   # Needed if PixelBoundaryTestType = 0

# The following parameters are used if PixelBoundaryTestType = 2
Sigmax = 10.0					 # Sigma of incoming light profile
Sigmay = 10.0					 # Sigma of incoming light profile
Xoffset = 0.0					 # Center offset of incoming light profile
Yoffset = 0.0					 # Center offset of incoming light profile
NumSteps = 1					 # Number of steps, each one adding NumElec electrons

NumElec = 0					 # Number of electrons to be traced between field recalculation

CalculateZ0 = 0				   # 0 - don't calculate - Use ElectronZ0
	      				   # 1 - calculate from filter and SED. 
#FilterBand = r				   # Filter band from LSST used to calculate Z0
#FilterFile = notebooks/gclef_pdf.dat	   # SED used to calculate Z0
ElectronZ0Fill = 95.0       	      	   # Starting z value of electron for tracking.
ElectronZ0Area = 95.0       	      	   # Starting z value of electron for Area/Vertex finding.

LogEField = 1	 	       	      	   # 0 - don't calculate E-Field, 1 - Calculate and store E-Field
LogPixelPaths = 0			   # 0 - only the final (z~0) point is logged, 1 - Entire path is logged
PixelAreas = 1				   # -1 - Don't calculate areas, N - calculate areas every nth step
NumVertices = 32 			   # Number of vertices per side for the pixel area calculation.
	      				   # Since there are also 4 corners, there will be:
					   # (4 * NumVertices + 4) vertices in each pixel

#  ------------------------------------------------------------------------------
# Electron tracking parameters

CCDTemperature = 173.0			   # Temp in Degrees K.  Used to calculate diffusion steps.

DiffMultiplier = 1.0			   # Used to adjust the amount of diffusion.
       	 				   # A value of 1.0 gives the theoretical amount of diffusion
					   # A value of 0.0 turns off diffusion completely
EquilibrateSteps = 10			   # Number of diffusion steps each electron takes after reaching the bottom,
					   # and before beginning to log the charge.
BottomSteps = 100			   # Number of diffusion steps each electron takes while 
                                           # logging final charge location					   
NumDiffSteps = 1			   # A speed/accuracy trade-off. A value of 1 uses the theoretical diffusion
	       				   # step.  A higher value takes larger steps. I have done a few tests
					   # but I recommend using a value of 1 unless you test larger values.
SaturationModel = 0			   # Saturation Model 1=On, 0=Off; Experimental!
#  ------------------------------------------------------------------------------
# These control the location and naming of the output
outputfiledir = data/run100K_center_6B
outputfilebase 	= Pixel
SaveData = 1 				# 0 - Save only Pts data, N - Save all data every Nth step
SaveElec = 1 				# 0 - Save only Pts data, N - Save Elec data every Nth step
SaveMultiGrids = 0			
#  ------------------------------------------------------------------------------
# These control the continuation if you want to save a simuation before it is complete

Continuation = 0			# Use this to continue an existing simulation and read in where you left off
	       				# 0 - No continuation
					# 1 Continue at step LastContinuationStep
LastContinuationStep = 0
\end{verbatim}
}
\end{minipage}

\end{document}